\documentclass[twocolumn,oneside,openany]{aastex631}

\usepackage[normalem]{ulem}
\usepackage{color, colortbl}
\usepackage{dirtytalk}
\usepackage [english]{babel}
\usepackage [autostyle, english = american]{csquotes}
\MakeOuterQuote{"}

\begin{document}

\title{On the properties of X-ray corona in Seyfert 1 galaxies}

\author[0000-0002-7825-1526]{Indrani Pal}
\affiliation{Indian Institute of Astrophysics, Bangalore, 560034, Karnataka, India}
\affiliation{Pondicherry University, R.V. Nagar, Kalapet, 605014, Puducherry, India}
\affiliation{Department of Physics and Astronomy, Clemson University, Kinard Lab of Physics, Clemson, SC 29634, USA}

\author{Anju A.}
\affiliation{Department of Physical Sciences, Indian Institute of Science Education And Research Kolkata, Mohanpur, Nadia-741246, West Bengal, India}

\author{H. Sreehari}
\affiliation{Indian Institute of Astrophysics, Bangalore, 560034, Karnataka, India}
\affiliation{Department of Physics, Faculty of Natural Sciences, University of Haifa, Mount Carmel, Haifa 3498838, Israel}

\author{Gitika Rameshan}
\affiliation{Cochin University of Science and Technology, South Kalamassery, Kochi, Kerala, 682022, India}

\author{C. S. Stalin}
\affiliation{Indian Institute of Astrophysics, Bangalore, 560034, Karnataka, India}

\author{Claudio Ricci}
\affiliation{Instituto de Estudios Astrof\'{\i}sicos, Facultad de Ingenier\'{\i}a y Ciencias, Universidad Diego Portales, Avenida Ejercito Libertador 441, Santiago, Chile}
\affiliation{Kavli Institute for Astronomy and Astrophysics, Peking University, Beijing 100871, China}

\author[0000-0001-5544-0749]{S. Marchesi}
\affiliation{Dipartimento di Fisica e Astronomia (DIFA), Università di Bologna, via Gobetti 93/2, I-40129 Bologna, Italy}
\affiliation{Department of Physics and Astronomy, Clemson University, Kinard Lab of Physics, Clemson, SC 29634, USA}
\affiliation{INAF - Osservatorio di Astrofisica e Scienza dello Spazio di Bologna, Via Piero Gobetti, 93/3, 40129, Bologna, Italy}



\begin{abstract}
We carried out a uniform and systematic analysis of a sample of 112 nearby bright Seyfert 1 type AGN, the observations of which were carried out by the {\it Nuclear Spectroscopic Telescope Array (NuSTAR)} between August 2013 and May 2022. The main goal of this analysis is to investigate the nature of the X-ray corona in Seyfert 1 galaxies. By fitting a physical model to the {\it NuSTAR} spectra, we were able to constrain the high-energy cut-off ($\rm{E_{cut}}$) for 73 sources in our sample. To estimate the temperature of the corona ($\rm{kT_{e}}$) in our sample of 112 sources, we used the Comptonization model to fit their spectra. We could constrain $\rm{kT_{e}}$ in 42 sources. We found a strong positive correlation between $\rm{E_{cut}}$ and $\rm{kT_{e}}$ with most of the sources lying above the empirical approximation of $\rm{E_{cut}}$ = 2$-$3 $\rm{kT_{e}}$. We investigated for possible correlations between various properties of the corona obtained from physical model fits to the observed spectra and between various coronal parameters and physical properties of the sources such as Eddington ratio and black hole mass. We found (a) a strong correlation between $\rm{E_{cut}}$ and the photon index and (b) a significant negative correlation between $\rm{kT_{e}}$ and the optical depth. From detailed statistical analysis of the correlation of coronal parameters with the Eddington ratio and black hole mass, we found no significant correlation. The correlations observed in this study indicate that an optically thin corona is needed to sustain a hotter corona with a steeper spectrum.
\end{abstract}



\section{Introduction} \label{sec:intro}

Most massive galaxies host supermassive black holes (SMBHs) at their centres with masses (M$_{BH}$) of the order of 10$^5$ to 10$^{10}$ M$_{\odot}$. These SMBHs power active galactic nuclei (AGN) by accretion of matter from their surroundings \citep{1964ApJ...140..796S,1969Natur.223..690L,1973A&A....24..337S, 1983MNRAS.205..593G,2008ARA&A..46..475H}. The observed optical, ultra-violet (UV) radiation from these accretion-powered systems is believed to be thermal emission from the standard optically thick, geometrically thin accretion disk \citep{1978Natur.272..706S,1982ApJ...254...22M,1989ApJ...346...68S} that surrounds the SMBHs.  These AGN are also sources of intense X-ray emission \citep{1978MNRAS.183..129E, 1993ARA&A..31..717M}. The X-ray emission in the radio-quiet category of  AGN is believed to originate from a compact region that contains hot electrons ($T_e \sim 10^{8-9} K$) called the corona situated close to the vicinity of the SMBH. Observations indicate that the corona is physically compact with size scales of the order of 3 $-$ 10 $R_G$ \citep{2005MNRAS.359.1469M, Risaliti_2005}, where $R_G$ is the gravitational radius defined as $R_G = GM_{BH}/c^2$, here, G is the gravitational constant and c is the speed of light.  The hot electrons in the corona, inverse Compton scatter the optical or UV thermal photons from the geometrically thin, optically thick accretion disk, thereby producing X-ray emission \citep{1991ApJ...380L..51H,1993ApJ...413..507H}. The emergent X-ray spectrum follows a power law with the high energy roll off of the form $\rm{N(E)} \propto \rm{E^{-\Gamma}exp(-E}/\rm{E_{cut}})$, where  $\Gamma$ is the power law photon index and $\rm{E_{cut}}$ is the high energy cut-off \citep{1980A&A....86..121S}. In this paradigm, expecting a connection between the accretion disk and the X-ray-emitting corona is natural. One piece of observational evidence for this accretion disk corona connection is the observed positive correlation \citep{2008ApJ...682...81S,2009ApJ...700L...6R,2012MNRAS.425..907J, 2021ApJ...910..103L,2022arXiv221206183T} between $\Gamma$ and the mass-normalized accretion rate usually represented by the Eddington ratio ($\rm{\lambda_{Edd}}$ = L$_{\rm{Bol}}$/L$_{\rm{Edd}}$). Here, L$_{\rm{Bol}}$ is the bolometric luminosity and L$_{\rm{Edd}}$ is the Eddington luminosity defined as L$_{\rm{Edd}}$ = 1.3 $\times$ $10^{38}$ $\rm{M_{BH}}/M_{\odot}$ erg s$^{-1}$. A possible explanation for this observed correlation is that at a higher $\rm{\lambda_{Edd}}$ the increased optical, UV photons from the accretion disk can lead to a more effective cooling of the corona, thereby leading to a decrease in the temperature of the corona ($\rm{kT_{e}}$) and larger $\Gamma$ or softening of the X-ray spectrum. Recently, \cite{2018MNRAS.480.1819R} proposed that another explanation for this is the pair thermostat, due to the changes in temperature across the compactness$-$temperature ($l-\theta$) plane, and they could successfully reproduce the slope of the $\Gamma$$-$$\rm{\lambda_{Edd}}$ correlation.

According to Comptonization models, for a corona with slab geometry, $\rm{E_{cut}}$ is related to the temperature of the corona as $\rm{E_{cut}}$ = 2$-$3 $\rm{kT_{e}}$ for optically thin and thick plasma respectively \citep{2001ApJ...556..716P}. However, according to \cite{2014ApJ...783..106L}, the relation between  $\rm{E_{cut}}$ and $\rm{kT_{e}}$ cannot be simple in the case of the non-static corona. Also, \cite{2019A&A...630A.131M} have shown that the relation of $\rm{E_{cut}}$ = 2$-$3 $\rm{kT_{e}}$ is only valid for low values of $\rm{kT_{e}}$ and $\tau$. Recently, for the source MR 2251$-$178, \cite{2022A&A...662A..78P} found $\rm{E_{cut}}$ = 4.84 $\pm$ 0.11 $\rm{kT_{e}}$, which deviates from the generally considered relation between  $\rm{E_{cut}}$ and $\rm{kT_{e}}$ \citep{2001ApJ...556..716P}. Also, $\Gamma$ is expected to depend on various parameters of the corona, such as its temperature $\rm{kT_{e}}$, the optical depth ($\tau$) as well as the seed photon temperature.  To understand the properties of AGN, it is important to have better constraints on the coronal parameters of AGN that characterise the X-ray emission, such as $\Gamma$ and $\rm{kT_{e}}$.

Earlier studies on the determination of the temperature of the corona in Seyfert galaxies used data from high-energy instruments such as the  CGRO \citep{2000ApJ...542..703Z,1997ApJ...482..173J}, {\it BeppoSAX} \citep{2000ApJ...536..718N,refId03}, {\it INTEGRAL} \citep{2014ApJ...782L..25M,2010MNRAS.408.1851L,2016MNRAS.458.2454L}, {\it Swift-BAT} \citep{2013ApJ...770L..37V, Ricci_2017, 2018MNRAS.480.1819R} and {\it Suzaku} \citep{2011ApJ...738...70T}. These studies have found that in Seyfert galaxies, the coronal temperature shows a wide range, with the values of $\rm{E_{cut}}$ ranging from 50$-$500 keV. These less sensitive observations were, however, limited to nearby bright Seyfert galaxies.  Increased interest in studies on the hard X-ray spectra of AGN, as well as the determination of its coronal temperature, happened after the launch of the {\it Nuclear Spectroscopic Telescope Array} ({\it NuSTAR}; \citealt{2013ApJ...770..103H}) in the year 2012, due to its wide spectral coverage of 3$-$79 keV and its high sensitivity beyond 10 keV. Since its launch, values of the temperature of the corona are known for many AGN, but most of those studies are restricted to the determination of $\rm{E_{cut}}$. Also, data from {\it NuSTAR }have led to the finding of the variation in  $\rm{kT_{e}}$ \citep{2020MNRAS.492.3041B,2021ApJ...921...46B,2021MNRAS.502...80K,2022A&A...662A..78P,2023MNRAS.518.2529P} as well as $\rm{E_{cut}}$ \citep{2017ApJ...836....2Z,2014ApJ...794...62B,2016MNRAS.463..382U,refId05,2016MNRAS.456.2722K,2018ApJ...863...71Z,2021MNRAS.502...80K}. 

In recent years, there have been a few studies on characterising the temperature of the corona ($\rm{E_{cut}}$ or $\rm{kT_{e}}$) in samples of AGN \citep{2018ApJ...866..124K, refId0,2019MNRAS.484.2735M,2019MNRAS.484.5113R,2020ApJ...905...41B,2020A&A...640A..31P, Kang_2020,2021MNRAS.506.4960H,refId01,2022ApJ...927...42K, Kang_2022, 2023MNRAS.518.2529P}. Most of these studies focused on the determination of $\rm{E_{cut}}$ from phenomenological model fits to the observed X-ray spectra. Though $\rm{E_{cut}}$ can serve as a good proxy for $\rm{kT_{e}}$, the recent findings of deviation from the $\rm{E_{cut}}$ = 2$-$3 $\rm{kT_{e}}$ in few sources, have necessitated the determination of $\rm{kT_{e}}$ in AGN based on physical model fit to the observed X-ray spectra. The literature also contains results on $\rm{E_{cut}}$/$\rm{kT_{e}}$ values and their correlation with other physical parameters. However, these correlations vary between studies, possibly due to small sample sizes and large error bars in the $\rm{E_{cut}}$/$\rm{kT_{e}}$ measurements. Therefore, it is crucial to increase the sample size, analyze data consistently, and explore various correlations. To address these, we carefully selected type 1 sources with good signal-to-noise spectra from the {\it NuSTAR} archive and homogeneously conducted the needed analysis with a larger sample size than before.

In this work, we carried out an analysis of 112 Seyfert\,1 type AGN to determine $\rm{E_{cut}}$ based on physical model fits to the {\it NuSTAR} data. Of these 112 sources, we could constrain $\rm{E_{cut}}$ in 73 sources. Further, physical model fits were carried out on the 112 sources to constrain $\rm{kT_{e}}$. We could constrain $\rm{kT_{e}}$ in 42 sources. We investigated the correlation between different physical parameters obtained from the physical model fits. The selection of our sample of sources and data reduction are given in Section 2. We describe in Section 3 the model fits carried out on the data; the results are given in Sections 4 and 5, and a comparison of our findings on $\rm{E_{cut}}$ and $\rm{kT_{e}}$ with those found from the literature are given in Section 6, followed by the discussion and the summary in the final two sections. In this work we adopted the cosmological parameters of $H_0$ = 70 km sec$^{-1}$ $\rm{Mpc^{-1}}$, $\Omega_M$ = 0.3 and $\Omega_{\lambda}$ = 0.7. All the quoted uncertainties in the derived parameters were calculated at the 90 per cent confidence level.

\begin{figure}
\hspace{-0.7 cm}
\includegraphics[scale=0.60]{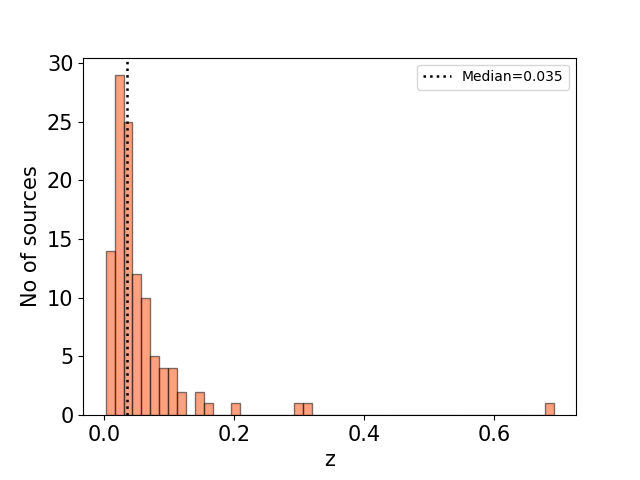}
\caption{The redshift distribution of the sources. The vertical dotted line is the median of the distribution (z=0.035).
} 
\label{figure-1}       
\end{figure}

\begin{figure*}
\hbox{
     \includegraphics[scale=0.55]{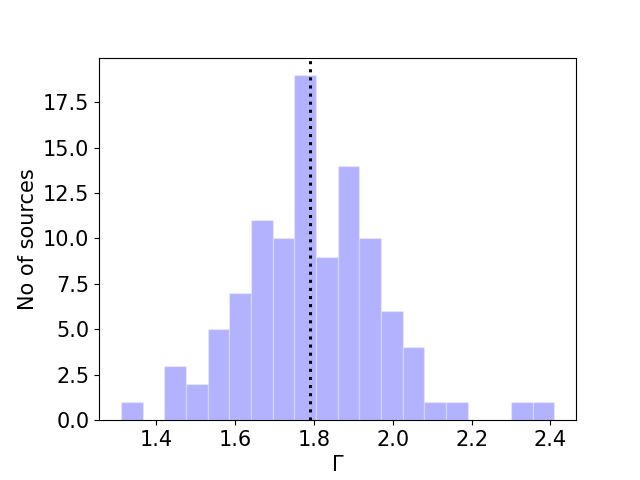}
     \includegraphics[scale=0.55]{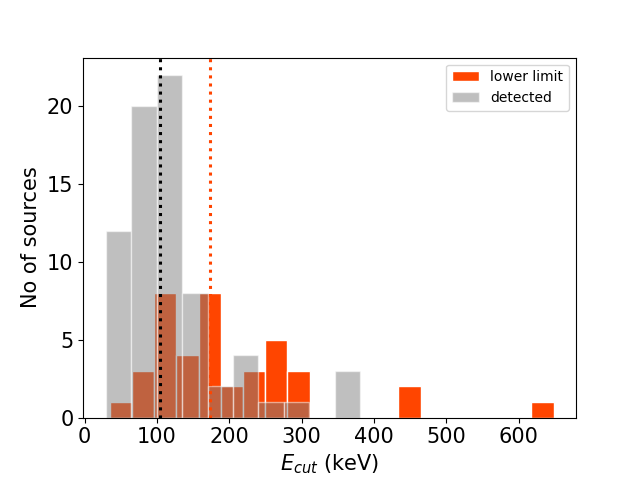}
     }
\caption{Distribution of $\Gamma$ (left panel) and distribution of $\rm{E_{cut}}$ (right panel) obtained from Model$-$1 fit to all 112 source spectra. The vertical dotted lines are the median of the distributions.}
\label{figure-2}       
\end{figure*}

\section{Sample selection and Data reduction}
\subsection{Sample selection}
Our sample of sources for this study was selected from the 
{\it NuSTAR} Master Catalog\footnote{\href{https://heasarc.gsfc.nasa.gov/W3Browse/nustar/numaster.html}{https://heasarc.gsfc.nasa.gov/W3Browse/nustar/numaster.html}}. We examined publicly accessible data for Seyfert galaxies sourced either from the {\it Swift-BAT} 105-month catalogue \citep{Oh_2018} or from the NASA/IPAC Extragalactic Database (NED) during the period spanning August 2013 to May 2022. We found a total of 850 Seyfert galaxies. We selected only Seyfert 1 galaxies with a net count rate greater than  0.1 counts/sec in the 3$-$79 keV band to have a sufficiently good signal-to-noise ratio spectrum for model fitting. Adopting the above-mentioned criteria, we arrived at a final sample of 130 Seyfert 1 galaxies spanning the redshift interval of 0.002 $<$ $z$ $<$  0.692. Of these 130 sources, around 90 per cent of the sources were studied in \cite{Ricci_2017}. \cite{Ricci_2017} carried out the broadband (0.3$-$150 keV) X-ray spectroscopic analysis of {\it Swift-BAT} selected sources by combining {\it XMM-Newton}, {\it Swift/XRT}, {\it ASCA}, {\it Chandra}, and {\it Suzaku} observations in the soft X-ray band with 70-month averaged {\it Swift/BAT} data. Based on the value of the line of sight column densities ($\rm{N_{H}}$) required in the absorption power law fit, 18 sources were classified as obscured AGN ($10^{22}$ $\leq$ ($\rm{N_{H}}$ $cm^{-2}$)$<$ $10^{24}$) in \cite{Ricci_2017}. For this study, we selected the 112 unobscured nearby AGN with a median redshift of 0.035.  We show in Fig. \ref{figure-1} the redshift distribution for our sample of sources. The redshifts are taken from SIMBAD\footnote{\href{http://simbad.cds.unistra.fr/simbad/}{http://simbad.cds.unistra.fr/simbad/}}. The full list of the Seyfert 1 galaxies and their {\it NuSTAR} observational details are given in Table \ref{table-1}. Among 112, about 50\% of sources were observed more than once by {\it NuSTAR}. The observations with the highest exposure were chosen for this study to ensure good signal-to-noise ratio spectra.

\begin{figure}
\hspace{-0.7 cm}
\includegraphics[scale=0.40]{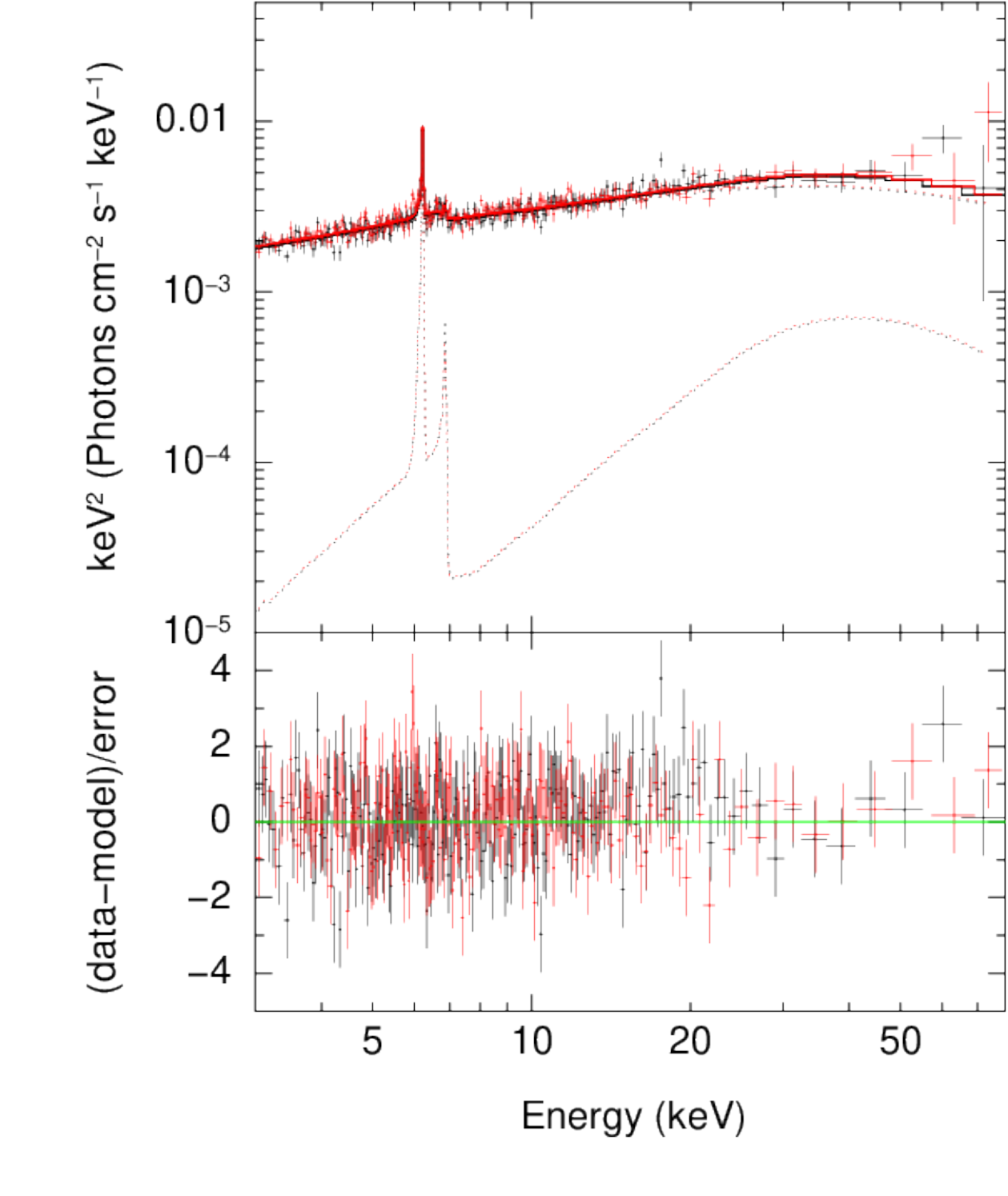}
\caption{The best-fit unfolded FPMA/FPMB spectra of Mrk 279 (upper panel) with data to model residue of the Model$-$1c fit to the source spectra (lower panel).} \label{figure-4}     
\end{figure}

\subsection{Data reduction}
For the 112 sources, we carried out the reduction of the raw event data taken from the HEASARC archive \footnote{\href{https://heasarc.gsfc.nasa.gov/db- perl/W3Browse/w3browse.pl}{https://heasarc.gsfc.nasa.gov/db- perl/W3Browse/w3browse.pl}}, using the standard {\it NuSTAR} data reduction software NuSTARDAS\footnote{\href{https://heasarc.gsfc.nasa.gov/docs/nustar/analysis/nustar swguide.pdf}{https://heasarc.gsfc.nasa.gov/docs/nustar/analysis/nustar swguide.pdf}} v1.9.3 distributed by HEASARC within HEASoft v6.26.1. We generated the calibrated and cleaned event files using the {\tt nupipeline} task and the instrument responses taken from the {\it NuSTAR} calibration database (CALDB release 20190607). To exclude the periods of elevated background, we selected the filtering options SAACALC=2, SAAMODE=OPTIMIZED and TENTACLE=YES to consider the passage of the satellite through the South Atlantic Anomaly (SAA). The source regions for the 112 Seyferts were extracted using circular radii between $30''$ $-$ $70''$ to maximize the S/N, depending on the source. A source-free circular area of the same radius on the same chip was selected to extract the background counts. All the science products, including energy spectra, response matrix files (RMFs) and auxiliary response files (ARFs), were generated using the task {\tt nuproducts} for both the focal plane modules FPMA and FPMB. For spectral analysis, we fitted the background subtracted spectra from FPMA and FPMB simultaneously using XSPEC version 12.10.1 \citep{1996ASPC..101...17A}, allowing the cross normalization factor to vary freely during spectral fits. The spectra were binned to have minimum counts of 20 per spectral energy bin. We note that, for faint sources, the binning of 20 counts/bin could be insufficient at the high energy end with E $>$ 50 keV for the $\chi^2$ statistics to be applicable. Also, it is likely that the choice of binning can have some effect on the derived $\rm{E_{cut}}$ or $\rm{kT_{e}}$ values. To verify this, we identified the 10 faintest sources in our sample and redid the analysis using a binning of 50 counts/bin. For those faint sources, we found that the $\rm{E_{cut}}$ values obtained using a binning of 50 counts/bin agree within errors to those obtained with a binning of 20 counts/bin. Therefore, it is likely that the binning choice adopted in this work has a negligible effect on the derived values of $\rm{E_{cut}}$ and/or $\rm{kT_{e}}$. To get an estimate of the model parameters that best describe the observed data, we used the chi-square ($\chi^2$) statistics, and for calculating the errors in the model parameters, we used the $\chi^2$ = 2.71 criterion, which is equivalent to the 90 per cent confidence range in XSPEC.

\begin{figure*}
\hbox{
     \includegraphics[scale=0.55]{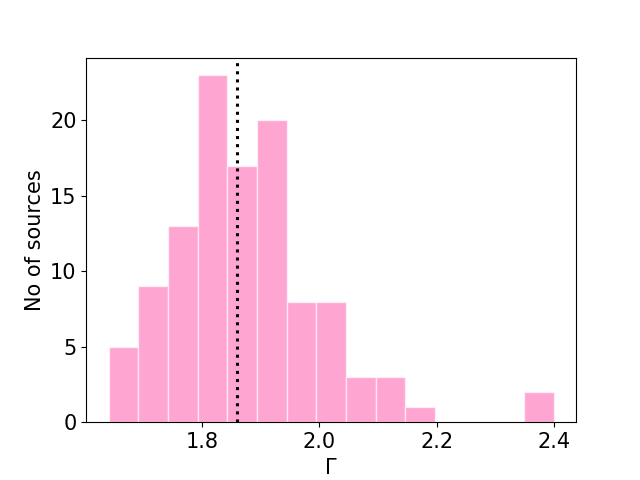}
     \includegraphics[scale=0.55]{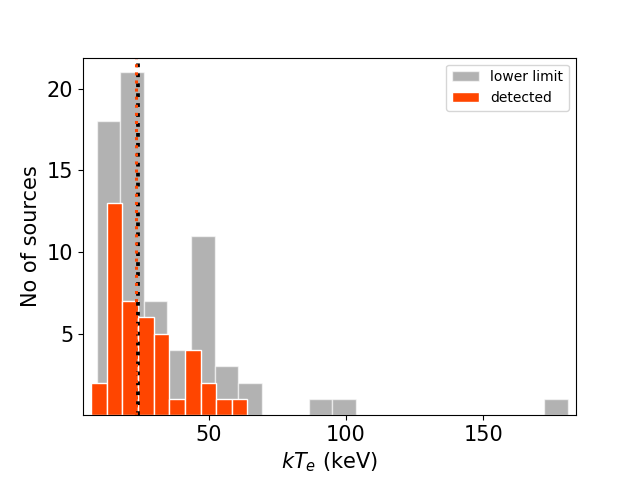}
     }
\caption{Distribution of $\Gamma$ (left panel) and $\rm{kT_{e}}$ (right panel) from Model$-$2 fit for 112 sources. The vertical dotted lines are the median of the distributions.} 
\label{figure-5}
\end{figure*}

\begin{figure*}
\hbox{
     \includegraphics[scale=0.55]{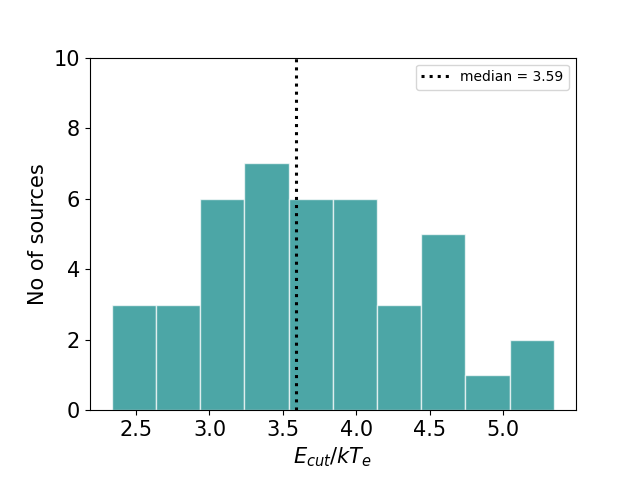}
     \includegraphics[scale=0.55]{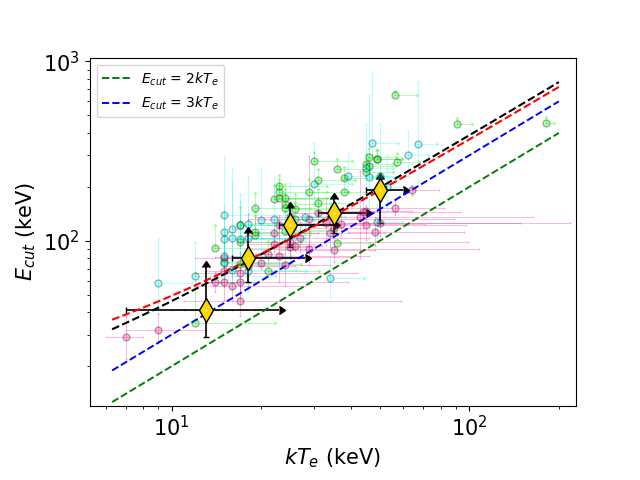}
     }
\caption{Distribution of the ratio between $\rm{E_{cut}}$ to $\rm{kT_{e}}$ for the 42 sources with $\rm{kT_{e}}$ measurements (left panel). The vertical dot line is the median of the ratio. The variation of $\rm{E_{cut}}$ against $\rm{kT_{e}}$ (right panel). Here, the green dashed line shows the $\rm{E_{cut}}$ = 2 $\rm{kT_{e}}$ relation, the blue dashed line shows the $\rm{E_{cut}}$ = 3 $\rm{kT_{e}}$ relation. The black dashed line is the linear least squares fit to the median values of $\rm{E_{cut}}$ (yellow diamond points) obtained from the KMPL survival analysis in each $\rm{kT_{e}}$ bins. The best fit is $\rm{E_{cut}}$ = (3.80$\pm$0.53)$\rm{kT_{e}}$ + (8.15$\pm$16.51). The red dashed line is the linear relation obtained from the statistical analysis package \textit{ASURV}. The pink dots with error bars are the constrained values, and the green dots represent the censored values of $\rm{E_{cut}}$ and $\rm{kT_{e}}$; the cyan dots denote the constrained values of $\rm{E_{cut}}$ and lower limits of $\rm{kT_{e}}$.}
\label{figure-6} 
\end{figure*}

\begin{figure*}
\hbox{
     \includegraphics[scale=0.55]{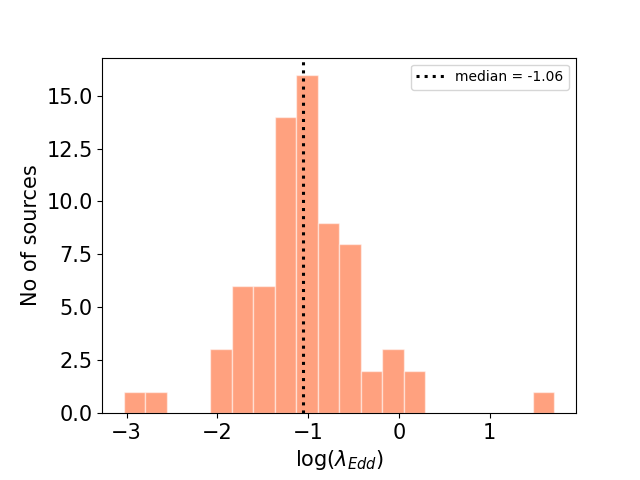}
     \includegraphics[scale=0.55]{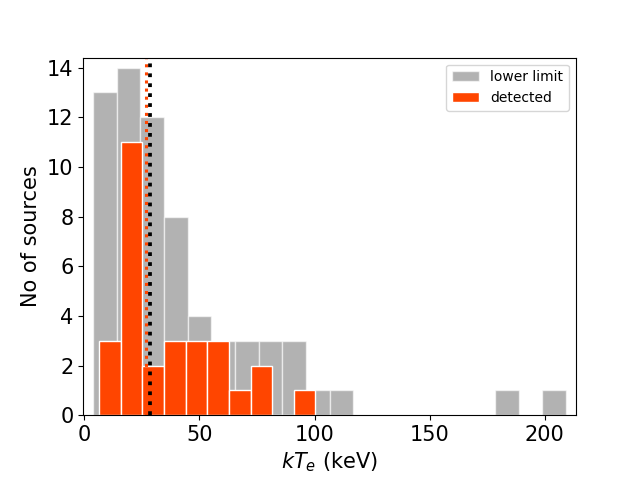}
     }
\caption{Left panel: Eddington ratio distribution for the 109 sources for which we could find the black hole mass from the literature. Right panel: Distribution of both constrained and censored values of $\rm{kT_{e}}$ as calculated using Equation \ref{equation1} from the $\rm{E_{cut}}$ values taken from \cite{2018MNRAS.480.1819R} for 96 sources which are common between this work and \cite{2018MNRAS.480.1819R}. The vertical dotted lines are the median of the distributions.}
\label{figure-3}       
\end{figure*}

\begin{figure*}
\hbox{
\includegraphics[scale=0.58]{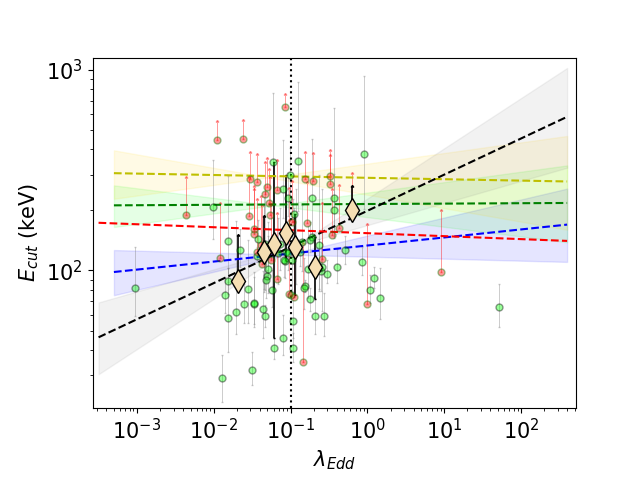}
\includegraphics[scale=0.58]{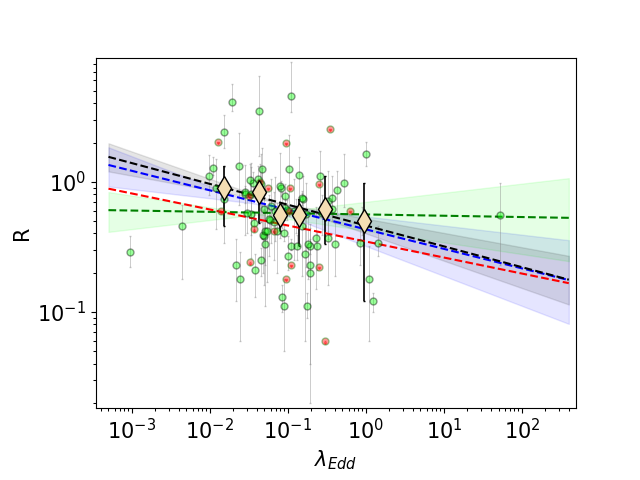}
}
\caption{The relation between $\rm{E_{cut}}$-$\lambda_{Edd}$ (left panel) and R-$\lambda_{Edd}$ (right panel). In the plots, the green dots denote the constrained values, and the red dots represent the lower/upper limits of the dependent parameters. In each panel, the blue dashed line is the linear fit to the constrained values, and the blue-shaded region indicates the errors in the slope and the intercept to the linear fit (Method I). The beige diamond points represent the median values of $\rm{E_{cut}}$ and R in each bin of $\lambda_{Edd}$ obtained from the survival analysis using the KMPL approach. The black dashed line and the black shaded region are the linear fit and the errors in the fit parameters to the median values (Method II). The red dashed line represents the linear fit using the Schmitt method obtained using the package \textit{ASURV}.} The green dashed line and the shaded region represent the linear fit and the errors in the slope and the intercept for the linear relation fit between two parameters considering both uncensored (including the corresponding asymmetric errors) and censored (upper bound of $\rm{E_{cut}}$ = 500 keV; lower bound of R = 0.01) values (Method III) in each panel. The yellow dashed line and the shaded region in the first panel are the linear fit, and the errors in the fit parameters are similar to the green shaded region but considering the upper bound of $\rm{E_{cut}}$ to be 1000 keV (Method IV).
\label{figure-19} 
\end{figure*}

\section{Spectral Analysis}
We carried out a detailed spectral analysis of the {\it NuSTAR} data in the energy range of 3$-$79 keV for the 112 sources, a few of which also have soft X-ray observations. Since these are unobscured Seyfert 1 galaxies, we do not have degeneracies between $\rm{N_{H}}$ and continuum parameters that have been found in obscured Seyfert 2 galaxies \citep{2018ApJ...854...49M}. Therefore, we chose to fit the {\it NuSTAR} data alone, and we do not expect our results to be significantly affected by the lack of information at energies $<$ 3 keV. In the past, too, a similar approach has been followed in several studies \citep{refId0, 2018ApJ...866..124K, 2020ApJ...905...41B, 2020MNRAS.495.3373E, Kang_2020, refId01} aimed at characterising the corona. For the completeness of our study, we compared our findings with those found in the literature where $\rm{E_{cut}}$/$\rm{kT_{e}}$ were obtained with and without the soft X-ray coverage (see Table \ref{table-2}).

We used the following two models
\begin{itemize}
\item Model$-$1: \textsc{const $\times$ TBabs $\times$ zTBabs $\times$ (xillver/relxill/(relxill+xillver))} 
\item Model$-$2: \textsc{const $\times$ TBabs $\times$ zTBabs $\times$ (xillverCP/relxillCP/(relxillCP+xillverCP))}
\end{itemize}
In both models, the Fe-K$\alpha$ line present in the source spectrum would be self-consistently taken care of. From our previous study \citep{2022A&A...662A..78P, 2023MNRAS.518.2529P}, we confirmed that the model parameters such as, $\rm{E_{cut}}$/$\rm{kT_{e}}$/R, obtained using these models did not differ significantly from the best-fit measurements found from fitting the spectra with a phenomenological power law with a cut-off in which the Fe-K$\alpha$ line is not coupled with the reflection continuum.

In our both Model$-$1 and Model$-$2,  \textsc{const} represents the 
calibration constant between the {\it NuSTAR} focal plane modules, FPMA and FPMB. 
\textsc{TBabs} was used to model the Milky Way Galactic hydrogen column density, 
which was taken from \cite{2013MNRAS.431..394W} for each source. The component
\textsc{zTBabs} represents the hydrogen column density ($\rm N_{H}^{INT}$) of the 
host galaxy. During the modelling of the source spectrum, the value of 
$\rm N_{H}^{INT}$ was allowed to vary freely.

\textsc{xillver}/\textsc{relxill} \citep{2010ApJ...718..695G, Garc_a_2011} was used to model the spectra with an absorbed cut-off power law along with the reflection features present in it.
In XSPEC Model$-$1 took the following forms,
\begin{itemize}
    \item Model$-$1a: \textsc{const $\times$ TBabs $\times$ zTBabs $\times$ (xillver)}
    \item Model$-$1b: \textsc{const $\times$ TBabs $\times$ zTBabs $\times$ (relxill)}
    \item Model$-$1c: \textsc{const $\times$ TBabs $\times$ zTBabs $\times$ (relxill+xillver)}
\end{itemize}

During the fit using Model$-$1a, the parameters that were kept free were $\Gamma$, $\rm{E_{cut}}$, $R$ and the normalization ($N_{\rm{xillver}}$) of the \textsc{xillver} component. The reflector was considered neutral; therefore, we fixed the ionization parameter ($log\xi$) to 0.0. The values of $AF_{e}$ and the inclination angle were fixed to the solar value (=1.0) and $30^{\circ}$ respectively.

In Model$-$1b, we replaced \textsc{xillver} with \textsc{relxill} to take care of the relativistic smeared Comptonization spectrum for a few sources. In addition to the parameters described in Model$-$1a, there are a few more parameters, such as the inner and outer emissivity indices ($\beta1$ and $\beta2$ respectively), inner and outer radii of the accretion disk ($r_{in}$ and $r_{out}$ respectively), break radius ($r_{br}$) between $r_{in}$ and $r_{out}$ and the spin of the black-hole ($a_{*}$). We tied $\beta1$ and $\beta2$ together during the fit and kept them as free parameters. $r_{br}$, $r_{in}$ and $r_{out}$ were kept frozen to their default values of 15$r_{g}$, 3$r_{g}$ and 400$r_{g}$ respectively. We considered a highly spinning SMBH and fixed $a_{*}$ to 0.998 \citep{1974ApJ...191..507T}. $AF_{e}$ was frozen to the solar value. The inclination angle was fixed to $30^{\circ}$. The other parameters that were kept free during the fit were $\Gamma$, $\rm{E_{cut}}$, $R$, $log\xi$ and the normalization ($N_{\rm{relxill}}$) of the \textsc{relxill} model. 

The spectra of a few sources could not be well-fitted using either \textsc{xillver} or \textsc{relxill}. In those sources, where significant narrow Fe-K$\alpha$ emission lines were detected, we used Model$-$1c, in which we fitted \textsc{relxill} and \textsc{xillver} together. Between these two components $\Gamma$, $\rm{kT_{e}}$ and $AF_{e}$ were tied together and kept as free parameters during the fitting. The other parameters were treated similarly as described earlier in Model$-$1a and Model$-$1b. We could constrain $\rm{E_{cut}}$ for 73 sources from the model fits. The summary of the spectral analysis from this model fits to the spectra given in Table \ref{table-2}. 

Out of 112 sources, we used Model$-$1a in 86 sources to estimate different coronal parameters. In 20 out of the remaining 26 Seyferts, the presence of a broad emission line was confirmed. To take care of the relativistic broadening of the Fe-K$\alpha$ line, we fitted the spectra of those sources with Model$-$1b. In the other six sources (ARK 564, MCG-06-30-15, Mrk 1044, Mrk 279, NGC 3783 and NGC 4051), we used a \textsc{xillver} component in addition to \textsc{relxill} (Model$-$1c), since one model alone could not fit the reflection spectra properly. The distributions of $\Gamma$ and $\rm{E_{cut}}$ as found from the Model$-$1 fits are given in Fig. \ref{figure-2}. The median value of $\Gamma$ as obtained from the analysis using Model$-$1 was found to 1.79$\pm$0.02, which is consistent with the median value of $\Gamma$ as found from the broad-band analysis of the unobscured sources by \cite{Ricci_2017}. Using only the constrained $\rm{E_{cut}}$, a median of 104$\pm$8 keV was obtained. The errors on the median values represent the statistical uncertainties, calculated as the standard deviation of the distribution divided by the square root of the sample size ($\sigma$/$\sqrt{N}$, where N is the number of data points). The broad-band spectral fit using Model$-$1c with the data to model residue for the source Mrk 279 is presented in Fig. \ref{figure-4}.

\begin{figure*}
\hbox{
\includegraphics[scale=0.58]{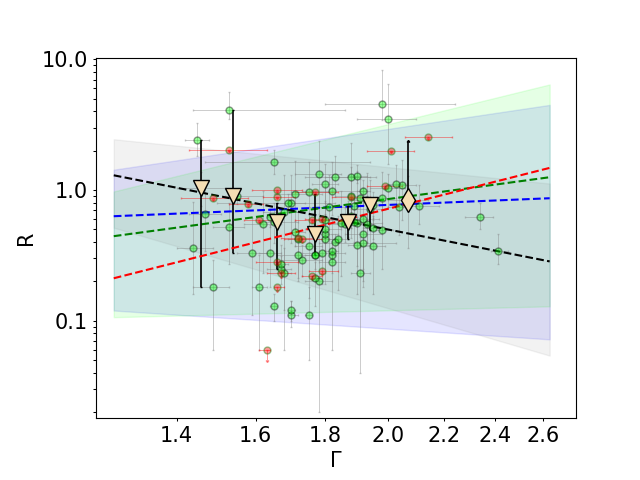}
\includegraphics[scale=0.58]{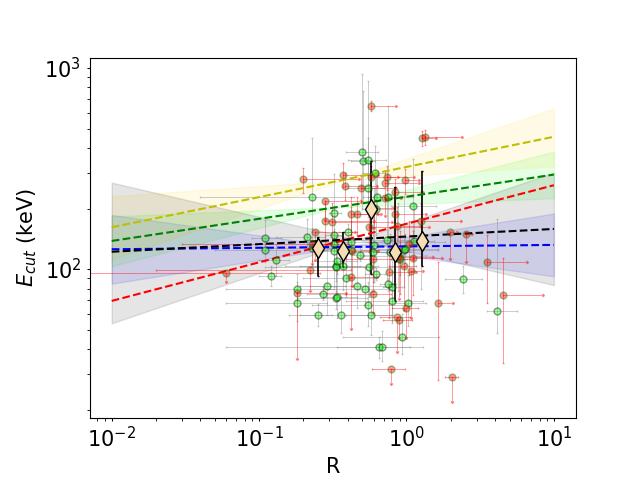}
}
\caption{The relation between R-$\Gamma$ (left panel) and $\rm{E_{cut}}$-R (right panel). The lines and the shaded regions have similar descriptions as of \ref{figure-19}.}
\label{figure-20} 
\end{figure*}

\begin{figure*}
\hbox{
\includegraphics[scale=0.58]{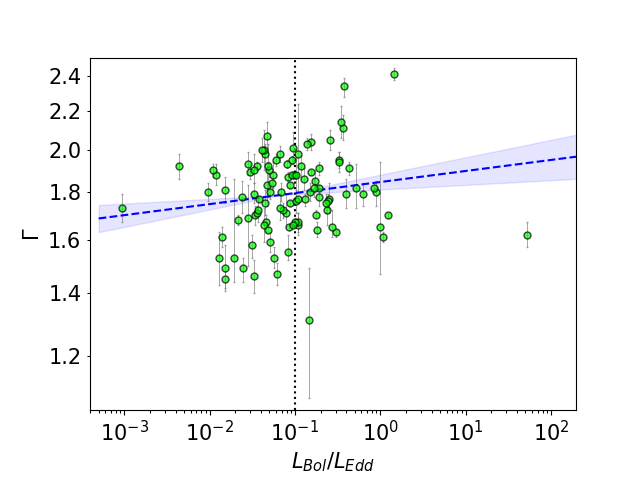}
\includegraphics[scale=0.58]{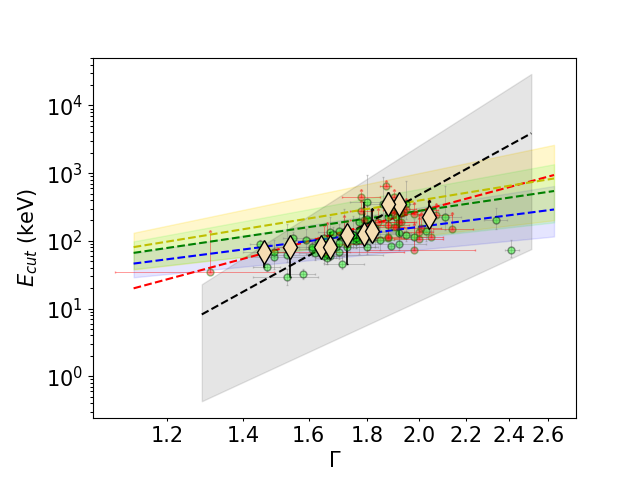}
}
\caption{The relation between $\Gamma$-$\lambda_{Edd}$ (left panel) and $\rm{E_{cut}}$-$\Gamma$ (right panel). The lines and the shaded regions have similar descriptions as of \ref{figure-19}.}
\label{figure-7} 
\end{figure*}

\begin{deluxetable*}{ccccccccc}
\caption{Results of the correlation analysis between different parameters. Provided are the Spearman's rank correlation coefficient ($\rho$) and the probability ($p$) for the null hypothesis (no correlation). We fail to reject the null hypothesis if $p$ is larger than 0.01. Here, Method I indicates the correlation study between two parameters with only uncensored values. Method IIa represents the correlation analysis performed between the median values of two parameters obtained from the survival analysis method using the non-parametric KMPL approach. Method IIb is the bivariate correlation analysis performed using the package \textit{ASURV}. The correlation analysis between two parameters considering both uncensored (including the corresponding asymmetric errors) and censored values($\rm{E_{cut}^{MAX}}$ = 1000 keV, $R^{MIN}$ = 0.01 and $\rm{kT_{e}^{MAX}}$ = 150 keV) is denoted by Method III, wherein Method IV represents the same as Method III except the $\rm{E_{cut}^{MAX}}$ is considered to be 500 keV here.} \label{table-4} 
\tablehead{Parameter 1 & Parameter 2 & Method &  \multicolumn{2}{c}{Full sample} & \multicolumn{2}{c}{Moderately accreting sources} & \multicolumn{2}{c}{Highly accreting sources}} 
\startdata
&&&&&\multicolumn{2}{c}{($\rm{\lambda_{Edd}}<0.1$)} & \multicolumn{2}{c}{($\rm{\lambda_{Edd}}>0.1$)} \\
\hline
&		&		&	$\rho$	&	p	&	$\rho$ &	p	&	$\rho$	&	p	\\
\hline
$\rm{E_{cut}}$	&	$\rm{\lambda_{Edd}}$	&	I	&	0.19	&	0.12	&	0.41	&	0.009	&	$-$0.02	&	0.89	\\
    &       &  IIa  &   0.54    &   0.22    &	$-$	&	$-$	&	$-$	&	$-$ \\
    &       &  IIb  &   0.06    &   0.55    &	0.11	&	0.39	&	$-$0.16	&	0.30 \\
	&		&	III	&	0.03	&	0.75	&	0.12	&	0.37	&	$-$0.08	&	0.59	\\
	&		&	IV	&	0.03	&	0.73	&	0.09	&	0.48	&	$-$0.08	&	0.61	\\
$R$	&	$\rm{\lambda_{Edd}}$	&	I	&	$-$0.28	&	0.01	&	$-$0.30	&	0.04	&	$-$0.07	&	0.68	\\
    &       &   IIa  &   $-$0.25 &   0.05    &	$-$	&	$-$	&	$-$	&	$-$	\\
    &       &   IIb  &   $-$0.22 &   0.02    &	$-$0.25	&	$-$0.06	&	$-$0.02	&	$-$0.88	\\
	&		&	III	&	$-$0.18	&	0.07	&	$-$0.21	&	0.11	&	$-$0.002	&	0.73	\\
$R$	&	$\Gamma$	&	I	&	0.28	&	0.009	&	$-$	&	$-$	&	$-$	&	$-$	\\
    &       &   IIa  &   $-$0.55 &   0.26    & $-$	&	$-$	&	$-$	&	$-$	\\
    &       &   IIb  &   0.59 &   0.03    & $-$	&	$-$	&	$-$	&	$-$	\\
	&		&	III	&	0.30	&	0.002	&	$-$	&	$-$	&	$-$	&	$-$	\\
$\rm{E_{cut}}$	&	$R$	&	I	&	$-$0.03	&	0.84	&	$-$	&	$-$	&	$-$	&	$-$	\\
    &       &   IIa  &   0.10 &   0.87    &	$-$	&	$-$	&	$-$	&	$-$	\\
    &       &   IIb  &   0.18 &   0.08    &	$-$	&	$-$	&	$-$	&	$-$	\\
	&		&	III	&	0.15	&	0.12	&	$-$	&	$-$	&	$-$	&	$-$	\\
	&		&	IV	&	0.15	&	0.12	&	$-$	&	$-$	&	$-$	&	$-$	\\
$\Gamma$	&	$\rm{\lambda_{Edd}}$	&	I	&	0.17	&	0.08	&	0.18	&	0.16	&	0.1	&	0.49	\\
    &       &   IIa  &   $-$     &   $-$ &	$-$	&	$-$	&	$-$	&	$-$	\\
    &       &   IIb  &   $-$     &   $-$ &	$-$	&	$-$	&	$-$	&	$-$	\\
	&		&	III	&	0.17	&	0.08	&	0.17	&	0.18	&	0.12	&	0.41	\\
$\rm{E_{cut}}$	&	$\Gamma$	&	I	&	0.69	&	1.75E-11	&	$-$	&	$-$	&	$-$	&	$-$	\\
    &       &   IIa  &   0.99    &   1.46E-05    &	$-$	&	$-$	&	$-$	&	$-$	\\
    &       &   IIb  &   0.57    &   0    &	$-$	&	$-$	&	$-$	&	$-$	\\
	&		&	III	&	0.60	&	4.11E-12	&	$-$	&	$-$	&	$-$	&	$-$	\\
	&		&	IV	&	0.61	&	1.75E-12	&	$-$	&	$-$	&	$-$	&	$-$	\\
$\tau$	&	$\rm{kT_{e}}$	&	I	&	$-$0.96	&	1.82E-23	&	$-$	&	$-$	&	$-$	&	$-$	\\
    &       &   IIa  &   $-$0.99 &   1.40E-24    &	$-$	&	$-$	&	$-$	&	$-$	\\
     &       &   IIb  &   $-$     &   $-$ &	$-$	&	$-$	&	$-$	&	$-$	\\
	&		&	III	&	$-0.66$	&	1.89E-10	&	$-$	&	$-$	&	$-$	&	$-$	\\
$\rm{E_{cut}}$	&	$\frac{M_{BH}}{M_{Sun}}$	&	I	&	$-$0.02	&	0.86	&	$-$	&	$-$	&	$-$	&	$-$	\\
    &       &   IIa  &   $-$0.54 &   0.27    &	$-$	&	$-$	&	$-$	&	$-$	\\
    &       &   IIb  &   $-$0.08 &   0.38    &	$-$	&	$-$	&	$-$	&	$-$	\\
	&		&	III	&	$-$0.20	&	0.04	&	$-$	&	$-$	&	$-$	&	$-$	\\
	&		&	IV	&	$-$0.20	&	0.03	&	$-$	&	$-$	&	$-$	&	$-$	\\
$\Gamma$	&	$\frac{M_{BH}}{M_{Sun}}$	&	I	&	$-$0.05	&	0.58	&	$-$	&	$-$	&	$-$	&	$-$	\\
    &       &  IIa  &   $-$     &   $-$ &	$-$	&	$-$	&	$-$	&	$-$	\\
    &       &  IIb  &   $-$     &   $-$ &	$-$	&	$-$	&	$-$	&	$-$	\\
	&		&	III	&	$-$0.05	&	0.58	&	$-$	&	$-$	&	$-$	&	$-$	\\
\enddata
\end{deluxetable*}

We carried out the Comptonization model fits (Model$-$2) to the 112 sources to estimate the coronal temperature. We could constrain $\rm{kT_{e}}$ for 42 sources using this model. 
In XSPEC Model$-$2 took the following forms,
\begin{itemize}
    \item Model$-$2a: \textsc{const $\times$ TBabs $\times$ zTBabs $\times$ (xillverCP)}
    \item Model$-$2b: \textsc{const $\times$ TBabs $\times$ zTBabs $\times$ (relxillCP)}
    \item Model$-$2c: \textsc{const $\times$ TBabs $\times$ zTBabs $\times$ (relxillCP+xillverCP)}
\end{itemize}

All the model parameters were handled similarly as described for Model$-$1. The best-fit values of various coronal parameters found from the Comptonization model fit (Model$-$2)  are given in Table \ref{table-3}. The distribution of the best-fit values of $\Gamma$ and $\rm{kT_{e}}$, as obtained from Model$-$2, are shown in Fig. \ref{figure-5}. The median value of $\Gamma$ was determined to be 1.86$\pm$0.01, and when considering only the constrained value of $\rm{kT_{e}}$, the median was found to be 24$\pm$2 keV.

\section{Relation between $\rm{E_{cut}}$ and $\rm{kT_{e}}$}
It is believed that the phenomenological high-energy cut-off could be related to the temperature as $\rm{E_{cut}}$ = 2$-$3 $\rm{kT_{e}}$ \citep{2001ApJ...556..716P}. However, recent studies do indicate that this simple relation between $\rm{E_{cut}}$ and $\rm{kT_{e}}$ may not be valid for all sources \citep{2014ApJ...783..106L,2019A&A...630A.131M,2022A&A...662A..78P}. The relation can be complicated in the case of a non-static corona, such as the one with outflows \citep{2014ApJ...783..106L}. Also, according to \cite{2019A&A...630A.131M}, the relation of $\rm{E_{cut}}$ = 2$-$3 $\rm{kT_{e}}$ is valid only for low values of $\tau$ and $\rm{kT_{e}}$. The authors also argued that if the origin of the X-ray emission is different than the thermal Comptonization, the typical relation between $\rm{E_{cut}}$ and $\rm{kT_{e}}$ may not hold. 
We show in the left panel of Fig. \ref{figure-6} the distribution of the ratio between $\rm{E_{cut}}$ to $\rm{kT_{e}}$ for 42 sources for which we could constrain both $\rm{E_{cut}}$ and $\rm{kT_{e}}$. We found the ratio to vary between 2.33 and 5.35, with a median of 3.59 in 42 sources. In the right panel of Fig. \ref{figure-6} is shown the distribution of the sources in the $\rm{E_{cut}}$ versus $\rm{kT_{e}}$ plane. We excluded the sources with $\rm{E_{cut}}$ $>$ 300 keV from this correlation analysis since at the high energy limit, the best fit $\rm{kT_{e}}$ values obtained using the Comptonization model produces comparative lower corona temperature than that obtained using a cut-off power law \citep{10.1046/j.1365-8711.2003.06556.x, 2015MNRAS.451.4375F}. To include the lower limits of $\rm{E_{cut}}$ and $\rm{kT_{e}}$ in our calculation we performed a survival analysis method using the non-parametric Kaplan-Maier Product Limit (KMPL) approach in \textit{Python}\footnote{\href{https://medium.com/the-researchers-guide/survival-analysis-in-python-km-estimate-cox-ph-and-aft-model-5533843c5d5d}{https://medium.com/the-researchers-guide/survival-analysis-in-python-km-estimate-cox-ph-and-aft-model-5533843c5d5d}}. The yellow points in the plot represent the median values of $\rm{E_{cut}}$ in different $\rm{kT_{e}}$ bins obtained from the survival analysis method using the KMPL approach. The errors in the median values of $\rm{E_{cut}}$ and $\rm{kT_{e}}$ were estimated in 95 per cent confidence. Also, shown in the same figure are the $\rm{E_{cut}}$ = 2 $\rm{kT_{e}}$ (red dashed) and $\rm{E_{cut}}$ = 3 $\rm{kT_{e}}$ (blue dashed) lines. From the linear least squares fit to the estimated median values of $\rm{E_{cut}}$ in different $\rm{kT_{e}}$ bins (black dashed line in Fig. \ref{figure-6}), we found
\begin{equation}
\label{equation1}
E_{cut} = (3.80 \pm 0.53) \rm{kT_{e}} + ( 8.15 \pm 16.51)
\end{equation}
We calculated the Pearson's correlation coefficient and the null hypothesis probability for no correlation to check the significance of the linear fit, and we found $r$ = 0.97 and $p$ = 0.006. 
We also used the \textit{Astronomy SURVival Analysis} (\textit{ASURV}) package \citep{1985ApJ...293..192F} to take into account the lower limits in the $\rm{E_{cut}}$ and $\rm{kT_{e}}$ measurements. We derived the bivariate correlation and linear regression parameters using Spearman's rho and the Schmitt method and found $\rm{E_{cut}}$ = 3.54 $\rm{kT_{e}}$ + 13.99
These observations thus indicate that for the sample of sources studied in this work, $\rm{E_{cut}}$ and $\rm{kT_{e}}$ maintain a strong correlation with most of our sources lying above the $\rm{E_{cut}}$ = 3 $\rm{kT_{e}}$ line.

\begin{figure}
\hspace{-0.7 cm}
\includegraphics[scale=0.60]{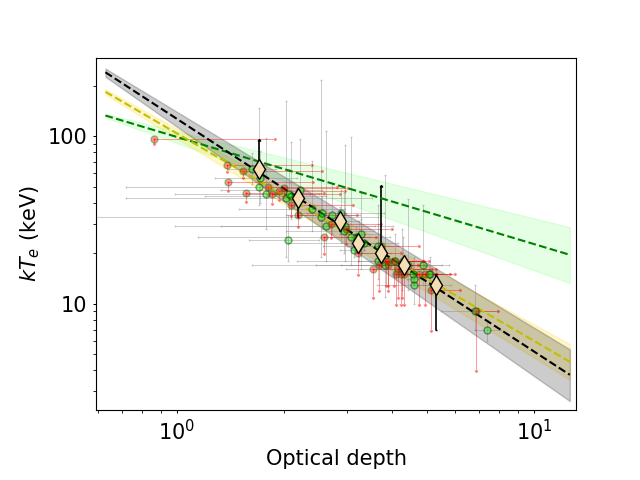}
\caption{The relation between $\rm{kT_{e}}$ and $\tau$. The green dots are the constrained values of the parameters, and the red dots denote the lower and upper limits of $\rm{kT_{e}}$ and $\tau$, respectively. The yellow dashed line and the yellow shaded region are the linear fit to the constrained values of $\rm{kT_{e}}$ and $\tau$, and the errors in the slope and the intercept of the linear fit (Method I). The beige diamond points represent the median values of $\rm{kT_{e}}$ in each bin of $\tau$ obtained from the survival analysis using the KMPL approach. The black dashed line and the black shaded region are the linear fit and the errors in the fit parameters to the median values of $\rm{kT_{e}}$ (Method II). The green dashed line and the green shaded region represent the linear fit and the errors in the slope and the intercept for the fit between two parameters considering both uncensored (including the corresponding asymmetric errors in $\rm{kT_{e}}$) and censored (upper bound of $\rm{kT_{e}}$ = 150 keV) values of two parameters (Method III).} \label{figure-8}
\end{figure}

\section{Correlation analysis}
This section presents the correlation analysis between different best-fitted model parameters obtained using Model$-$1. We also discussed the correlation between the coronal properties and the physical properties of the sources, such as $\rm{\lambda_{Edd}}$ and $\rm{M_{BH}}$. For the latter, we had to exclude three sources from the correlation analysis, namely, ESO 416$-$G002, IRAS F12397+3333 and UGC 10120, as we did not find a black hole mass ($\rm{M_{BH}}$) measurement for them from the literature. We adopted black hole mass estimates from the second data release of optical broad emission line measurements from the BASS survey \citep{2022ApJS..261....5M} except for ARK 564. The black hole mass for this source was taken from \cite{2009ApJ...702.1353D}.

For getting $L_{bol}$, we used the 2$-$10 keV intrinsic luminosity. The absorption and $k-$ corrected intrinsic luminosities were converted to bolometric luminosities using the relation log(L$_{\rm{Bol}}$) = 20 $\times$ log($L_{2-10 keV}$) \citep{10.1111/j.1365-2966.2007.12328.x}. The distribution of the logarithm of the Eddington ratio ($\rm{L_{bol}/L_{Edd}}$=$\rm{\lambda_{Edd}}$) for 109 sources is given in Fig. \ref{figure-3}.

The analysis of 112 sources using Model$-$1 revealed that $\rm{E_{cut}}$ could be constrained in 73 sources, while in the remaining 39 sources, only lower limits could be determined. These controlled $\rm{E_{cut}}$ measurements in 73 sources exhibited asymmetric errors, and $\rm{E_{cut}}$ in the remaining 39 sources only provided lower limits. From the measurements of $\rm{kT_{e}}$ also, we found constrained values in 42 sources, and the remaining 70 sources produced lower limits. In a few cases, we only found the upper limits of R. To account for the lower and upper limits in our correlation analysis, we used the KMPL approach, as described in Section 4. The median values of Parameter 2 (the dependent variable) were calculated using the KMPL method for each bin of Parameter 1 (the independent variable), in cases where only the dependent variable had upper or lower limits. When both the dependent and independent variables had upper and lower limits, the KMPL estimator calculated the survival function based on the data status of the independent variable (censored = 0, constrained = 1) and determined the median value for each bin. For the dependent variable, the probability function and median value were calculated for the corresponding bins of independent variables using the KMPL estimator. To tackle the unconstrained values of the independent and dependent variables in the correlation analysis, we used the survival statistics within the \textit{ASURV} \textit{FORTRAN} package as well to calculate Spearman's rho correlation coefficient and the probability of no correlation. We derived the linear regression intercepts and slops using the Schmitt method in \textit{ASURV}. 

Though the survival analysis is well-suited to take care of the limits in the correlation analysis, it does not consider the asymmetric errors in the analysis. Thus, it is essential to consider both the asymmetric errors and lower limits in the correlation analysis. Therefore, we employed a similar approach as described in \cite{2022A&A...662A..78P} to perform various correlation analyses and find the median of the parameters. The results of this analysis are given in Table \ref{table-4}.

\begin{figure*}
\hbox{
\includegraphics[scale=0.58]{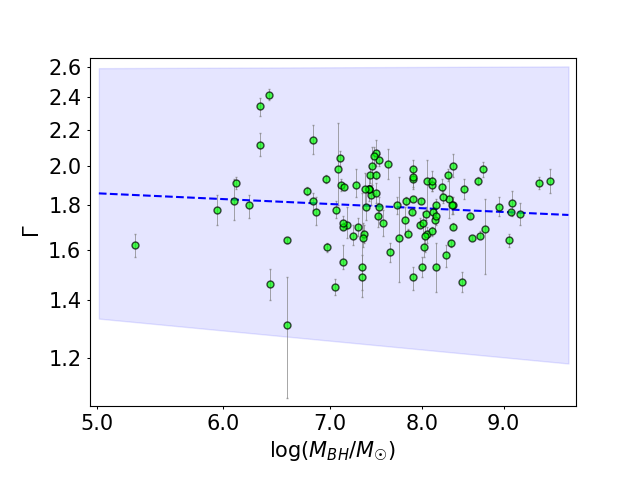}
\includegraphics[scale=0.58]{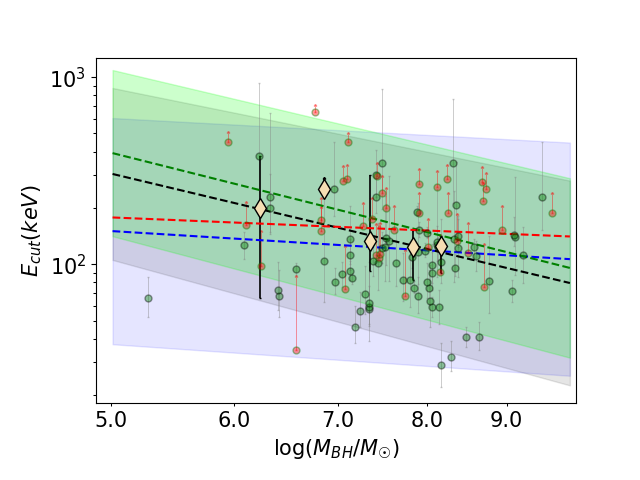}
}
\caption{The relation between $\Gamma$ and $\rm{E_{cut}}$ with the black hole mass. The lines and the shaded regions have similar descriptions as of \ref{figure-19}.}
\label{figure-22} 
\end{figure*}

We neglected the asymmetric errors associated with the controlled $\rm{E_{cut}}$ measurements in the initial approach and the lower limits. Instead, we solely considered the controlled best-fit values of $\rm{E_{cut}}$ and calculated the median values. We also performed correlation analysis using only those best-fit values between $\rm{E_{cut}}$ and other parameters by employing a logarithmic scale for fitting the parameters with a linear relation:
\begin{equation}
\label{equ22}
    log(y) = alog(x)+b
\end{equation}
In the second approach, we included both controlled and limited values in the correlation analysis by incorporating the survival analysis using the KMPL test and found the median values of the dependent parameters in each bin of the independent variable. We also used the \textit{ASURV} package to handle both constrained and unconstrained values of the parameters in the correlation analysis. To assess the strength of the linear correlation in both cases (Method IIa and IIb in Table \ref{table-4}), we computed Spearman's rank correlation coefficient ($\rho$) and the null hypothesis probability ($p$) for no correlation. We considered a correlation to be significant if $p$ was less than 0.01. 

In the third approach, we incorporated the asymmetric errors related to the controlled $\rm{E_{cut}}$ measurements and the lower limits of $\rm{E_{cut}}$ by simulating $10^{5}$ random points within the range of ($\rm{E_{cut}^{min}}$, $\rm{E_{cut}^{max}}$) and ($\rm{E_{cut}^{LL}}$, $\rm{E_{cut}^{MAX}}$) respectively. For constrained $\rm{E_{cut}}$ with asymmetric errors, $\rm{E_{cut}^{min}}$ and $\rm{E_{cut}^{max}}$ represented the respective lower and upper bounds, while $\rm{E_{cut}^{LL}}$ denoted the lower limit obtained from the model fit. The $10^{5}$ random points were generated between the lower limit ($\rm{E_{cut}^{LL}}$) and a hypothetical upper bound ($\rm{E_{cut}^{MAX}}$) of 1000 keV. Following this approach, we calculated the median of $\rm{E_{cut}}$ for each run and then determined the mean of the median distribution. The statistical errors associated with the median values were calculated as the standard deviations of the $10^{5}$ simulated median $\rm{E_{cut}}$ values.

In the fourth case, we handled the asymmetric errors corresponding to the constrained $\rm{E_{cut}}$ similarly as discussed in the third case. However, here, the upper bound $\rm{E_{cut}^{MAX}}$ was set to 500 keV for cases where only lower limits were available. It was necessary to consider $\rm{E_{cut}^{MAX}}$ = 1000 keV in three sources for which the lower limit of $\rm{E_{cut}}$ exceeded 450 keV. The median and standard deviation of each run were calculated, and the mean of their distributions was determined. In both cases, the linear relation was fitted (using Equation \ref{equ22}) between $\rm{E_{cut}}$ and other parameters for each run, resulting in distributions of the slope (a), the intercept (b), the Spearman's rank correlation coefficient ($\rho$), and the probability of no correlation ($p$). The median values from these distributions were used to represent the best-fit values of the correlation. All values and errors for the unweighted and simulated correlations are presented in Table \ref{table-4}.

We followed a similar approach for the other parameters also ($\Gamma$, $R$, and $\rm{kT_{e}}$), simulating $10^{5}$ points between the minimum and maximum bounds for the correlation analysis. In the case of the reflection fraction ($R$) and the coronal temperature ($\rm{kT_{e}}$), respective upper and lower limits were obtained. During the correlation analysis, the lower bound of $R$ was set at 0.01, and the upper bound of $\rm{kT_{e}}$ was set at 150 keV.

Using the constrained values only, the median of $\rm{E_{cut}}$ as obtained from Model$-$1 was 104$\pm$8 keV. The median of the unconstrained $\rm{E_{cut}}$ using the same model was 173$\pm$18 keV.  Considering the upper limit of 1000 keV for the censored values and the asymmetric errors associated with the controlled $\rm{E_{cut}}$ measurements, we obtained a median of 153$\pm$8 keV for the full sample, 158$\pm$11 keV for the moderately accreting system ($\rm{\lambda_{Edd}}<0.1$) and 150$\pm$10 keV for the systems with higher accretion ($\rm{\lambda_{Edd}}>0.1$). Considering the upper limit of 500 keV, a median of 151$\pm$7 keV was obtained for the entire sample. For the moderate ($\rm{\lambda_{Edd}}<0.1$) and high ($\rm{\lambda_{Edd}}>0.1$) accreting systems we obtained a median of 152$\pm$10 keV and 147$\pm$10 keV, respectively. Our result is consistent with the measurements of the median $\rm{E_{cut}}$ value obtained from the literature. For example, using a sample of unobscured Seyfert galaxies from {\it Swift-BAT} 70-month catalogue,  \cite{Ricci_2017} reported a median value of 210$\pm$36 keV considering both censored and uncensored measurements. Using the \textsc{xillver} model to a total number of 195 Seyfert1 galaxies, \cite{2022ApJ...927...42K} found a median $\rm{E_{cut}}$ of 156$\pm$13 keV.

We determined the median of $\rm{kT_{e}}$ as 24$\pm$2 keV and 24$\pm$3 keV respectively for the constrained and unconstrained best-fit values obtained using Model$-$2. Including the asymmetric errors and the lower limits by considering the upper bound of 150 keV, a median of 48$\pm$5 keV was obtained. The distribution of $\rm{E_{cut}}$ and $\rm{kT_{e}}$ (constraints and the lower limits) are given in Fig. \ref{figure-2} and Fig. \ref{figure-5} respectively.

We also determined the median value of $\rm{kT_{e}}$ by analyzing a sample of 96 sources common to both this study and \cite{2018MNRAS.480.1819R}. We obtained the estimate for $\rm{E_{cut}}$ from \cite{2018MNRAS.480.1819R} and computed $\rm{kT_{e}}$ using Equation \ref{equation1}. The median value of $\rm{kT_{e}}$ for the subset with constrained values was 27$\pm$7 keV, consistent with our findings of the median for only the controlled measurements. For the unconstrained best-fit value of $\rm{kT_{e}}$ as obtained from \cite{2018MNRAS.480.1819R} the median was 29$\pm$5 keV. For 67 sources where only lower limits were reported in \cite{2018MNRAS.480.1819R}, we considered both the asymmetric errors associated with the best-fit $\rm{E_{cut}}$ and an upper limit of 500 keV for cases where only a lower limit was reported, yielding a median $\rm{kT_{e}}$ of 85$\pm$6 keV. This value is higher than what we observed for our sample but aligns with the results presented in \cite{2022ApJ...927...42K}. The variance in median $\rm{kT_{e}}$ values between this study and \cite{2018MNRAS.480.1819R} may be attributed to the higher proportion of unconstrained $\rm{E_{cut}}$ values in the latter work. Additionally, setting an upper limit of 500 keV for unconstrained cases biases the median value towards higher temperatures. The distribution of $\rm{kT_{e}}$, as calculated using Equation \ref{equation1} based on $\rm{kT_{e}}$  values from \cite{2018MNRAS.480.1819R}, is illustrated in Figure \ref{figure-3}.

\begin{figure*}
\hbox{
     \includegraphics[scale=0.28]{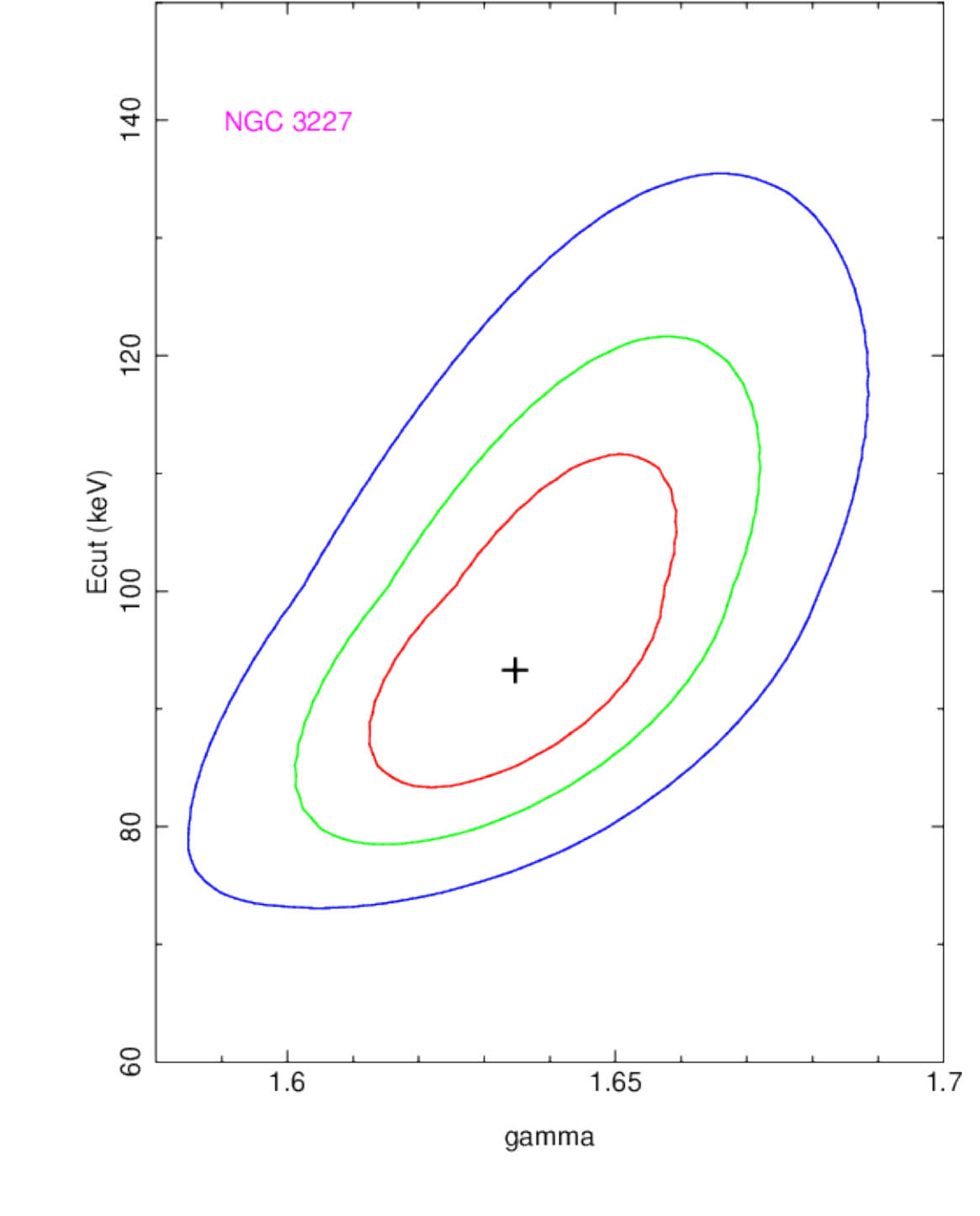}
     \includegraphics[scale=0.28]{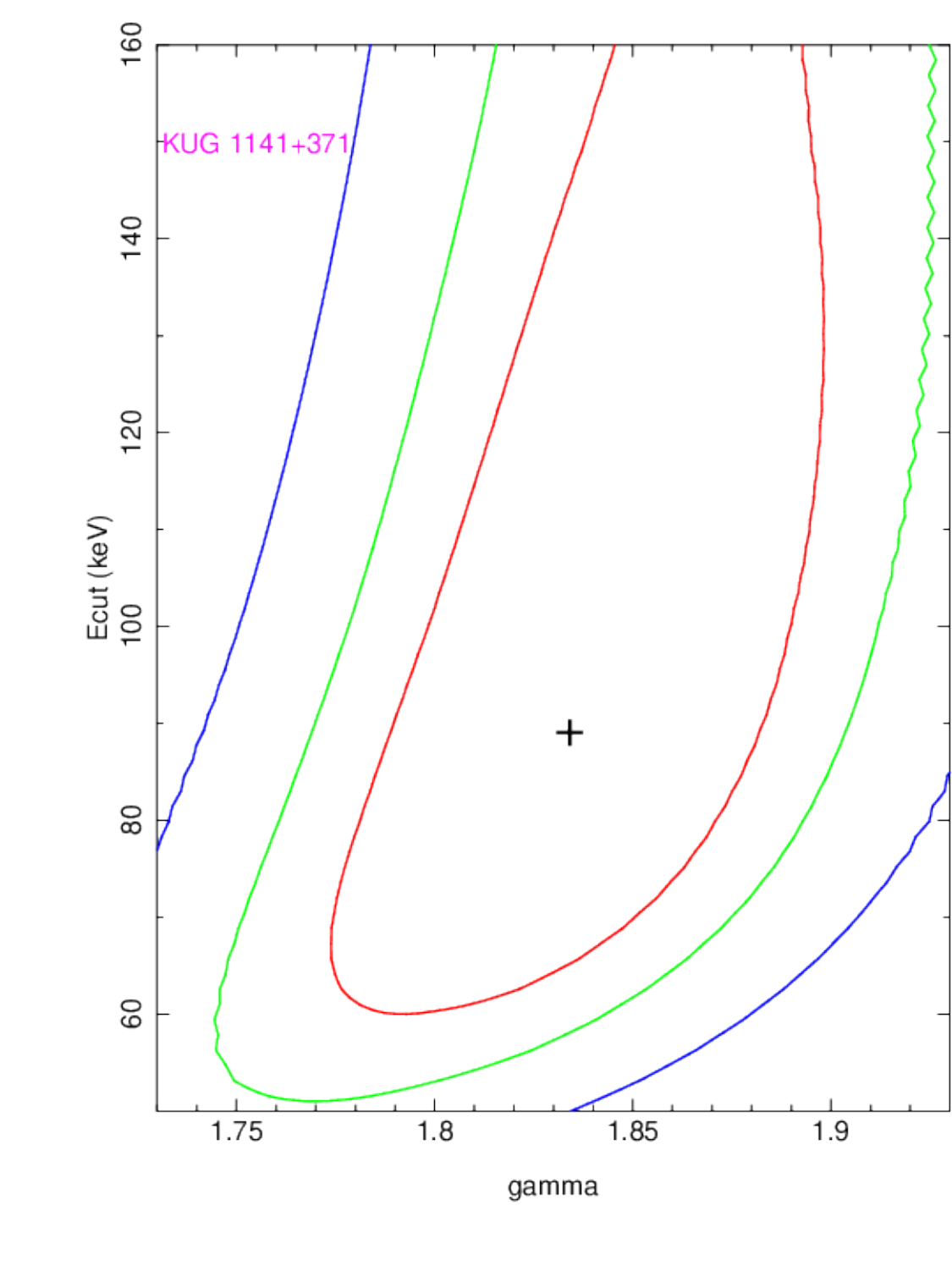}
     \includegraphics[scale=0.28]{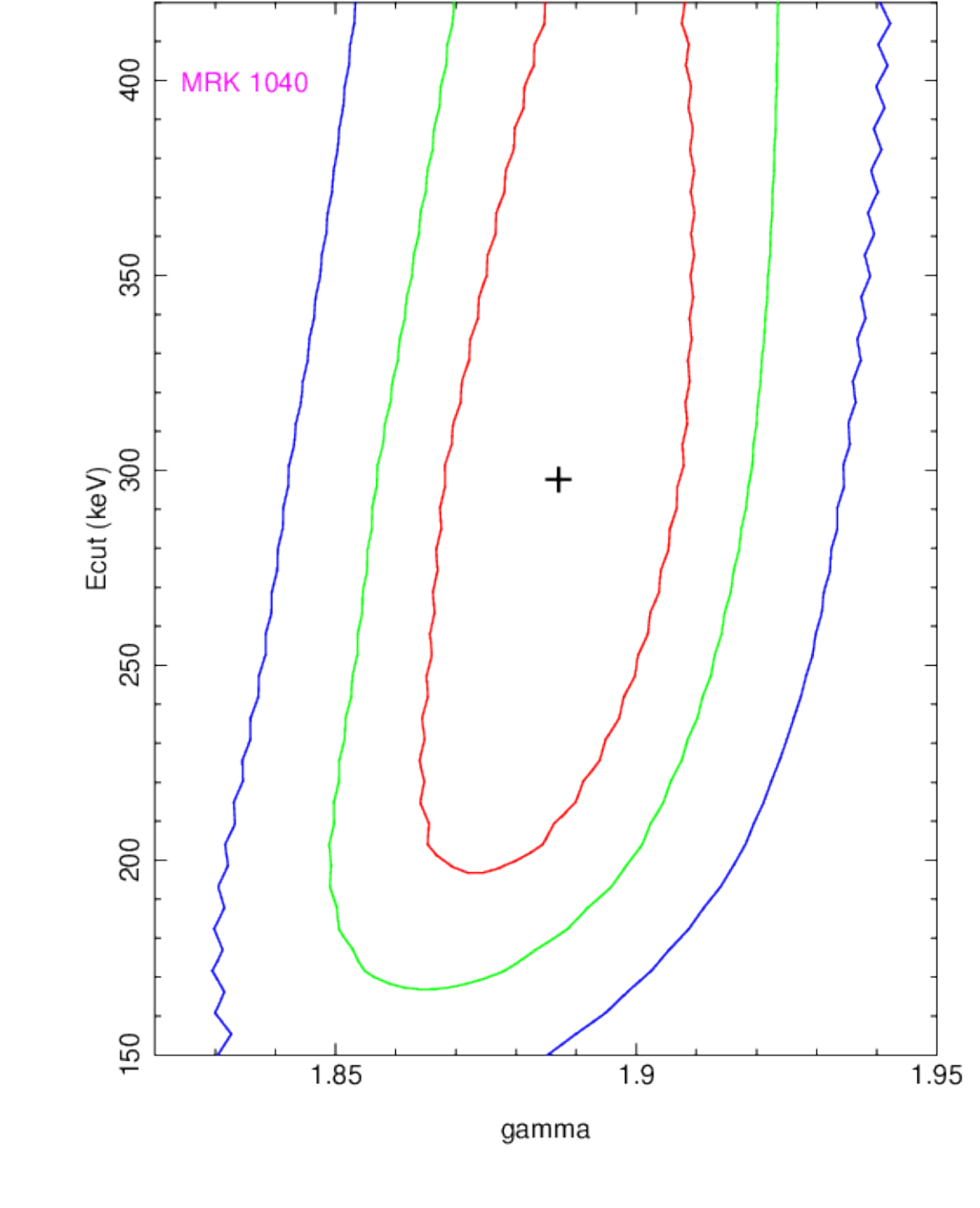}
     }
\hbox{
     \includegraphics[scale=0.28]{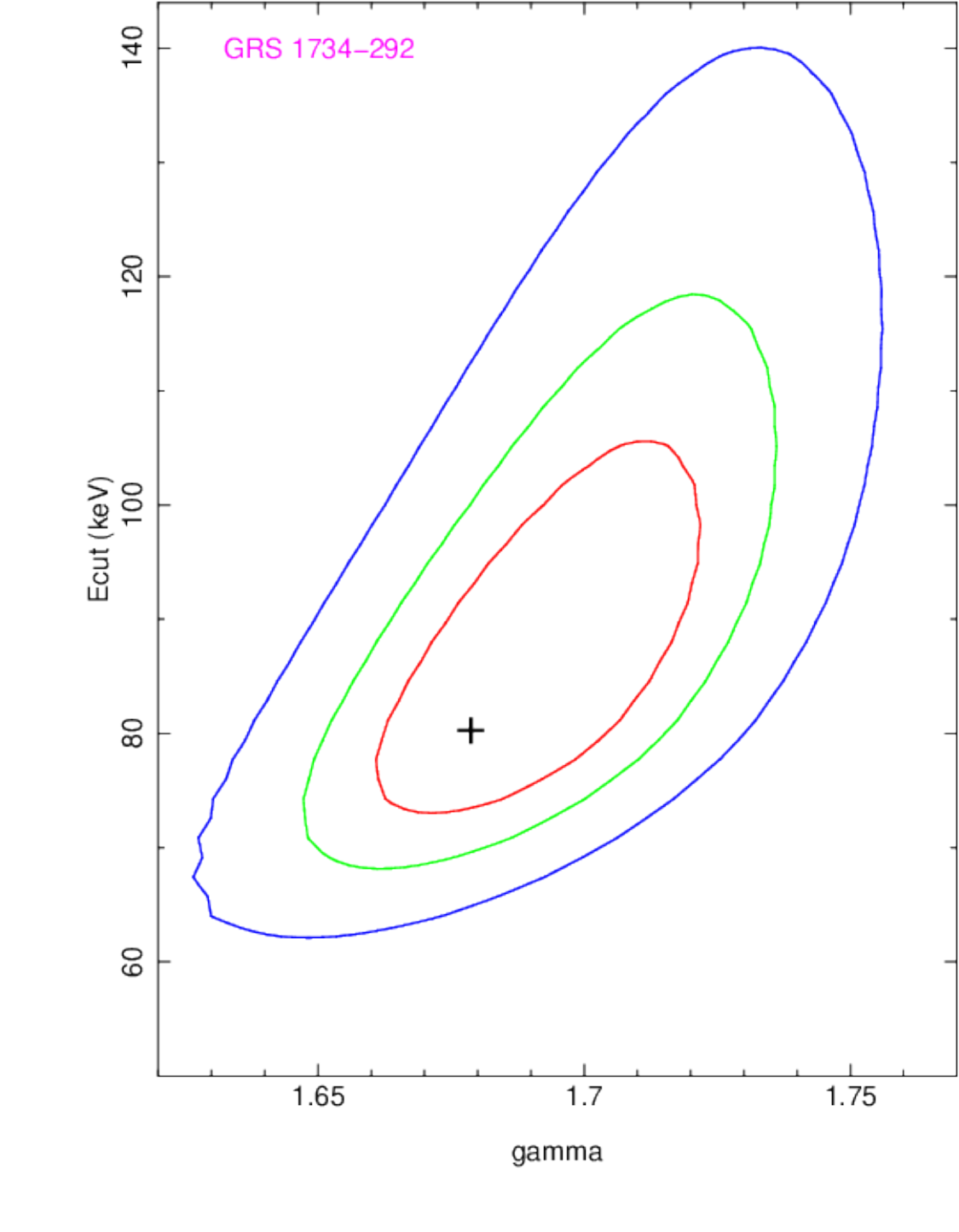}
     \includegraphics[scale=0.28]{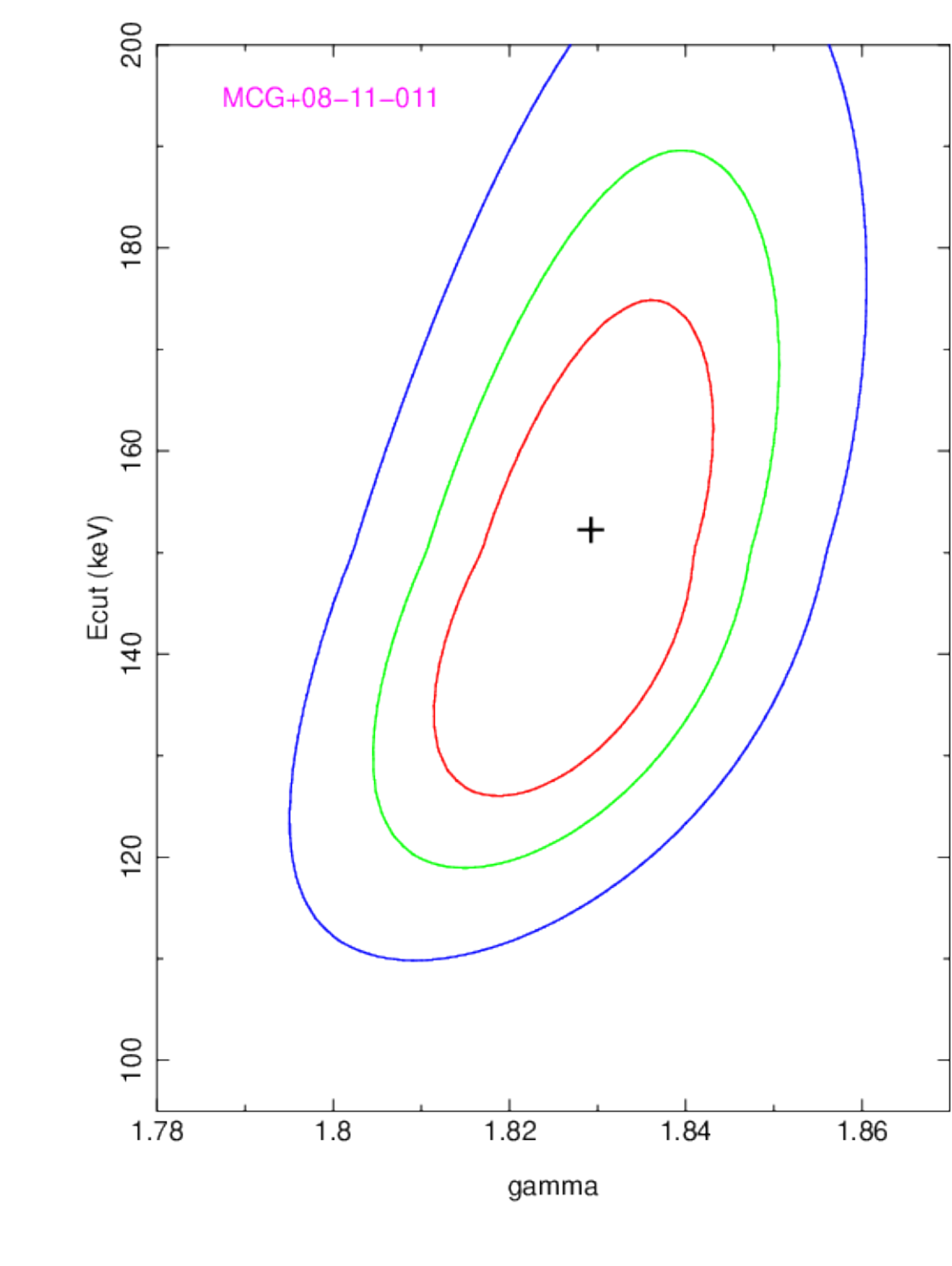}
     \includegraphics[scale=0.28]{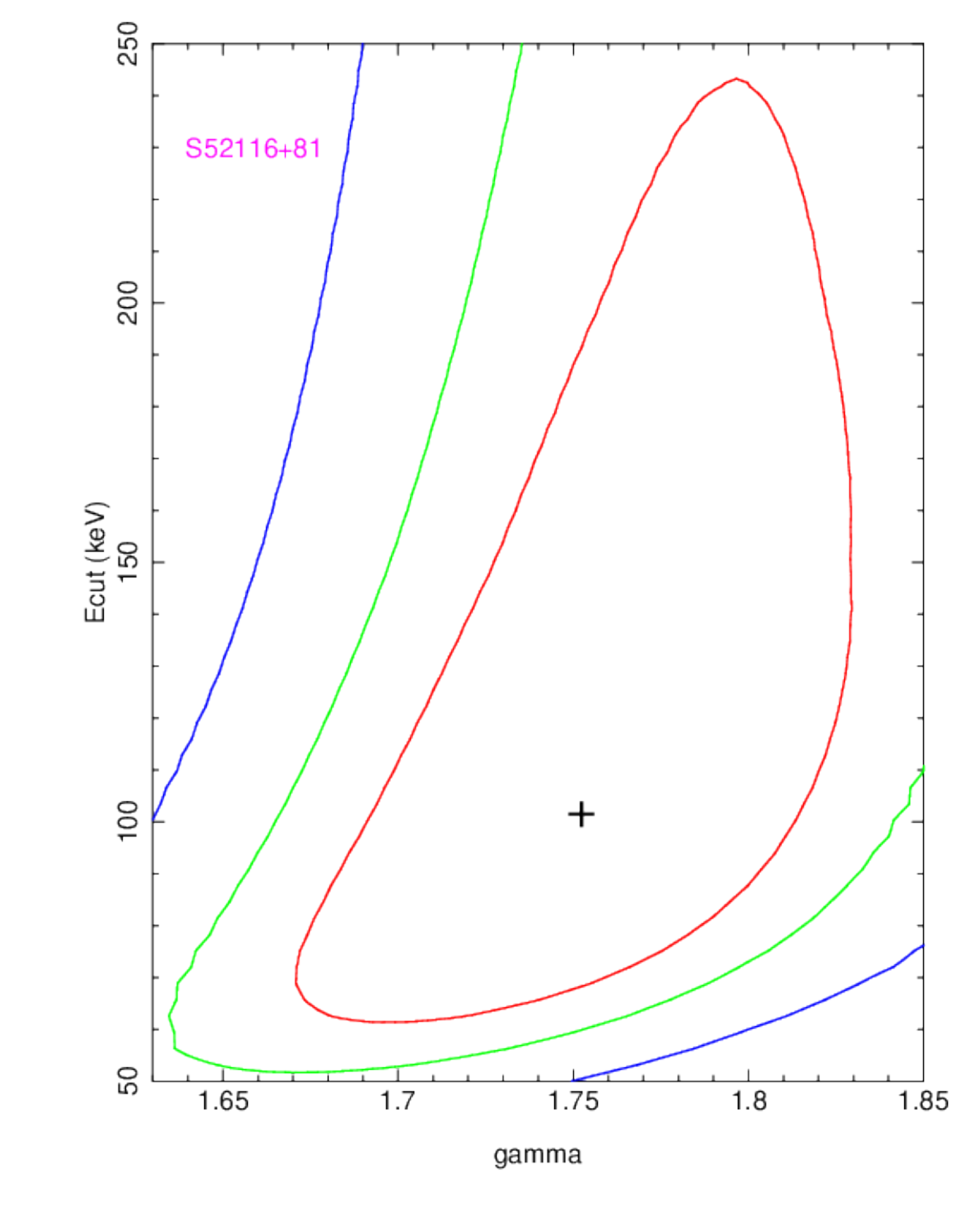}
     }
\caption{Confidence contours between $\rm{E_{cut}}$ and $\Gamma$ obtained from Model$-$1 for six sources. The colour codes for the contours are as follows: red, green, and blue for 68\% ($\Delta$$\chi^2$ = 2.30), 90\% ($\Delta$$\chi^2$ = 4.61), and 99\% ($\Delta$$\chi^2$ = 11.8) confidence levels}, respectively. In each panel, the best-fit location is marked by the black plus.
\label{figure-11}       
\end{figure*}

\subsection{Correlation between $\rm{E_{cut}}$ and $\rm{\lambda_{Edd}}$}
We looked for a relation between $\rm{E_{cut}}$  and $\rm{\lambda_{Edd}}$. Using all four methods of correlation analysis, we could not find any significant correlation between them. This is in agreement with the recent results in the literature \citep{refId0,2019MNRAS.484.2735M, 2021MNRAS.506.4960H,2022ApJ...927...42K}. 

For AGN with moderate accretion ($\rm{\lambda_{Edd}}<$0.1) \citep{2018MNRAS.480.1819R}, the observed spectral energy distribution can be explained by the standard optically thick geometrically thin accretion disk with H/R $<<$ 1, where H is the height of the disk at a radius R \citep{1973A&A....24..337S}. But for AGN with higher $\rm{\lambda_{Edd}}$, the accretion disk becomes geometrically thick with H$\leq$R and therefore, the accretion flow nature is expected to differ from the moderately accreting ones \citep{1981AcA....31..283P, 1982AcA....32....1M, 1988ApJ...332..646A, Wang_2014, 2022MNRAS.509.3599T}. The emergent X-ray spectrum from AGN with thick and thin accretion disks is likely to be different, and hence, the connection between the accretion disk and corona in low and high accretion AGN could be different. To look for any differences in the corona between low and high accreting AGN, we divided our sample into moderately accreting AGN ($\rm{\lambda_{Edd}}$ $<$ 0.1) and highly accreting AGN ($\rm{\lambda_{Edd}}$ $>$ 0.1) and carried out linear fit to the data using Equation \ref{equ22} in the $\rm{E_{cut}}$ versus $\rm{\lambda_{Edd}}$ plane. For the highly accreting sub-sample, we found no correlation between $\rm{E_{cut}}$ and $\rm{\lambda_{Edd}}$, which is expected for the sources with higher accretion rate \citep{2003A&A...398..927W}. From the Spearman's rank correlation test, we found $\rho$ of 0.41 and a $p$ of 0.009 considering only the controlled best-fit values $\rm{E_{cut}}$ in the moderately accreting systems ($\rm{\lambda_{Edd}}$ $<$ 0.1), but including the asymmetric errors and the lower limits into consideration we did not find any significant relation between these two parameters (see Table \ref{table-4} and fig. \ref{figure-19}). Due to a limited number of median data points in both regions, we could not perform a survival analysis test. Using the \textit{ASURV} package, we obtained $\rho$ of 0.06 and a $p$ of 0.55 for the entire sample; for the moderately accreting sub-sample, we obtained $\rho$ of 0.11 and a $p$ of 0.39 and for the highly accreting sub-sample a $\rho$ of $-$0.16 and a $p$ of 0.30 were calculated.



%

\begin{figure}
\hspace{-0.7 cm}
\includegraphics[scale=0.60]{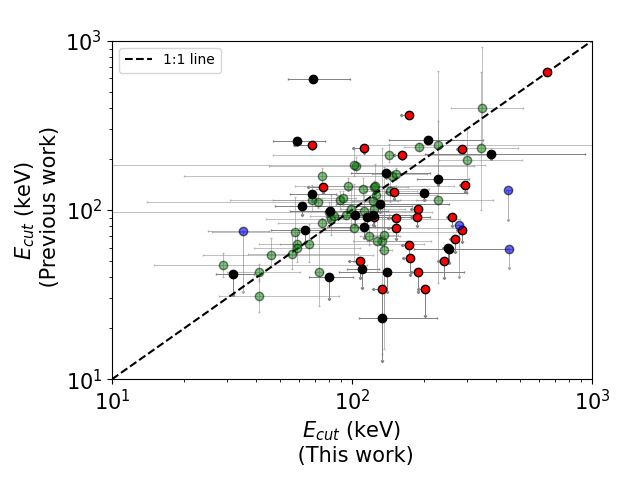}
\caption{Plot of the estimated $\rm{E_{cut}}$ values from this work and those obtained from the literature. The green dots indicate the constrained $\rm{E_{cut}}$ values obtained from both this study and prior research. The red dots, signify the lower limits of $\rm{E_{cut}}$ values derived from both this study and the existing literature. The black dots denote the constrained $\rm{E_{cut}}$ values from this study and the lower limits of $\rm{E_{cut}}$ from the literature. The blue dots represent the constrained $\rm{E_{cut}}$ values from the literature and the lower limits of $\rm{E_{cut}}$ values from this study. The 1:1 line is also shown (the black dotted line) for reference.} \label{figure-9}       
\end{figure}

\subsection{Correlation between $R$ and $\rm{\lambda_{Edd}}$}
We performed a simple linear fit to the data using Equation \ref{equ22} to look for the correlation between $R$ and $\rm{\lambda_{Edd}}$. In a few cases, we could not constrain $R$; rather, we found an upper limit. We considered both controlled values and the upper limits during the correlation analysis. Using only the controlled best-fit values, Spearman's correlation analysis yielded a $\rho$ of $-$0.28 and a $p$ value of 0.01. The survival analysis using the KMPL test indicated no correlation between $R$ and $\rm{\lambda_{Edd}}$ with $\rho$ = $-$0.25 and $p$ = 0.05. Similar outcomes were also derived using the \textit{ASURV} statistical analysis package. A Spearman's correlation coefficient of $-$0.22 and $p$ = 0.02 was obtained from the \textit{ASURV} analysis.  Using the third approach of the correlation analysis, we considered the asymmetric errors associated with the uncensored best-fit values of $R$ and a lower bound of 0.01 in the cases where only the upper limit was found; we performed the linear fit for $10^5$ times. From the distribution of $\rho$ and $p$, we obtained the median of $\rho$ = $-$0.18 and a $p$ = 0.07. We thus conclude that using both uncensored and censored values of $R$, we could not get a meaningful correlation between $R$ and $\rm{\lambda_{Edd}}$. For the sources with moderate ($\rm{\lambda_{Edd}}<$0.1) and high ($\rm{\lambda_{Edd}}>$0.1) accretion rates, we did not notice any significant correlation using methods I and III. Due to limited median data points, we could not perform the survival analysis for separate accreting regions. Using the \textit{ASURV} package we obtained a $\rho$ = $-$0.25 and a $p$ = 0.06 for moderately accreting sub-sample and $\rho$ = $-$0.02 and a $p$ = 0.88 for highly accreting sub-samples. Thus, considering our sample of objects and dividing them into two different accreting systems, we conclude that the relation between $R$ and $\rm{\lambda_{Edd}}$ is insignificant (see Fig. \ref{figure-19}).

\subsection{Correlation between $R$ and $\Gamma$}
We obtained an indication of a positive correlation between $\Gamma$ and the reflection fraction in Table \ref{table-4}. We used Equation \ref{equ22} to perform a linear fit between these two parameters. The fit gave us a $\rho$ of 0.30 and a p of 0.002, considering the controlled and upper limits of $R$. We obtained a positive correlation using only the constrained values with a $\rho$ of 0.28 and a $p$ of 0.009. However, the KMPL and \textit{ASURV} survival analysis denied the correlation with a $\rho$ of $-$0.55 and $0.59$ and a $p$ of 0.26 and 0.03, respectively. The correlation is plotted in the left panel of the Fig. \ref{figure-20}

A study on the dependence of $\Gamma$ on $R$ has been done several times in past \citep{1999MNRAS.303L..11Z, 1999ApJ...510L.123B, 2001MNRAS.326..417M, Mattson_2007,refIDadd0,2009MNRAS.399.1293M, refIBoid0, 2016MNRAS.458.2454L, Del_Moro_2017, Zappacosta_2018,refIpand0, 2020MNRAS.495.3373E, Kang_2020}.  In most studies, the authors found a strong correlation between $R$ and $\Gamma$. Although \cite{Mattson_2007} argued that this strong positive correlation could be due to the model degeneracies rather than any physical act, \cite{1999MNRAS.303L..11Z} suggested that the observed correlation could be explained by considering an internal feedback mechanism, where the medium emitting seed photons for the primary X-ray emission also serves as the medium for reflection. Recently, from the analysis of 14 nearby bright Seyfert galaxies, \cite{2020MNRAS.495.3373E} confirmed a strong correlation between $R$ and $\Gamma$. The authors argued that the observed correlation could be either due to the Compton cooling process or the changing geometry of the disk-corona system. Recently, \cite{Kang_2020} also reported a strong correlation between $R$ and $\Gamma$. According to the authors, a stronger reflection and a softer X-ray spectrum could be predicted in the case of an outflowing corona.

\subsection{Correlation between $\rm{E_{cut}}$ and $R$}
We investigated the correlation between $\rm{E_{cut}}$ and $R$, performing the correlation analysis using the four methods mentioned above. Since the survival analysis with the \textit{ASURV} package does not handle cases where the dependent variable has lower limits and the independent variable has upper limits simultaneously, we performed the survival analysis under the following conditions: (1) only the dependent variable ($\rm{E_{cut}}$) has a lower limit, (2) only the independent variable ($R$) has an upper limit, and (3) both dependent and independent have detected points. From the correlation analysis using the above-described methods i.e., considering only the constrained best-fit values of $\rm{E_{cut}}$ and $R$; using the survival analysis test (KMPL and \textit{ASURV} package); considering the asymmetric errors and the constant upper and lower limits of $\rm{E_{cut}^{MAX}}$ to be 1000 keV and $R^{MIN}$ = 0.01 and $\rm{E_{cut}^{MAX}}$ = 500 keV and $R^{MIN}$ = 0.01, we obtained $p$ = 0.84, 0.66, 0.08, 0.12 and 0.12 respectively, suggesting no significant correlation between these two parameters (see Fig. \ref{figure-20}). Previously, \cite{2021MNRAS.506.4960H} reported a mild anti-correlation between $\rm{E_{cut}}$ and $R$.

\subsection{Correlation between $\Gamma$ and $\rm{\lambda_{Edd}}$}
Next we examined the correlation between $\Gamma$ and $\rm{\lambda_{Edd}}$. Considering the whole sample, we noticed a positive trend, though insignificant, between these two parameters. The correlation analysis produced a $\rho$ of 0.17 and a $p$ value of 0.08. We also conducted an analysis accounting for the errors in the measurements of $\Gamma$. Still, it did not yield any significant correlation, which agrees with what was reported by \cite{refId0}. No significant correlation was found in the moderate ($\rm{\lambda_{Edd}}<$0.1) and high ($\rm{\lambda_{Edd}}>$0.1) Eddington ratio subsets. It is worth noting that previous studies \citep{2006ApJ...646L..29S, 2008ApJ...682...81S, 2009ApJ...700L...6R, 2017MNRAS.470..800T, 2018MNRAS.480.1819R} have reported a positive correlation between these two parameters. The correlation is plotted in Fig \ref{figure-7}.


\subsection{Correlation between $\rm{E_{cut}}$ and $\Gamma$}
The correlation between $\rm{E_{cut}}$ and $\Gamma$ is shown in Fig. \ref{figure-7}. From correlation analysis of the linear fit, we obtained a $\rho$ of 0.69 and a $p$ value of 1.75$\times 10^{-11}$ considering only the controlled measurements of $\rm{E_{cut}}$ and $\Gamma$, suggesting a significant correlation between these two parameters. The KMPL test of the survival analysis also indicated a strong correlation between $\rm{E_{cut}}$ and $\Gamma$ with $\rho$ = 0.99 and $p$ = 1.46$\times 10^{-5}$. The statistical analysis using the package \textit{ASURV} derived a $\rho$ = 0.57 with $p$ = 0.0, suggesting a strong correlation between these two parameters. Using the third approach, we found a $\rho$ of 0.60 and a $p$ value of 4.11$\times 10^{-12}$. We also obtained a significant correlation between them using our fourth approach and got a $\rho$ of 0.61 and a $p$ value of 1.75$\times 10^{-12}$. In the third and fourth cases, the errors in $\Gamma$ are taken care of by producing $10^5$ random points between $\Gamma^{min}$ and $\Gamma^{max}$ at each run, where $\Gamma^{min}$ and $\Gamma^{max}$ are the respective lower and upper bounds of $\Gamma$. 

Similar studies on the correlation analysis between the temperature of the corona and $\Gamma$ are available in the literature. From a study of 19 Seyfert galaxies using data from {\it NuSTAR}, \cite{refId0}  found no significant correlation between $\rm{kT_{e}}$ and $\Gamma$. In contrast, \cite{2019MNRAS.484.2735M} found a positive correlation between $\rm{E_{cut}}$ and $\Gamma$ based on spectral analysis of 18 Seyfert galaxies using data from {\it Swift-XRT} and {\it NuSTAR}. The authors suggested that the correlation observed in their sample could result from the systematic uncertainties affecting one of the two parameters or the lack of high-quality data in the soft X-ray regime. A few recent studies reported  positive correlation between $\rm{E_{cut}}$ and $\Gamma$ \citep{2018ApJ...866..124K,2021MNRAS.506.4960H, Kang_2022}. Of these, \cite{2018ApJ...866..124K} analysed a total of 46 Seyfert 1 galaxies, while \cite{2021MNRAS.506.4960H} and \cite{Kang_2022} carried out spectral analysis of 33 and 60 sources, respectively. While \cite{Kang_2022} found a relatively strong correlation between $\rm{E_{cut}}$ and $\Gamma$, it is rather weak, as reported in \cite{2018ApJ...866..124K} and \cite{2021MNRAS.506.4960H}. This may be due to the small sample size and domination of the lower limits in the $\rm{E_{cut}}$ measurements in their studies. While our study produced a stronger correlation between these two parameters for a large number of sources in which we could tightly constrain $\Gamma$ in all of them and $\rm{E_{cut}}$ in most of them, the influence of degeneracy between these two parameters could not be neglected. 
To check the degeneracy between $\rm{E_{cut}}$ and $\Gamma$ in our sample, we generated contours between these two parameters for all the sources, six of which are presented in Figure \ref{figure-11}. The contours are elliptical and have smooth levels, suggesting the parameters were well-constrained. While we noticed weak degeneracy in the contour plots, we managed to constrain both parameters effectively in the majority of cases. This implies that any potential artefacts between them may be minor, given the substantial number of sources analyzed in this study.

From an analysis of multiple epochs of observations on a source SWIFT J2127.4+5654, \cite{2021MNRAS.502...80K} found a $\Lambda$ shaped pattern. According to the authors, up to  $\Gamma$ $<$ 2.05, the source showed a "steeper-when-hotter" behaviour, while beyond $\Gamma$ $>$ 2.05, the source showed a "softer-when-cooler" behaviour. Though the finding of \cite{2021MNRAS.502...80K} is from multiple observations of a single source, we attempted to check the prevalence of such a trend in our sample of sources. There are only a few sources in our sample with $\Gamma$ $>$ 2.05, and the statistical test resulted in a negative trend between $\rm{E_{cut}}$ and $\Gamma$. However, we could not conclude about the significance of the anti-correlation noticed since very few sources were found in the $\Gamma$ $>$ 2.05 region. A systematic and homogeneous analysis of many sources is needed to confirm this finding. 

\subsection{Correlation between $\rm{kT_{e}}$ and $\tau$}
We calculated $\tau$ using the following Equation \citep{1996MNRAS.283..193Z,
1999MNRAS.309..561Z}

\begin{equation}
\label{equ2}
    \tau = \sqrt{\frac{9}{4} + \frac{3}{\theta\Big[\Big(\Gamma + \frac{1}{2}\Big)^2 - \frac{9}{4}\Big]}} - \frac{3}{2}  \\
\end{equation}   

where $\theta = {kT_e}/{m_{e}c^2}$. Considering only the constrain values of $\rm{kT_{e}}$ we found a strong negative correlation between $\rm{kT_{e}}$ and $\tau$ (see Fig. \ref{figure-8}). Spearman's rank correlation analysis yielded a $\rho$ of $-0.96$ with a $p$ value of 1.82$\times 10^{-23}$. Earlier, \cite{refId0} also found a strong anti-correlation between these parameters for slab and spherical geometries. The authors fitted a similar linear relation in the $\rm{kT_{e}}$ vs $\tau$ plane and reported,
\begin{equation}
   a = -0.7 \pm 0.2; b = 1.8 \pm 0.1. 
\end{equation}
for the spherical geometry.
We also found similar values of a and b from our linear fit to the data points,
\begin{equation}
   a = -1.24 \pm 0.07; b = 2.02 \pm 0.03.
\end{equation}
From the survival analysis, we obtained a $\rho$ of $-$0.99 and $p$ = 1.40$\times 10^{-24}$ indicating a strong anti-correlation between these two parameters. We could not perform the statistical test using the \textit{ASURV} package since it does not calculate the correlation when the dependent variable ($\rm{kT_{e}}$) has a lower limit, and the independent variable ($\tau$) has an upper limit simultaneously. Using an upper limit of 150 keV for unconstrained $\rm{kT_{e}}$, we confirmed a strong negative correlation between $\rm{kT_{e}}$ and $\tau$ with a $\rho$ of $-$0.66 and $p$ = 1.89$\times 10^{-10}$ (see Table \ref{table-4}).

\section{Comparison with previous work}
This section compares the best-fit values of $\rm{E_{cut}}$ from this work with those available in the literature. 
\label{sec:source cut-off description}
Of the 112 sources analysed in this work, we could constrain $\rm{E_{cut}}$ for 73 sources. For all these 112 sources, $\rm{E_{cut}}$ measurements were carried out using the most recent physical models ({\it xillver/relxill/(relxill+xillver)}) available. In the past, most of these nearby unobscured AGN were analysed vividly using mostly phenomenological models such as {\it pexrav/pexmon} etc.  Here, we present a comparison of the $\rm{E_{cut}}$ measurements obtained from our analysis with those available in the literature in Table \ref{table-2}. 


For the majority of the sources in literature, $\rm{E_{cut}}$ was reported using the broad-band spectral analysis of the {\it NuSTAR} data in conjunction with the soft X-ray data from several other instruments, such as {\it XMM-Newton}, {\it Swift-XRT} etc. \citep{2015MNRAS.451.4375F, 2018MNRAS.476.1258T, 2019MNRAS.484.2735M, refId04, 2021MNRAS.506.4960H, 2022ApJ...927...42K, refId02}. In a few references $\rm{E_{cut}}$ was obtained from the analysis of the {\it Swift-BAT}, {\it BeppoSAX} and {\it INTEGRAL} broad-band X-ray data \citep{refId03, Ricci_2017, 2013MNRAS.433.1687M, 2014ApJ...782L..25M}. As seen from Table \ref{table-2}, our results from the analysis of only the {\it NuSTAR} data agrees with the previous analysis (see Fig. \ref{figure-9}). Our derived $\rm{E_{cut}}$ also matches with those already reported in the literature using only the {\it NuSTAR} data \citep{refId0, 2018ApJ...866..124K, 2020ApJ...905...41B, 2020MNRAS.495.3373E, Kang_2020, refId01}. Since these sources are known to be variable, we noticed a mismatch in the $\rm{E_{cut}}$ values in a few cases where the epoch of observations differs from those used in this work. In Fig. \ref{figure-9}, we plotted the constrained $\rm{E_{cut}}$ values obtained both from this work and the previous work with green dots; the red dots represent the lower limit of $\rm{E_{cut}}$ from both this work and the literature, The black ones represent the constrained $\rm{E_{cut}}$ from this work and the lower limit of $\rm{E_{cut}}$ from the literature. Finally, we plotted the constrained $\rm{E_{cut}}$ from literature and the lower limit of $\rm{E_{cut}}$ values from this work using blue dots. The grey lines indicated the errors and the lower limits. Most of the sources lie around the 1:1 line (black dotted line), except for a few red dots, representing the lower limit obtained from this work is lower than that found in the literature.

\section{Discussion}
We examined the correlations between various coronal properties, as well as between coronal parameters and the physical properties of the sources studied in this work. We also examined whether moderately accreting sources ($\rm{\lambda_{Edd}}<$0.1) have different X-ray emission characteristics relative to the highly accreting sources ($\rm{\lambda_{Edd}}>$0.1).

From Table \ref{table-4}, we noticed a significant correlation between $\rm{E_{cut}}$ and $\Gamma$ (see Fig. \ref{figure-7}) for the entire sample of sources. Such positive correlation between $\rm{E_{cut}}$ and $\Gamma$ was also reported in the past \citep{2002A&A...389..802P,2018ApJ...866..124K,2019A&A...630A.131M,2021MNRAS.506.4960H,Kang_2022}. However, there are instances where the observed correlation between $\rm{E_{cut}}$ and $\Gamma$ was not definitively established \citep{2009MNRAS.399.1293M, Ricci_2017,refId0,2022ApJ...927...42K}. According to \cite{2019A&A...630A.131M}, the observed correlation might result from potential systematic uncertainties associated with one of the two parameters. The observed correlation could also be accounted for by the presence of an optically thin corona. Furthermore, the relationship between these two variables remains incomprehensible even for individual AGN \citep{2021MNRAS.502...80K}. 

From our analysis, another strong anti-correlation was found between $\rm{kT_{e}}$ and $\tau$ (see Fig. \ref{figure-8}). Such negative correlation between $\rm{kT_{e}}$ and $\tau$ is already known in literature and is attributed to either the fact that the cooling rate is more efficient in corona with higher opacity or to the variation in the intrinsic disk emission from the sources \citep{refId0, Kang_2022}. 
The positive correlation between $\rm{E_{cut}}$ and $\Gamma$  suggests that a steeper spectrum typically corresponds to a hotter corona. The corona must become optically thinner for a spectrum to steepen, indicating a hotter environment. Consequently, the inverse correlation between $\rm{kT_{e}}$ and $\tau$ indicates that a sustainable hotter corona has lower opacity and a softer spectrum. These findings challenge the traditional notion that a corona in a higher accreting system tends to be cooler due to its rapid interaction with the seed disk photons, resulting in a softer X-ray spectrum.  Therefore, the observed correlation is likely driven by variations in the intrinsic disk emission across different sources.

We therefore examined the correlations between $\rm{E_{cut}}$ and $\Gamma$ with the physical parameters of the sources ($\rm{\lambda_{Edd}}$ and $\rm{M_{BH}}$) (see Table \ref{table-4} and Fig. \ref{figure-22}). In the past, several authors have reported a positive correlation between $\Gamma$ and $\rm{\lambda_{Edd}}$ \citep{1997MNRAS.285L..25B,2006ApJ...646L..29S,2008ApJ...682...81S,2009ApJ...700L...6R,2013MNRAS.433.2485B, Brightman_2016,2017MNRAS.470..800T}.
Most authors have utilized a similar linear relationship, as described in Equation \ref{equ22}, to investigate the connection between these two parameters. We found a slope (b) of 0.26, with an $\rho$-value of 0.17 and a p-value of 0.08 from Spearman's correlation analysis. Our findings align with the slope reported by \cite{2008ApJ...682...81S} and \cite{2013MNRAS.433.2485B}, both of whom identified a similar slope of approximately 0.3 from the correlation analyses between $\Gamma$ and $\rm{\lambda_{Edd}}$. However, \cite{2009ApJ...700L...6R}, in their examination of {\it SDSS} quasars with archival {\it XMM–Newton} observations, reported a steeper slope (b$\sim$0.6). More recently, \cite{2017MNRAS.470..800T} employed {\it BASS} data and found a considerably weaker and flatter correlation (b$\sim$0.15).  \cite{2017MNRAS.470..800T} argued that as $\rm{M_{BH}}$ decreases, the number of optical-UV seed photons increases and due to the production of a larger amount of seed photons, the corona interacts with it rapidly, and that in turns cools the corona down resulting a softer spectrum. Therefore, based on this argument, one should expect a positive correlation between $\Gamma$ and $\rm{M_{BH}}$. From the analysis of our sample of sources, we could not confirm such a trend. We also explored the relationship between $\rm{E_{cut}}$ and $\lambda_{Edd}$. When analyzing the entire sample, as well as the higher and lower accretion regimes separately, no significant correlation was observed between these two parameters. These findings are consistent with similar results reported in the literature \citep{refId0, 2019A&A...630A.131M, 2021MNRAS.506.4960H, 2022ApJ...927...42K}. Additionally, we did not find any significant correlation between $\rm{E_{cut}}$ and $\rm{M_{BH}}$. Therefore, the lack of significant correlations between the coronal parameters and the physical properties of the sources suggests that the observed significant correlation between $\rm{E_{cut}}$ and $\Gamma$, with $\tau$ decreasing significantly as $\rm{kT_{e}}$ increases, may not be due to intrinsic changes in the sources themselves or due to the Compton cooling effect (i.e., cooler corona has steeper spectrum), but rather due to morphological variations in the corona across different sources. A larger dataset is needed to better understand the nature of these observed correlations.




\section{Summary}

In this study, we analysed the 3$-$79 {\it NuSTAR} spectra of a sample of 112 Seyfert 1 galaxies, the data for which were publicly available between August 2013 and May 2022 in {\it NuSTAR} Master Catalog. The motivation is to carry out a systematic study of the coronal properties of Seyfert 1 type AGN. From the physical model fits to the spectrum of  112 sources, we could constrain $\rm{E_{cut}}$ in 73 sources. We used physically motivated Comptonization models to derive various physical coronal properties. We could constrain $\rm{kT_{e}}$ in 42 sources. The results of this study are summarized below: 
\begin{enumerate}
\item Using Model$-$1, we estimated the median value of $\Gamma$ to be 1.79$\pm$0.02. Using Model$-$2, we derived a median of 1.86$\pm$0.01 for $\Gamma$.

\item The median values, as calculated using Model-1, were 104$\pm$8 keV and 173$\pm$18 keV for the best-fitted constrained and unconstrained values of $\rm{E_{cut}}$, respectively. When considering an upper limit of 1000 keV for the censored $\rm{E_{cut}}$ measurements and accounting for the asymmetric errors in the constrained $\rm{E_{cut}}$ values, the median value for $\rm{E_{cut}}$ across the entire sample was determined to be 153$\pm$8 keV. In the sub-sample with moderate accretion rates ($\rm{\lambda_{Edd}}<$0.1), the median $\rm{E_{cut}}$ was found to be 158$\pm$11 keV, while in the sub-sample with high accretion rates ($\rm{\lambda_{Edd}}>$0.1), the median $\rm{E_{cut}}$ is 150$\pm$10 keV.

\item The median values of constrained and unconstrained $\rm{kT_{e}}$ as estimated using Model$-$2 were 24$\pm$2 keV and 24$\pm$3 keV respectively. Using both the controlled and censored $\rm{kT_{e}}$ (considering an upper bound of 150 keV), the median value of $\rm{kT_{e}}$ was calculated to be 48$\pm$5 keV.

\item For our sample of sources we found $\rm{E_{cut}}$ is strongly correlated with $\rm{kT_{e}}$ as $\rm{E_{cut}}$ = (3.80 $\pm$ 0.53) $\rm{kT_{e}}$ + 
(8.15 $\pm$ 16.51). This is in agreement with the notion that
the X-ray spectra of AGN are related to the temperature of the corona as $\rm{E_{cut}}$ = 2$-$3 $\rm{kT_{e}}$. However, a large number of sources analysed in this work lie above the $\rm{E_{cut}}$ = 3 $\rm{kT_{e}}$ line, which is deviant from the general notion of $\rm{E_{cut}}$ = 2$-$3 $\rm{kT_{e}}$. Analysis of more sources is needed to confirm this.

\item For our entire sample, we found a strong correlation between $\rm{E_{cut}}$ and $\Gamma$.  

\item We found a significant anti-correlation between $\rm{kT_{e}}$ and $\tau$. The best-fit relation yielded a slope and intercept of $-1.24 \pm 0.07$ and 2.02 $\pm$ 0.03.

\item All these correlations indicate that an optically thin corona is necessary to sustain a hotter corona with a steeper spectrum. With the increasing accretion rate, the hotter corona could move vertically away from the central engine and become optically thinner. A systematic and homogeneous analysis of a larger sample of sources is needed to establish the correlation observed between various physical quantities, thereby enhancing our understanding of AGN corona. 


\end{enumerate}

\begin{acknowledgments}
We thank the anonymous referee for their very helpful
comments on the manuscript. We thank the {\it NuSTAR} Operations, Software and Calibration teams for support with the execution and analysis of these observations. This research has made use of the {\it NuSTAR} Data Analysis Software (NuSTARDAS) jointly developed by the ASI Science Data Center (ASDC, Italy) and the California Institute of Technology (USA). C.R. acknowledges support from the Fondecyt Regular (grant 1230345) and ANID BASAL (project FB210003). This research has used data and/or software provided by the High Energy Astrophysics Science Archive Research Center (HEASARC), a service of the Astrophysics Science Division at NASA/GSFC.
\end{acknowledgments}

%

\vspace{5mm}
\facilities{{\it NuSTAR}}


\software{FTOOLS \footnote{\href{https://heasarc.gsfc.nasa.gov/ftools/}{https://heasarc.gsfc.nasa.gov/ftools/}}
          }



\appendix
\section{Additional table}
\label{sec:tables}
\begin{center}
\startlongtable
\begin{longrotatetable}
\begin{deluxetable*}{lllccccccc}
\tablecaption{Details of the sources analysed in this work. The columns are (1) the name of the source, (2) right ascension (h:m:s), (3) declination (d:m:s), (4) redshift, (5) type of the source, (6) observation ID (OBSID), (7) count rate (counts/sec) (8) exposure time in sec (9) black hole mass, and (10) the Eddington ratio. Some of the information, including the right ascension, declination, and $z$, are from SIMBAD\footnote{\href{http://simbad.cds.unistra.fr/simbad/}{http://simbad.cds.unistra.fr/simbad/}}.} \label{table-1} 
\tablehead{\colhead{Source} & \colhead{$\alpha_{2000}$} & \colhead{$\delta_{2000}$} & \colhead{$z$} & \colhead{Type} & \colhead{OBSID} & \colhead{Count rate} & \colhead{Exposure} & \colhead{$\rm{M_{BH}}$/M$_{\odot}$} & \colhead{$L_{\rm{bol}}$/$L_{\rm{edd}}$} \\
\hline
}
\startdata
1H 0419-577      	&	 04 26 00.71 	&	 $-$57 12 01.76 	&	0.104	&	  Sy1.0 	&	60101039002	&	0.4	&	169462	&	8.06	&	0.05 \\
1H1934-063       	&	 19 37 33.02 	&	 $-$06 13 04.80 	&	0.01	&	  Sy1.0 	&	60702018006	&	0.52	&	65521	&	6.33	&	0.37	\\
2E1739.1-1210    	&	 17 41 55.25 	&	 $-$12 11 56.58 	&	0.037	&	 Sy1.2      	&	60160670002	&	0.3	&	21366	&	8.23	&	0.03	\\
2MASS J1830231+731310 	&	 18 30 23.16 	&	 +73 13 10.71  	&	0.123	&	 Sy1.0      	&	60464150002	&	0.16	&	26019	&	$-$	&	$-$	\\
2MASSJ17485512$-$3254521 	&	 17 48 55.13 	&	 $-$32 54 52.10  	&	0.02	&	  Sy1.0 	&	60160677002	&	0.27	&	21801	&	8.02	&	0.01	\\
2MASXJ04372814$-$4711298 	&	 04 37 28.16 	&	 $-$47 11 29.48 	&	0.053	&	 Sy1.0 	&	30001061002	&	0.12	&	73821	&	7.89	&	0.04	\\
2MASXJ11324928+1017473 	&	  11 32 49.27 	&	 +10 17 47.27 	&	0.044	&	 Sy1.0 	&	60061212002	&	0.05	&	20469	&	7.44	&	0.04	\\
2MASXJ12313717-4758019  	&	 12 31 37.14 	&	 $-$47 58 02.00  	&	0.028	&	 Sy1.0      	&	60160498002	&	0.14	&	19356	&	7.41	&	0.06	\\
2MASXJ15295830$-$1300397 	&	 15 29 58.33 	&	 $-$13 00 39.78 	&	0.104	&	  Sy1.0 	&	60160617002	&	0.15	&	24227	&	7.52	&	0.63	\\
2MASXJ1802473-145454 	&	 18 02 47.30 	&	 $-$14 54 55.00 	&	0.035	&	 Sy1.0      	&	60160680002	&	0.59	&	19958	&	7.56	&	0.24	\\
2MASXJ18470283-7831494   	&	 18 47 02.69 	&	 $-$78 31 49.60 	&	0.074	&	  Sy1.0 	&	60160699002	&	0.22	&	21505	&	8.37	&	0.07	\\
2MASXJ18560128+1538059 	&	 18 56 01.28 	&	 +15 38 05.90  	&	0.084	&	 Sy1.0 	&	60160701002	&	0.22	&	21352	&	8.47	&	0.06	\\
2MASXJ19380437$-$5109497 	&	 19 38 04.39 	&	 $-$51 09 49.38 	&	0.04	&	  Sy1.0 	&	60160716002	&	0.24	&	21830	&	7.43	&	0.17	\\
2MASXJ21192912+3332566   	&	 21 19 29.12 	&	 +33 32 56.67   	&	0.051	&	  Sy1.5 	&	60061358002	&	0.23	&	21483	&	7.71	&	0.15	\\
2MASXJ21355399+4728217  	&	  21 35 54.02 	&	 +47 28 21.89  	&	0.025	&	 Sy1.0      	&	60160761002	&	0.24	&	18704	&	7.24	&	0.11	\\
2MASXJ23013626-5913210 	&	 23 01 36.23 	&	 $-$59 13 21.08 	&	0.15	&	 Sy1.8 	&	60160814002	&	0.16	&	19500	&	$-$	&	$-$	\\
3C 109 	&	 04 13 40.34 	&	 +11 12 14.78 	&	0.306	&	 Sy1.8 	&	60301011004	&	0.17	&	89150	&	9.07	&	0.18	\\
3C 111	&	 04 18 21.27 	&	 +38 01 35.80 	&	0.05	&	  Sy1.0 	&	60202061004	&	0.74	&	49361	&	8.57	&	0.09	\\
3C 120           	&	 04 33 11.09 	&	 +05 21 15.61   	&	0.034	&	  Sy1.0 	&	60001042003	&	1.31	&	127716	&	7.99	&	0.19	\\
3C 206           	&	 08 39 50.58 	&	 $-$12 14 34.32  	&	0.198	&	 Sy1.2      	&	60160332002	&	0.29	&	17390	&	9.22	&	0.10	\\
3C 227   	&	 09 47 45.14 	&	 +07 25 20.59 	&	0.086	&	  Sy1.5     	&	60061329002	&	0.3	&	17195	&	8.94	&	0.03	\\
3C 380          	&	 18 29 31.78 	&	 +48 44 46.16 	&	0.692	&	 Sy1.0       	&	60160690002	&	0.13	&	19610	&	$-$	&	$-$	\\
3C 382           	&	 18 35 03.38 	&	 +32 41 46.85  	&	0.058	&	 Sy1.0      	&	60001084002	&	0.82	&	82583	&	8.6	&	0.08	\\
3C 390.3         	&	 18 42 08.99 	&	 +79 46 17.12 	&	0.06	&	  Sy1.0 	&	60001082003	&	1.03	&	47557	&	9.1	&	0.04	\\
6dFJ1254564-265702 	&	 12 54 56.37 	&	 $-$26 57 02.10  	&	0.059	&	 Sy1.0 	&	60363001002	&	0.14	&	20296	&	8.28	&	0.03	\\
ARK 120          	&	 05 16 11.40   	&	 $-$00 08 59.15 	&	0.03	&	  Sy1.0 	&	60001044004	&	0.99	&	65453	&	8.31	&	0.06	\\
Ark 241 	&	  10 21 40.25 	&	 $-$03 27 13.75 	&	0.041	&	 Sy1.0 	&	60160392002	&	0.18	&	20329	&	8.5	&	0.01	\\
ARK 564        	&	22 42 39.35	&	+29 43 31.31	&	0.025	&	Sy1.8	&	60401031004	&	0.28	&	408958	&	6.41	&	1.44	\\
CGCG229$-$015    	&	 19 05 25.94 	&	 +42 27 39.76 	&	0.028	&	  Sy1.0 	&	60160705002	&	0.13	&	21992	&	7.18	&	0.08	\\
ESO 025$-$G002   	&	 18 54 40.26 	&	 $-$78 53 54.10 	&	0.029	&	 Sy1.0 	&	60160700002	&	0.24	&	27978	&	7.35	&	0.11	\\
ESO 031$-$G008   	&	 03 07 35.34 	&	 $-$72 50 02.50  	&	0.028	&	  Sy1.0 	&	60160141002	&	0.19	&	31655	&	7.1	&	0.15	\\
ESO 209$-$G012   	&	 08 01 57.97 	&	 $-$49 46 42.39 	&	0.04	&	  Sy1.5 	&	60160315002	&	0.29	&	23715	&	8.11	&	0.05	\\
ESO 323$-$G077   	&	 13 06 26.12 	&	 $-$40 24 52.59 	&	0.015	&	  Sy1.5 	&	60202021006	&	0.13	&	43403	&	7.05	&	0.02	\\
ESO 416$-$G002 	&	 02 35 13.45 	&	 $-$29 36 17.25 	&	0.059	&	 Sy1.9 	&	60061340002	&	0.1	&	20606	&	$-$	&	$-$	\\
ESO 511$-$G030   	&	 14 19 22.40 	&	 $-$26 38 41.13 	&	0.022	&	  Sy1.0 	&	60502035008	&	0.12	&	41807	&	7.29	&	0.03	\\
ESO381-G007 	&	 12 40 46.96 	&	 $-$33 34 11.84 	&	0.055	&	 Sy1.5 	&	60160508002	&	0.12	&	21250	&	8.03	&	0.04	\\
FAIRALL 1146 	&	 08 38 30.77 	&	 $-$35 59 33.33 	&	0.032	&	 Sy1.5 	&	60061082002	&	0.34	&	21278	&	7.52	&	0.14	\\
Fairall 51       	&	 18 44 53.98  	&	  $-$62 21 52.87 	&	0.014	&	  Sy1.0 	&	60402014002	&	0.24	&	63532	&	7.33	&	0.02	\\
GRS 1734-292     	&	 17 37 28.38 	&	 $-$29 08 02.11 	&	0.021	&	 Sy1.0      	&	60301010002	&	0.15	&	26020	&	7.84	&	0.10	\\
H1821+643        	&	 18 21 57.21 	&	 +64 20 36.22 	&	0.297	&	  Sy1.0 	&	60160683002	&	0.37	&	22173	&	9.48	&	0.19	\\
HE 1143$-$1810   	&	 11 45 40.46 	&	 $-$18 27 14.96  	&	0.033	&	  Sy1.0 	&	60302002006	&	0.69	&	23096	&	7.38	&	0.40	\\
HE1136$-$2304    	&	 11 38 51.00  	&	  $-$23 21 35.34 	&	0.027	&	  Sy    	&	80002031003	&	0.26	&	63565	&	6.97	&	1.09	\\
IC 1198 	&	 16 08 36.38 	&	 +12 19 51.60 	&	0.033	&	 Sy1.5 	&	60361014002	&	0.11	&	26973	&	7.51	&	0.04	\\
IC 4329A         	&	 13 49 19.26 	&	 $-$30 18 34.21 	&	0.016	&	  Sy1.2 	&	60001045002	&	2.61	&	162390	&	7.88	&	0.11	\\
IGR J14471-6414 	&	 14 46 28.20 	&	 $-$64 16 24.00 	&	0.053	&	 Sy1.2 	&	60061257002	&	0.1	&	15042	&	7.61	&	0.09	\\
IGRJ14552$-$5133 	&	 14 55 17.51 	&	 $-$51 34 15.18 	&	0.016	&	  Sy1.0 	&	60401022002	&	0.23	&	100942	&	6.96	&	0.08	\\
IGRJ19378-0617   	&	 19 37 33.02   	&	 $-$06 13 04.80 	&	0.01	&	  Sy1.0 	&	60101003002	&	0.52	&	65521	&	6.33	&	0.37	\\
IRAS 05589+2828  	&	 06 02 10.47 	&	 +28 28 19.40   	&	0.033	&	  Sy1.0 	&	60061062002	&	0.78	&	29276	&	8.32	&	0.05	\\
IRAS 09149-6206 	&	 09 16 09.36 	&	 $-$62 19 29.56  	&	0.057	&	  Sy1.0     	&	90401630002	&	0.4	&	112121	&	8.76	&	0.03	\\
IRAS F12397+3333 	&	 12 42 10.60 	&	 +33 17 02.66 	&	0.044	&	  Sy1.0 	&	60501007002	&	0.16	&	48709	&	$-$	&	$-$	\\
IRAS04124$-$0803 	&	 04 14 52.66 	&	 $-$07 55 39.68 	&	0.039	&	 Sy1.0  	&	60761001002	&	0.32	&	18345	&	8	&	0.06	\\
IRAS04392-2713 	&	 04 41 22.53 	&	 $-$27 08 19.33  	&	0.084	&	 Sy1.0     	&	60160201002	&	0.19	&	19553	&	9.63	&	0.00	\\
KUG 1141+371 	&	 11 44 29.87 	&	 +36 53 08.61 	&	0.038	&	 Sy1.0 	&	90601618002	&	0.28	&	38562	&	8.06	&	0.05	\\
MCG-06-30-15 	&	 13 35 53.76 	&	 $-$34 17 44.16 	&	0.008	&	  Sy1.2 	&	60001047005	&	0.8	&	23267	&	6.09	&	0.52	\\
MCG+05-40-026    	&	 17 01 07.77 	&	 +29 24 24.58 	&	0.036	&	 Sy1.0 	&	60061276002	&	0.12	&	21000	&	6.86	&	0.25	\\
MCG+08-11-011 	&	05 54 53.61	&	+46 26 21.61	&	0.02	&	Sy1.5	&	60201027002	&	1.23	&	97921	&	7.9	&	0.09	\\
MR 2251$-$178    	&	 22 54 05.88 	&	 $-$17 34 55.40 	&	0.064	&	  Sy1.0 	&	60102025002	&	1.22	&	23112	&	8.34	&	0.30	\\
MRK 1040         	&	 02 28 14.46   	&	 +31 18 41.46 	&	0.017	&	  Sy1.5 	&	60101002004	&	0.69	&	64242	&	7.41	&	0.10	\\
MRK 1044         	&	 02 30 05.52 	&	 $-$08 59 53.20 	&	0.016	&	  Sy1.0 	&	60401005002	&	0.22	&	267078	&	6.23	&	0.90	\\
MRK 110          	&	 09 25 12.84 	&	 +52 17 10.38   	&	0.036	&	  Sy1.0 	&	60201025002	&	0.98	&	184563	&	7.13	&	1.24	\\
MRK 1148         	&	 00 51 54.76 	&	 +17 25 58.50 	&	0.064	&	  Sy1.0 	&	60160028002	&	0.5	&	22087	&	8.05	&	0.25	\\
MRK 1310 	&	 12 01 14.35 	&	 $-$03 40 41.01 	&	0.02	&	 Sy1.0 	&	60160465002	&	0.23	&	21131	&	6.83	&	0.17	\\
MRK 1383         	&	 14 29 06.57 	&	 +01 17 06.15   	&	0.086	&	  Sy1.0 	&	60501049002	&	0.18	&	95955	&	8.67	&	0.04	\\
MRK 1392 	&	 15 05 56.55 	&	 +03 42 26.33 	&	0.036	&	 Sy1.0 	&	60160605002	&	0.14	&	21084	&	7.9	&	0.03	\\
MRK 1393         	&	  15 08 53.95 	&	 $-$00 11 49.00 	&	0.054	&	  Sy1.5 	&	60376005002	&	0.21	&	30816	&	7.42	&	0.32	\\
MRK 205 	&	 12 21 44.07 	&	 +75 18 38.24 	&	0.071	&	 Sy1.0 	&	60160490002	&	0.21	&	20372	&	8.11	&	0.12	\\
Mrk 279          	&	 13 53 03.43  	&	  +69 18 29.41 	&	0.031	&	  Sy1.5 	&	60601011004	&	0.16	&	200632	&	7.89	&	0.02	\\
MRK 290          	&	 15 35 52.40 	&	 +57 54 09.51  	&	0.03	&	  Sy1.0 	&	60061266004	&	0.2	&	26348	&	7.64	&	0.05	\\
MRK 335          	&	 00 06 19.53 	&	 +20 12 10.61 	&	0.025	&	  Sy1.2 	&	60001041005	&	0.17	&	93022	&	7.08	&	0.11	\\
MRK 359          	&	 01 27 32.52 	&	 +19 10 43.83 	&	0.017	&	  Sy1.5 	&	60402021002	&	0.15	&	52526	&	6.11	&	0.43	\\
Mrk 509          	&	 20 44 09.75 	&	 $-$10 43 24.72 	&	0.034	&	  Sy1.5 	&	60101043002	&	1.19	&	165885	&	8.13	&	0.13	\\
MRK 590          	&	 02 14 33.56   	&	 $-$00 46 00.18 	&	0.026	&	  Sy1.2 	&	80502630002	&	0.33	&	68123	&	8.12	&	0.02	\\
MRK 595 	&	 02 41 34.87 	&	 +07 11 13.85 	&	0.027	&	 Sy1.5 	&	60160119002	&	0.06	&	21298	&	6.58	&	0.15	\\
MRK 684 	&	  14 31 04.78 	&	 +28 17 14.12 	&	0.045	&	 Sy1.0 	&	60160586002	&	0.08	&	20497	&	6.83	&	0.34	\\
MRK 704          	&	 09 18 25.99 	&	 +16 18 19.63   	&	0.029	&	  Sy1.5 	&	60061090002	&	0.27	&	21524	&	8.35	&	0.01	\\
MRK 732 	&	 11 13 49.75 	&	 +09 35 10.58 	&	0.029	&	 Sy1.5 	&	60061208002	&	0.21	&	26359	&	7.06	&	0.19	\\
MRK 79           	&	 07 42 32.82   	&	 +49 48 34.78 	&	0.022	&	  Sy1.2 	&	60601010002	&	0.58	&	65805	&	7.48	&	0.13	\\
MRK 813          	&	 14 27 25.05 	&	 +19 49 52.26 	&	0.11	&	  Sy1.0 	&	60160583002	&	0.21	&	24562	&	8.73	&	0.07	\\
MRK 817          	&	 14 36 22.08   	&	 +58 47 39.39  	&	0.031	&	  Sy1.5 	&	60601007002	&	0.21	&	135300	&	7.74	&	0.99	\\
MRK 841          	&	 15 04 01.19 	&	 +10 26 15.78 	&	0.036	&	  Sy1.5 	&	60101023002	&	0.44	&	23419	&	8.16	&	0.05	\\
MRK 876 	&	 16 13 57.18 	&	 +65 43 09.95 	&	0.121	&	 Sy1.0 	&	60160633002	&	0.1	&	29969	&	9.11	&	0.02	\\
MRK 885 	&	 16 29 48.38 	&	 +67 22 41.98 	&	0.025	&	 Sy1.5 	&	60160641002	&	0.08	&	28304	&	7.27	&	0.03	\\
MRK 915	&	 22 36 46.50 	&	 $-$12 32 42.89 	&	0.024	&	 Sy1.0 	&	60002060004	&	1.53	&	54249	&	7.13	&	0.07	\\
MRK 926 	&	 23 04 43.48 	&	 $-$08 41 08.62 	&	0.047	&	 Sy1.5 	&	60201029002	&	1.53	&	106201	&	8.37	&	0.18	\\
Mrk739E         	&	 11 36 29.30 	&	 +21 35 45.00    	&	0.03	&	    Sy1.0   	&	60260008002	&	0.12	&	18547	&	7.48	&	0.05	\\
NGC 0985          	&	 02 34 37.88 	&	 $-$08 47 17.02 	&	0.043	&	  Sy1.0 	&	60761008002	&	0.39	&	21326	&	8.25	&	0.05	\\
NGC 3227 	&	10 23 30.57	&	+19 51 54.28	&	0.004	&	Sy1.5	&	60202002002	&	0.96	&	49800	&	6.58	&	0.05	\\
NGC 3516         	&	 11 06 47.46   	&	 +72 34 07.29   	&	0.009	&	  Sy1.5 	&	60002042004	&	0.17	&	72088	&	7.11	&	0.01	\\
NGC 3783         	&	 11 39 01.71 	&	 $-$37 44 19.00 	&	0.009	&	  Sy1.0 	&	60101110002	&	1.11	&	41265	&	7.13	&	0.08	\\
NGC 4051         	&	 12 03 09.61  	&	 +44 31 52.68 	&	0.002	&	  Sy1.5 	&	60401009002	&	0.43	&	311139	&	5.95	&	0.02	\\
NGC 4579	&	12 37 43.52 	&	+11 49 05.49	&	0.005	&	Sy1.9	&	60201051002	&	0.17	&	117843	&	7.8	&	0.00	\\
NGC 4593         	&	 12 39 39.44 	&	 $-$05 20 39.03 	&	0.008	&	  Sy1.0 	&	60001149002	&	0.63	&	23317	&	6.77	&	0.08	\\
NGC 5273 	&	 13 42 08.38 	&	 +35 39 15.46 	&	0.004	&	 Sy1.9 	&	60061350002	&	0.46	&	21117	&	6.42	&	0.03	\\
NGC 5548         	&	 14 17 59.54 	&	 +25 08 12.60 	&	0.016	&	  Sy1.5 	&	60002044006	&	0.99	&	51460	&	7.97	&	0.04	\\
NGC 7469         	&	 23 03 15.67 	&	 +08 52 25.28   	&	0.017	&	  Sy1.2 	&	60101001002	&	0.75	&	21579	&	7.48	&	0.09	\\
NGC 931 	&	 02 28 14.46 	&	 +31 18 41.46 	&	0.017	&	  Sy1.0     	&	60101002004	&	0.74	&	64242	&	7.41	&	0.09	\\
PG0026+129  	&	 00 29 13.70 	&	 +13 16 03.94   	&	0.142	&	  Sy1.0 	&	60663003002	&	0.19	&	147374	&	7.82	&	0.85	\\
PG0052+251       	&	 00 54 52.11 	&	 +25 25 38.98 	&	0.155	&	  Sy1.2 	&	60661001002	&	0.13	&	24392	&	8.7	&	0.09	\\
PG0804+761 	&	 08 10 58.66 	&	 +76 02 42.45 	&	0.101	&	 Sy1.0 	&	60160322002	&	0.18	&	17315	&	7.9	&	0.33	\\
RBS 1037 	&	 11 49 18.68 	&	 $-$04 16 50.79 	&	0.085	&	 Sy1.0 	&	60061215002	&	0.1	&	40679	&	8.36	&	0.04	\\
RBS0295          	&	 02 14 37.40 	&	 $-$64 30 05.06 	&	0.074	&	  Sy1.0 	&	60061021002	&	0.13	&	23366	&	8.15	&	0.07	\\
RBS0770 	&	  09 23 43.00 	&	 +22 54 32.57 	&	0.033	&	 Sy1.2 	&	60602018002	&	0.57	&	42960	&	7.34	&	0.27	\\
S52116+81  	&	 21 14 01.17 	&	 +82 04 48.35 	&	0.084	&	 Sy1.0      	&	60061303002	&	0.36	&	18542	&	8.16	&	0.23	\\
SDSS J114921.52+532013.4 	&	 11 49 21.53 	&	  +53 20 13.29 	&	0.095	&	 Sy1.0 	&	60260009002	&	0.06	&	24886	&	8.16	&	0.01	\\
SDSSJ104326.47+110524.2 	&	  10 43 26.47 	&	 +11 05 24.26 	&	0.047	&	  Sy1.0 	&	60376004002	&	0.13	&	31062	&	8.01	&	0.04	\\
SWIFTJ2127.4+5654 	&	21 27 45.39	&	+56 56 34.91	&	0.0147	&	Sy1.0	&	60001110005	&	0.712	&	74578	&	7.15	&	0.15	\\
UGC 10120 	&	 15 59 09.62 	&	 +35 01 47.56 	&	0.031	&	 Sy1.0 	&	60560027002	&	0.05	&	62881	&	$-$	&	$-$	\\
UGC 3478         	&	 06 32 47.17 	&	 +63 40 25.28   	&	0.013	&	  Sy1.2 	&	60061068002	&	0.13	&	21680	&	$-$	&	$-$	\\
UGC03601 	&	 06 55 49.53 	&	 +40 00 01.12 	&	0.017	&	 Sy1.5 	&	60160278002	&	0.1	&	19674	&	7.33	&	0.02	\\
UGC06728     	&	 11 45 15.94  	&	 +79 40 53.37 	&	0.067	&	  Sy1.2 	&	60160450002	&	0.14	&	22615	&	5.28	&	51.96	\\
VII ZW 653       	&	 16 25 25.95 	&	 +85 29 41.69   	&	0.063	&	  Sy1.0  	&	60160639002	&	0.14	&	27580	&	7.46	&	0.26	\\
VII ZW 742 	&	 17 46 59.94 	&	 +68 36 39.59 	&	0.063	&	 Sy1.0 	&	60160676004	&	0.05	&	31393	&	7.37	&	0.10	\\
\hline
\enddata
\end{deluxetable*} 
\end{longrotatetable}
\end{center}

\begin{center}
\startlongtable
\begin{longrotatetable}
\begin{deluxetable*}{lcccccc}
\tablecaption{Best-fit parameters obtained from the model $\textsc{const} \times \textsc{TBabs} \times \textsc{zTBabs} \times ({\textsc{xillver/relxill/(relxill+xillver}}))$ to the source spectra. $\rm{E_{cut}}$ is in units of KeV.}\label{table-2}  
\tablehead{\colhead{Source} & \colhead{$\Gamma$} & \colhead{$\rm{E_{cut}}$} & \colhead{$R$} & \colhead{$\chi^2/dof$} & \colhead{$\rm{E_{cut}}$ from the literature} & \colhead{References} \\
\hline
}
\startdata
1H 0419-577      	&	 1.67$^{+0.03}_{-0.04}$ 	&	 59$^{+8}_{-7}$ 	&	0.25$^{+0.06}_{-0.06}$	&	 1318/1255  	&	83$^{+78}_{-31}$	&	{\cite{Ricci_2017}}	\\
	&		&		&		&		&	63$^{+8}_{-9}$	&	 {\cite{2018MNRAS.476.1258T}}	\\
	&		&		&		&		&	49$^{+7}_{-5}$	&	{\cite{2022ApJ...927...42K}}	\\
1H1934-063       	&	 2.34$^{+0.05}_{-0.06}$ 	&	200$^{+102}_{-52}$	&	0.62$^{+0.17}_{-0.12}$	&	1223/1215 	&	$\geq$126	&	{\cite{Ricci_2017}}	\\
2E1739.1-1210    	&	 1.89$^{+0.04}_{-0.03}$ 	&	 $>$286	&	0.57$^{+0.29}_{-0.17}$	&	443/405 	&	$\geq$230	&	{\cite{Ricci_2017}}	\\
2MASS J18302317+731310	&	 1.44$^{+0.05}_{-0.04}$ 	&	59$^{+14}_{-11}$	&	0.36$^{+0.45}_{-0.20}$	&	 298/280       	&	60$^{+49}_{-20}$	&	{\cite{refId01}}	\\
2MASSJ17485512$-$3254521 	&	1.61$^{+0.04}_{-0.04}$	&	75$^{+18}_{-14}$	&	$<$0.59	&	414/427  	&	159$^{+66}_{-55}$	&	{\cite{Ricci_2017}}	\\
2MASXJ04372814$-$4711298 	&	1.98$^{+0.05}_{-0.05}$	&	116$^{+96}_{-38}$	&	0.87$^{+0.49}_{-0.38}$	&	272/297 	&	$\geq$91	&	{\cite{Ricci_2017}}	\\
	&		&		&		&		&	$>$142	&	{\cite{refId01}}	\\
2MASXJ11324928+1017473 	&	 2.00$^{+0.10}_{-0.10}$ 	&	 $>$108	&	3.50$^{+3.00}_{-1.93}$	&	63/58  	&	$\geq$50	&	{\cite{Ricci_2017}}	\\
2MASXJ12313717-4758019  	&	 1.88$^{+0.06}_{-0.06}$ 	&	 $>$112	&	$<$0.89	&	136/178 	&	$\geq$231	&	{\cite{Ricci_2017}}	\\
2MASXJ15295830$-$1300397 	&	 1.79$^{+0.05}_{-0.05}$ 	&	 $>$201 	&	$<$0.60	&	224/214 	&	$\geq$34	&	{\cite{Ricci_2017}}	\\
2MASXJ1802473-145454 	&	 1.72$^{+0.06}_{-0.06}$ 	&	133$^{+165}_{-52}$	&	0.32$^{+0.09}_{-0.11}$	&	603/569 	&	$\geq$74	&	{\cite{Ricci_2017}}	\\
	&		&		&		&		&	66$^{+36}_{-18}$	&	{\cite{refId01}}	\\
2MASXJ18470283-7831494   	&	 1.80$^{+0.04}_{-0.04}$ 	&	122$^{+60}_{-36}$	&	0.59$^{+0.49}_{-0.24}$	&	 250/276  	&	$\geq$93	&	{\cite{Ricci_2017}}	\\
2MASXJ18560128+1538059 	&	 1.47$^{+0.04}_{-0.04}$ 	&	41$^{+5}_{-5}$	&	0.65$^{+0.40}_{-0.31}$	&	287/307 	&	43$^{+20}_{-11}$	&	{\cite{refId01}} \\
2MASXJ19380437$-$5109497 	&	 1.85$^{+0.05}_{-0.05}$ 	&	102$^{+64}_{-29}$	&	0.56$^{+0.42}_{-0.34}$	&	243/255 	&	78$^{+191}_{-42}$	&	{\cite{Ricci_2017}}	\\
	&		&		&		&		&	$>$195	&	{\cite{2022ApJ...927...42K}}	\\
2MASXJ21192912+3332566   	&	 1.80$^{+0.04}_{-0.04}$ 	&	82$^{+33}_{-18}$	&	0.46$^{+0.36}_{-0.27}$	&	351/344  	&	62$^{+150}_{-32}$	&	{\cite{Ricci_2017}}	\\
	&		&		&		&		&	89$^{+199}_{-38}$	&	{\cite{refId01}}	\\
2MASXJ21355399+4728217  	&	 1.66$^{+0.05}_{-0.04}$ 	&	56$^{+15}_{-10}$	&	$<$0.89	&	287/292  	&	67$^{+96}_{-23}$	&	{\cite{Ricci_2017}}	\\
	&		&		&		&		&	55$^{+50}_{-19}$	&	{\cite{refId01}} 	\\
2MASXJ23013626-5913210 	&	 1.68$^{+0.06}_{-0.06}$ 	&	 41$^{+8}_{-6}$ 	&	0.68$^{+0.62}_{-0.62}$	&	174/179 	&	31$^{+47}_{-13}$	&	{\cite{Ricci_2017}}	\\
	&		&		&		&		&	59$^{+150}_{-26}$	&	{\cite{2022ApJ...927...42K}}	\\
3C 109 	&	 1.64$^{+0.03}_{-0.03}$ 	&	72$^{+10}_{-9}$	&	0.33$^{+0.21}_{-0.15}$	&	587/627 	&	$\geq$115	&	{\cite{Ricci_2017}}	\\
	&		&		&		&		&	112$^{+62}_{-58}$	&	{\cite{2021MNRAS.506.4960H}}	\\
3C 111	&	 1.75$^{+0.01}_{-0.01}$ 	&	 124$^{+13}_{-16}$ 	&	0.11$^{+0.07}_{-0.06}$	&	 910/890   	&	$\geq$144	&	{\cite{Ricci_2017}}	\\
	&		&		&		&		&	136$^{+47}_{-29}$	&	{\cite{2014ApJ...782L..25M}}	\\
3C 120           	&	 1.82$^{+0.01}_{-0.01}$ 	&	147$^{+12}_{-9}$	&	0.32$^{+0.04}_{-0.04}$	&	1591/1594 	&	$\geq$193	&	{\cite{Ricci_2017}}	\\
	&		&		&		&		&	158$^{+8}_{-7}$	&	{\cite{2021MNRAS.506.4960H}}	\\
3C 206           	&	 1.76$^{+0.05}_{-0.05}$ 	&	112$^{+47}_{-33}$	&	$<$0.59	&	272/264  	&	$\geq$272	&	{\cite{Ricci_2017}}	\\
	&		&		&		&		&	$>$68	&	{\cite{Kang_2020}}	\\
	&		&		&		&		&	$>$53	&	{\cite{2022ApJ...927...42K}}	\\
	&		&		&		&		&	$>$79	&	{\cite{refId01}}	\\
3C 227   	&	 1.79$^{+0.05}_{-0.04}$ 	&	 $>$152	&	$<$0.24	&	 264/290        	&	$\geq$90	&	{\cite{Ricci_2017}}	\\
	&		&		&		&		&	$>$87	&	{\cite{Kang_2020}}	\\
3C 380          	&	 1.66$^{+0.06}_{-0.06}$ 	&	 $>$217 	&	$<$0.28	&	198/178 	&	$-$	&	$-$	\\
3C 382           	&	 1.65$^{+0.01}_{-0.01}$ 	&	 111$^{+22}_{-19}$ 	&	0.13$^{+0.03}_{-0.03}$	&	1188/1244 	&	158$^{+39}_{-76}$	&	{\cite{2022ApJ...927...42K}}	\\
	&		&		&		&		&	133$^{+98}_{-40}$	&	{\cite{2020MNRAS.495.3373E}}	\\
	&		&		&		&		&	215$^{+150}_{-60}$	&	{\cite{refId0}}	\\
3C 390.3         	&	 1.77$^{+0.01}_{-0.01}$ 	&	 144$^{+34}_{-19}$ 	&	0.21$^{+0.07}_{-0.08}$	&	997/1017  	&	166$^{+64}_{-37}$	&	{\cite{Ricci_2017}}	\\
	&		&		&		&		&	130$^{+42}_{-32}$	&	{\cite{2019MNRAS.484.2735M}}	\\
	&		&		&		&		&	120$\pm$20	&	{\cite{refId0}}	\\
6dFJ1254564-265702 	&	 1.58$^{+0.04}_{-0.05}$ 	&	 32$^{+7}_{-5}$ 	&	$<$0.78	&	 164/186       	&	$\geq$42	&	{\cite{Ricci_2017}}	\\
	&		&		&		&		&	91$^{+100}_{-50}$	&	{\cite{refId01}}	\\
ARK 120          	&	 1.95$^{+0.03}_{-0.03}$ 	&	346$^{+422}_{-133}$	&	 0.51$^{+0.10}_{-0.09}$ 	&	1147/1146  	&	$\geq$292	&	{\cite{Ricci_2017}}	\\
	&		&		&		&		&	$>$763	&	{\cite{2022ApJ...927...42K}}	\\
	&		&		&		&		&	233$^{+147}_{-67}$	&	{\cite{refId01}}	\\
Ark 241 	&	1.88$^{+0.05}_{-0.05}$	&	$>$115	&	0.90$^{+0.60}_{-0.47}$	&	197/214 	&	$-$	&	$-$	\\
ARK 564        	&	2.41$^{+0.04}_{-0.03}$	&	73$^{+30}_{-16}$	&	0.34$^{+0.12}_{-0.07}$	&	582/574 	&	43$^{+3}_{-3}$	&	{\cite{2017MNRAS.468.3489K}}	\\
CGCG229$-$015    	&	 1.71$^{+0.08}_{-0.06}$ 	&	 46$^{+14}_{-8}$  	&	0.93$^{+0.74}_{-0.56}$	&	 164/173 	&	$\geq$38	&	{\cite{Ricci_2017}}	\\
	&		&		&		&		&	54$^{+76}_{-22}$	&	{\cite{refId01}}	\\
ESO 025$-$G002   	&	 1.67$^{+0.04}_{-0.04}$ 	&	133$^{+93}_{-26}$	&	$>$0.23	&	 385/417   	&	$\geq$23	&	{\cite{Ricci_2017}}	\\
ESO 031$-$G008   	&	 2.04$^{+0.04}_{-0.04}$ 	&	 $>$286	&	0.74$^{+0.37}_{-0.29}$	&	305/354 	&	$\geq$76	&	{\cite{Ricci_2017}}	\\
ESO 209$-$12   	&	 1.90$^{+0.04}_{-0.03}$ 	&	 $>$260	&	 0.38$^{+0.24}_{-0.22}$ 	&	 399/427    	&	$\geq$91	&	{\cite{Ricci_2017}}	\\
ESO 323$-$G077   	&	 1.45$^{+0.03}_{-0.03}$ 	&	89$^{+14}_{-13}$	&	2.40$^{+0.85}_{-0.57}$ 	&	409/386 	&	115$^{+114}_{-42}$	&	{\cite{2022ApJ...927...42K}}	\\
ESO 416$-$G002 	&	 1.81$^{+0.07}_{-0.07}$ 	&	 $>$172 	&	0.31$^{+0.47}_{-0.23}$	&	119/125  	&	$\geq$366	&	{\cite{Ricci_2017}}	\\
ESO 511$-$G030   	&	 1.70$^{+0.04}_{-0.04}$ 	&	69$^{+29}_{-15}$	&	0.80$^{+0.50}_{-0.41}$	&	308/289 	&	$\geq$591	&	{\cite{Ricci_2017}}	\\
ESO381-G007 	&	 1.66$^{+0.07}_{-0.07}$ 	&	64$^{+35}_{-18}$	&	$<$1.00	&	163/161 	&	$\geq$76	&	{\cite{Ricci_2017}}	\\
FAIRALL 1146 	&	 2.03$^{+0.03}_{-0.03}$ 	&	138$^{+73}_{-39}$	&	1.12$^{+0.42}_{-0.35}$	&	433/423 	&	$\geq$72	&	{\cite{Ricci_2017}}	\\
	&		&		&		&		&	$>$166	&	{\cite{2022ApJ...927...42K}}	\\
Fairall 51       	&	 1.53$^{+0.33}_{-0.09}$ 	&	62$^{+115}_{-14}$	&	 4.11$^{+1.49}_{-0.62}$ 	&	780/762   	&	$\geq$105	&	{\cite{Ricci_2017}}	\\
GRS 1734-292     	&	 1.67$^{+0.02}_{-0.01}$ 	&	 75$^{+6}_{-5}$ 	&	0.27$^{+0.11}_{-0.06}$	&	 894/923      	&	84$^{+38}_{-26}$	&	{\cite{Ricci_2017}}	\\
	&		&		&		&		&	53$^{+13}_{-9}$	&	{\cite{2019MNRAS.484.2735M}} 	\\
	&		&		&		&		&	53$\pm$10	&	{\cite{refId0}}	\\
H1821+643        	&	 1.91$^{+0.03}_{-0.03}$ 	&	229$^{+221}_{-77}$	&	0.23$^{+0.21}_{-0.19}$	&	450/454 	&	$\geq$130	&	{\cite{Ricci_2017}}	\\
	&		&		&		&		&	114$^{+159}_{-44}$	&	{\cite{refId01}}	\\
HE 1143$-$1810   	&	 1.79$^{+0.07}_{-0.06}$ 	&	 104$^{+24}_{-17}$ 	&	0.33$^{+0.15}_{-0.14}$	&	 524/617   	&	183$^{+219}_{-59}$	&	{\cite{Ricci_2017}}	\\
HE 1136$-$2304    	&	1.61$^{+0.02}_{-0.02}$	&	80$^{+15}_{-11}$	&	 0.18$^{+0.14}_{-0.12}$ 	&	648/650 	&	$\geq$63	&	{\cite{Ricci_2017}}	\\
	&		&		&		&		&	 97$^{+136}_{-77}$ 	&	 {\cite{refId02}} 	\\
IC 1198 	&	1.75$^{+0.06}_{-0.05}$	&	124$^{+107}_{-43}$	&	0.96$^{+0.68}_{-0.47}$	&	191/195  	&	$-$	&	$-$	\\
IC 4329A         	&	 1.77$^{+0.01}_{-0.01}$ 	&	191$^{+14}_{-10}$	&	0.32$^{+0.04}_{-0.02}$	&	2211/2088 	&	236$^{+42}_{-26}$	&	{\cite{Ricci_2017}}	\\
IGR J14471-6414 	&	 2.01$^{+0.08}_{-0.08}$ 	&	 $>$153 	&	$<$1.98	&	115/106  	&	$\geq$78	&	{\cite{Ricci_2017}}	\\
IGRJ14552$-$5133 	&	 1.93$^{+0.02}_{-0.02}$ 	&	254$^{+194}_{-72}$	&	0.55$^{+0.15}_{-0.15}$	&	741/775 	&	$\geq$59	&	{\cite{Ricci_2017}}	\\
IGRJ19378-0617   	&	 2.11$^{+0.07}_{-0.06}$ 	&	228$^{+419}_{-83}$	&	 0.75$^{+0.35}_{-0.20}$ 	&	757/786  	&	$\geq$126	&	{\cite{Ricci_2017}}	\\
	&		&		&		&		&	241$^{+1377}_{-114}$	&	{\cite{2022ApJ...927...42K}}	\\
IRAS 05589+2828  	&	 1.83$^{+0.11}_{-0.09}$ 	&	 136$^{+109}_{-56}$ 	&	1.25$^{+0.57}_{-0.36}$	&	843/771 	&	71$^{+20}_{-14}$	&	{\cite{Ricci_2017}}	\\
IRAS 09149-6206 	&	 1.69$^{+0.14}_{-0.19}$ 	&	 81$^{+60}_{-26}$ 	&	0.80$^{+0.13}_{-0.11}$	&	1000/1005 	&	$\geq$99	&	{\cite{Ricci_2017}}	\\
IRAS F12397+3333 	&	 2.34$^{+0.04}_{-0.04}$ 	&	 $>$97 	&	0.76$^{+0.38}_{-0.36}$	&	453/399 	&	$-$	&	$-$	\\
IRAS04124$-$0803 	&	 1.53$^{+0.04}_{-0.04}$ 	&	 80$^{+21}_{-14}$ 	&	0.52$^{+0.27}_{-0.25}$	&	298/330 	&	$\geq$40	&	{\cite{Ricci_2017}}	\\
IRAS04392-2713 	&	 1.92$^{+0.06}_{-0.06}$  	&	 $>$188 	&	0.46$^{+0.20}_{-0.28}$	&	172/185  	&	$\geq$43	&	{\cite{Ricci_2017}}	\\
KUG 1141+371 	&	 1.92$^{+0.11}_{-0.14}$ 	&	90$^{+27}_{-17}$	&	0.39$^{+0.22}_{-0.19}$	&	 470/514 	&	$-$	&	$-$	\\
MCG-06-30-15 	&	 1.82$^{+0.11}_{-0.09}$ 	&	126$^{+23}_{-19}$	&	0.98$^{+0.66}_{-0.34}$	&	1574/1516 	&	123$^{+101}_{-39}$	&	{\cite{Ricci_2017}}	\\
	&		&		&		&		&	170$^{+240}_{-53}$	&	{\cite{refId03}}	\\
	&		&		&		&		&	63$^{+24}_{-14}$	&	{\cite{2014ApJ...782L..25M}}	\\
	&		&		&		&		&	$>$110	&	{\cite{refId0}}	\\
MCG+05-40-026    	&	 1.77$^{+0.07}_{-0.06}$ 	&	104$^{+151}_{-41}$	&	$<$0.96	&	161/155 	&	$-$	&	$-$	\\
MCG+08-11-011 	&	1.83$^{+0.01}_{-0.01}$	&	153$^{+15}_{-13}$	&	0.40$^{+0.06}_{-0.05}$	&	1506/1419 	&	252$^{+131}_{-60}$	&	{\cite{Ricci_2017}}	\\
	&		&		&		&		&	163$^{+53}_{-32}$	&	{\cite{2019MNRAS.484.2735M}}	\\
	&		&		&		&		&	171$^{+44}_{-30}$	&	{\cite{2014ApJ...782L..25M}}	\\
	&		&		&		&		&	175$^{+110}_{-50}$	&	{\cite{refId0}}	\\
MR 2251$-$178    	&	1.63$^{+0.01}_{-0.02}$	&	 96$^{+17}_{-9}$ 	&	$<$0.06	&	818/859 	&	$\geq$59	&	{\cite{Ricci_2017}}	\\
	&		&		&		&		&	132$^{+130}_{-68}$	&	{\cite{refId03}}	\\
	&		&		&		&		&	138$^{+57}_{-38}$	&	{\cite{2014ApJ...782L..25M}}	\\
MRK 1040         	&	 1.88$^{+0.01}_{-0.01}$ 	&	 300$^{+108}_{-70}$ 	&	0.61$^{+0.11}_{-0.10}$	&	1025/1007 	&	$\geq$152	&	{\cite{Ricci_2017}}	\\
	&		&		&		&		&	$>$356	&	{\cite{2020MNRAS.495.3373E}}	\\
	&		&		&		&		&	198$^{+212}_{-70}$	&	{\cite{2022ApJ...927...42K}}	\\
MRK 1044         	&	1.80$^{+0.05}_{-0.06}$ 	&	381$^{+553}_{-179}$	&	0.50$^{+0.14}_{-0.11}$	&	1004/1004 	&	$\geq$99	&	{\cite{Ricci_2017}}	\\
	&		&		&		&		&	$>$120	&	{\cite{refId01}}	\\
	&		&		&		&		&	$\geq$214	&	{\cite{2018ApJ...866..124K}}	\\
MRK 110          	&	1.70$^{+0.01}_{-0.01}$	&	92$^{+12}_{-10}$	&	0.12$^{+0.02}_{-0.02}$	&	1624/1576  	&	191$^{+207}_{-57}$	&	{\cite{Ricci_2017}}	\\
	&		&		&		&		&	117$^{+12}_{-17}$	&	{\cite{refId04}}	\\
MRK 1148         	&	 1.76$^{+0.03}_{-0.03}$ 	&	 99$^{+30}_{-20}$ 	&	$<$0.22	&	545/532  	&	$\geq$71	&	{\cite{Ricci_2017}}	\\
	&		&		&		&		&	101$^{+11}_{-9}$	&	{\cite{2021MNRAS.506.4960H}}	\\
MRK 1310 	&	1.82$^{+0.04}_{-0.04}$	&	$>$173	&	0.28$^{+0.31}_{-0.22}$	&	304/293  	&	$\geq$62	&	{\cite{Ricci_2017}}	\\
MRK 1383         	&	 1.92$^{+0.02}_{-0.02}$ 	&	 $>$276 	&	0.98$^{+0.21}_{-0.21}$	&	818/768 	&	$-$	&	$-$	\\
MRK 1392 	&	1.93$^{+0.06}_{-0.05}$	&	$>$187	&	0.84$^{+0.50}_{-0.45}$	&	185/193 	&	$\geq$91	&	{\cite{Ricci_2017}}	\\
MRK 1393         	&	1.95$^{+0.04}_{-0.04}$	&	$>$295	&	0.37$^{+0.26}_{-0.21}$	&	352/376 	&	$\geq$140	&	{\cite{Ricci_2017}}	\\
MRK 205 	&	 1.92$^{+0.05}_{-0.05}$ 	&	 131$^{+122}_{-45}$ 	&	0.60$^{+0.44}_{-0.35}$	&	259/255 	&	$\geq$56	&	{\cite{Ricci_2017}}	\\
	&		&		&		&		&	$>$108	&	{\cite{refId01}}	\\
	&		&		&		&		&	$>$365	&	{\cite{2018ApJ...866..124K}}	\\
MRK 279          	&	 1.49$^{+0.04}_{-0.05}$ 	&	68$^{+18}_{-13}$	&	 0.18$^{+0.11}_{-0.12}$ 	&	989/994  	&	$\geq$125	&	{\cite{Ricci_2017}}	\\
MRK 290          	&	 1.59$^{+0.04}_{-0.04}$ 	&	102$^{+46}_{-25}$	&	0.33$^{+0.28}_{-0.22}$	&	316/364 	&	184$^{+256}_{-100}$	&	{\cite{Ricci_2017}}	\\
	&		&		&		&		&	$>$53	&	{\cite{2020MNRAS.495.3373E}}	\\
MRK 335          	&	 1.98$^{+0.26}_{-0.18}$ 	&	 $>$74	&	 4.52$^{+3.73}_{-1.10}$ 	&	764/771 	&	$\geq$185	&	{\cite{Ricci_2017}}	\\
MRK 359          	&	1.91$^{+0.03}_{-0.03}$	&	$>$163	&	 0.86$^{+0.35}_{-0.31}$ 	&	485/433 	&	$\geq$40	&	{\cite{Ricci_2017}}	\\
Mrk 509          	&	1.77$^{+0.02}_{-0.03}$	&	123$^{+17}_{-18}$	&	0.32$^{+0.07}_{-0.07}$	&	1673/1603  	&	102$^{+43}_{-19}$	&	{\cite{Ricci_2017}}	\\
	&		&		&		&		&	60$^{+71}_{-23}$	&	{\cite{refId03}}	\\
MRK 590          	&	 1.68$^{+0.02}_{-0.02}$ 	&	127$^{+33}_{-23}$	&	 0.23$^{+0.13}_{-0.11}$ 	&	818/775 	&	$\geq$112	&	{\cite{Ricci_2017}}	\\
	&		&		&		&		&	66$^{+86}_{-26}$	&	{\cite{refId01}}	\\
MRK 595 	&	1.31$^{+0.18}_{-0.23}$	&	$>$35	&	frozen	&	104/100	&	75$^{+408}_{-42}$	&	{\cite{Ricci_2017}}	\\
MRK 684 	&	 2.14$^{+0.09}_{-0.08}$ 	&	 $>$150 	&	$<$2.52	&	99/103 	&	$\geq$127	&	{\cite{Ricci_2017}}	\\
MRK 704          	&	 1.80$^{+0.04}_{-0.04}$ 	&	207$^{+146}_{-64}$	&	1.11$^{+0.48}_{-0.32}$	&	374/342 	&	$\geq$261	&	{\cite{Ricci_2017}}	\\
MRK 732 	&	 1.78$^{+0.04}_{-0.04}$ 	&	 $>$279	&	0.20$^{+0.27}_{-0.18}$	&	324/352 	&	81$^{+200}_{-40}$	&	{\cite{Ricci_2017}}	\\
MRK 79           	&	 1.86$^{+0.05}_{-0.05}$ 	&	349$^{+516}_{-152}$	&	 0.55$^{+0.18}_{-0.15}$ 	&	993/968  	&	224$^{+366}_{-97}$	&	{\cite{Ricci_2017}}	\\
	&		&		&		&		&	402$^{+165}_{-90}$	&	{\cite{2022ApJ...927...42K}}	\\
MRK 813          	&	 1.98$^{+0.04}_{-0.04}$ 	&	 $>$252	&	 0.49$^{+0.35}_{-0.24}$ 	&	345/321 	&	$\geq$60	&	{\cite{Ricci_2017}}	\\
MRK 817          	&	1.65$^{+0.29}_{-0.18}$	&	$>$68	&	1.64$^{+0.39}_{-0.32}$	&	1013/1007  	&	$\geq$242	&	{\cite{Ricci_2017}}	\\
MRK 841          	&	1.80$^{+0.03}_{-0.03}$	&	125$^{+49}_{-30}$	&	 0.42$^{+0.23}_{-0.19}$ 	&	467/508 	&	$\geq$152	&	{\cite{Ricci_2017}}	\\
	&		&		&		&		&	139$^{+142}_{-49}$	&	{\cite{2021MNRAS.506.4960H}}	\\
MRK 876 	&	 1.81$^{+0.06}_{-0.06}$  	&	140$^{+154}_{-52}$	&	0.74$^{+0.55}_{-0.40}$	&	193/187 	&	$\geq$43	&	{\cite{Ricci_2017}}	\\
MRK 885 	&	1.90$^{+0.08}_{-0.06}$	&	$>$161	&	0.56$^{+0.83}_{-0.30}$	&	172/151 	&	$\geq$212	&	{\cite{Ricci_2017}}	\\
MRK 915	&	1.72$^{+0.03}_{-0.03}$	&	136$^{+68}_{-36}$	&	0.42$^{+0.26}_{-0.22}$	&	519/547 	&	$\geq$79	&	{\cite{Ricci_2017}}	\\
	&		&		&		&		&	58$^{+11}_{-7}$	&	{\cite{2021MNRAS.506.4960H}}	\\
MRK 926 	&	1.70$^{+0.02}_{-0.01}$	&	142$^{+33}_{-19}$	&	0.11$^{+0.03}_{-0.02}$	&	1491/1493  	&	320$^{+166}_{-79}$	&	{\cite{Ricci_2017}}	\\
	&		&		&		&		&	211$^{+235}_{-95}$	&	{\cite{refId03}}	\\
Mrk739E         	&	 2.07$^{+0.07}_{-0.07}$ 	&	 $>$241 	&	0.84$^{+0.43}_{-0.48}$	&	 133/132       	&	$\geq$50	&	{\cite{Ricci_2017}}	\\
NGC 0985          	&	 1.84$^{+0.03}_{-0.03}$ 	&	$>$188	&	0.42$^{+0.21}_{-0.25}$	&	393/469 	&	$\geq$102	&	{\cite{Ricci_2017}}	\\
NGC 3227 	&	1.64$^{+0.01}_{-0.01}$	&	94$^{+7}_{-6}$	&	0.62$^{+0.10}_{-0.10}$	&	1224/1163 	&	94$^{+16}_{-12}$	&	{\cite{Ricci_2017}}	\\
	&		&		&		&		&	60$^{+5}_{-4}$	&	{\cite{2021MNRAS.506.4960H}}	\\
	&		&		&		&		&	87$^{+16}_{-12}$	&	{\cite{2022ApJ...927...42K}}	\\
NGC 3516         	&	 1.90$^{+0.03}_{-0.03}$ 	&	 $>$448 	&	1.27$^{+0.29}_{-0.14}$	&	696/655 	&	132$^{+87}_{-43}$	&	{\cite{Ricci_2017}}	\\
NGC 3783         	&	 1.55$^{+0.07}_{-0.03}$ 	&	 112$^{+24}_{-19}$ 	&	0.90$^{+0.11}_{-0.12}$	&	1137/1145 	&	77$^{+16}_{-11}$	&	{\cite{2022ApJ...927...42K}}	\\
	&		&		&		&		&	98$^{+79}_{-34}$	&	{\cite{2015MNRAS.451.4375F}}	\\
NGC 4051         	&	1.78$^{+0.07}_{-0.07}$	&	$>$452	&	 1.33$^{+1.02}_{-0.67}$ 	&	1674/1647 	&	59$^{+25}_{-13}$	&	{\cite{Ricci_2017}}	\\
NGC 4579 	&	 1.73$^{+0.06}_{-0.06}$ 	&	82$^{+49}_{-23}$	&	0.29$^{+0.09}_{-0.07}$	&	819/739  	&	$-$	& $-$	\\
NGC 4593         	&	 1.87$^{+0.02}_{-0.02}$ 	&	 $>$648 	&	 0.57$^{+0.26}_{-0.13}$ 	&	 523/571   	&	$\geq$655	&	{\cite{Ricci_2017}}	\\
NGC 5273 	&	 1.46$^{+0.06}_{-0.06}$ 	&	68$^{+25}_{-16}$	&	1.03$^{+0.33}_{-0.28}$	&	593/574 	&	$\geq$294	&	{\cite{Ricci_2017}}	\\
	&		&		&		&		&	115$^{+91}_{-37}$	&	{\cite{refId01}}  	\\
	&		&		&		&		&	$>$220	&	{\cite{2022ApJ...927...42K}}	\\
NGC 5548         	&	 1.71$^{+0.03}_{-0.01}$ 	&	 118$^{+12}_{-8}$ 	&	0.48$^{+0.10}_{-0.06}$	&	1219/1143 	&	$\geq$281	&	{\cite{Ricci_2017}}	\\
	&		&		&		&		&	70$^{+40}_{-10}$	&	{\cite{refId05}}	\\
NGC 7469         	&	 1.95$^{+0.02}_{-0.02}$ 	&	122$^{+27}_{-21}$	&	0.77$^{+0.19}_{-0.18}$	&	647/692 	&	$\geq$316	&	{\cite{Ricci_2017}}	\\
	&		&		&		&		&	113$^{+33}_{-22}$	&	{\cite{2021MNRAS.506.4960H}}	\\
NGC 931 	&	1.88$^{+0.01}_{-0.01}$	&	229$^{+78}_{-42}$	&	0.63$^{+0.12}_{-0.12}$	&	974/954 	&	$\geq$152	&	{\cite{Ricci_2017}}	\\
PG0026+129  	&	 1.82$^{+0.02}_{-0.02}$ 	&	 110$^{+20}_{-15}$ 	&	0.34$^{+0.12}_{-0.11}$	&	903/856 	&	$\geq$45	&	{\cite{Ricci_2017}}	\\
PG0052+251       	&	 1.66$^{+0.02}_{-0.01}$ 	&	$>$76	&	$<$0.18	&	167/190 	&	$\geq$137	&	{\cite{Ricci_2017}}	\\
PG0804+761 	&	 1.94$^{+0.05}_{-0.05}$ 	&	 $>$269	&	0.71$^{+0.49}_{-0.35}$	&	217/221 	&	$\geq$67	&	{\cite{Ricci_2017}}	\\
RBS 1037 	&	2.00$^{+0.06}_{-0.06}$	&	 $>$133	&	1.04$^{+0.28}_{-0.30}$	&	208/200 	&	$\geq$34	&	{\cite{Ricci_2017}}	\\
RBS0295          	&	 1.73$^{+0.06}_{-0.05}$ 	&	 $>$91 	&	$<$0.42	&	212/218  	&	$-$	&	$-$	\\
RBS0770 	&	 1.65$^{+0.07}_{-0.04}$ 	&	59$^{+18}_{-12}$	&	0.57$^{+0.22}_{-0.18}$	&	755/713  	&	$\geq$256	&	{\cite{Ricci_2017}}	\\
	&		&		&		&		&	$\geq$267	&	{\cite{2018ApJ...866..124K}}	\\
S52116+81  	&	 1.75$^{+0.04}_{-0.04}$ 	&	103$^{+40}_{-24}$	&	0.37$^{+0.27}_{-0.22}$	&	370/409 	&	$\geq$175	&	{\cite{Ricci_2017}}	\\
	&		&		&		&		&	$>$93	&	{\cite{2019MNRAS.484.2735M}}	\\
SDSS J114921.52+532013.4 	&	 1.53$^{+0.10}_{-0.10}$ 	&	29$^{+9}_{-7}$	&	$<$2.02	&	88/75 	&	47$^{+86}_{-14}$	&	{\cite{Ricci_2017}}	\\
SDSSJ104326.47+110524.2 	&	 1.72$^{+0.05}_{-0.05}$ 	&	 $>$123 	&	$<$0.43	&	 220/236    	&	$\geq$91	&	{\cite{Ricci_2017}}	\\
SWIFTJ2127.4+5654 	&	1.89$^{+0.01}_{-0.01}$	&	84$^{+6}_{-6}$	&	0.75$^{+0.10}_{-0.09}$	&	1094/1089 	&	62$^{+25}_{-15}$	&	{\cite{Ricci_2017}}	\\
	&		&		&		&		&	92$^{+26}_{-17}$	&	{\cite{2021MNRAS.502...80K}}	\\
UGC 10120 	&	 1.91$^{+0.07}_{-0.06}$ 	&	 $>$225	&	$<$0.71	&	185/192 	&	$-$	&	$-$	\\
UGC 3478         	&	 1.99$^{+0.06}_{-0.06}$ 	&	 $>$98 	&	$<$1.07	&	212/196 	&	$-$	&	$-$	\\
UGC03601 	&	 1.49$^{+0.09}_{-0.08}$ 	&	58$^{+45}_{-19}$	&	$<$0.86	&	127/109 	&	74$^{+240}_{-32}$	&	{\cite{Ricci_2017}}	\\
UGC06728     	&	 1.62$^{+0.05}_{-0.05}$ 	&	 66$^{+20}_{-14}$ 	&	0.55$^{+0.43}_{-0.32}$	&	 223/241  	&	73$^{+31}_{-19}$	&	{\cite{Ricci_2017}}	\\
	&		&		&		&		&	63$^{+133}_{-25}$	&	{\cite{2022ApJ...927...42K}}	\\
VII ZW 653       	&	 2.05$^{+0.05}_{-0.05}$ 	&	 $>$114 	&	1.10$^{+0.61}_{-0.51}$	&	187/213 	&	$-$	&	$-$	\\
VII ZW 742 	&	 1.88$^{+0.08}_{-0.08}$ 	&	 $>$174	&	1.25$^{+1.04}_{-0.71}$	&	99/108 	&	$\geq$52	&	{\cite{Ricci_2017}}	\\
\hline
\enddata
\end{deluxetable*} 
\end{longrotatetable}
\end{center}

\begin{center}
\startlongtable
\begin{longrotatetable}
\begin{deluxetable*}{lccccccc}
\tablecaption{Best-fit parameters obtained from the model $\textsc{const} \times \textsc{TBabs} \times \textsc{zTBabs} \times ({\textsc{xillverCP/relxillCP/(relxillCP+xillverCP}}))$ to the source spectra. $\rm N_{H}^{INT}$ is the host galaxy hydrogen column density in units of $10^{22}$ atoms $\rm{cm^{-2}}$, $\rm{kT_{e}}$ is in units of KeV; the model normalization is in units of $10^{-4}$ photons keV$^{-1}$ cm$^{-2}$ s$^{-1}$.}\label{table-3}  
\tablehead{\colhead{Source} & \colhead{$\rm N_{H}^{INT}$} & \colhead{$\Gamma$} & \colhead{$\rm{kT_{e}}$} & \colhead{$R$} & \colhead{$\rm N_{\rm{relxillCP}}$} & \colhead{$\rm N_{\rm{xillverCP}}$} & \colhead{$\chi^2/dof$}} \label{table-3}  
\tablehead{\colhead{Source} & \colhead{$\rm N_{H}^{INT}$} & \colhead{$\Gamma$} & \colhead{$\rm{kT_{e}}$} & \colhead{$R$} & \colhead{$\rm N_{\rm{relxillCP}}$} & \colhead{$\rm N_{\rm{xillverCP}}$} & \colhead{$\chi^2/dof$} \\
\hline
}
\startdata
1H 0419-577      	&	 1.95$^{+0.36}_{-0.36}$ 	&	 1.82$^{+0.01}_{-0.01}$ 	&	 14$^{+1}_{-1}$ 	&	 0.15$^{+0.06}_{-0.06}$ 	&	$-$	&	 0.79$^{+0.01}_{-0.01}$ 	&	 1321/1255 	\\
1H1934-063       	&	$-$	&	 2.14$^{+0.03}_{-0.03}$ 	&	$>$53	&	 0.62$^{+0.16}_{-0.12}$ 	&	 0.97$^{+0.13}_{-0.13}$ 	&	$-$	&	1221/1215	\\
2MASS J1830231+731310	&	$-$	&	1.64$^{+0.03}_{-0.03}$	&	17$^{+22}_{-4}$	&	$<$0.55	&	$-$	&	0.31$^{+0.01}_{-0.01}$	&	307/280	\\
2MASSJ17485512$-$3254521 	&	$-$	&	 1.74$^{+0.03}_{-0.03}$ 	&	$>$15	&	$<$0.43	&	$-$	&	 0.68$^{+0.02}_{-0.02}$ 	&	420/427	\\
2MASXJ04372814$-$4711298 	&	$-$	&	2.03$^{+0.04}_{-0.04}$	&	$>$16	&	0.90$^{+0.50}_{-0.41}$	&	$-$	&	0.18$^{+0.01}_{-0.01}$	&	271/297	\\
2MASXJ1802473-145454 	&	$-$	&	1.79$^{+0.03}_{-0.04}$	&	29$^{+60}_{-11}$	&	0.27$^{+0.13}_{-0.10}$	&	1.11$^{+0.18}_{-0.12}$	&	$-$	&	604/569	\\
2MASXJ18470283-7831494   	&	$-$	&	1.87$^{+0.04}_{-0.04}$	&	$>$17	&	$<$0.95	&	$-$	&	0.44$^{+0.02}_{-0.02}$	&	194/263	\\
2MASXJ18560128+1538059 	&	$-$	&	1.72$^{+0.03}_{-0.03}$	&	13$^{+3}_{-2}$	&	0.49$^{+0.40}_{-0.41}$	&	$-$	&	0.49$^{+0.01}_{-0.01}$	&	295/307	\\
2MASXJ19380437$-$5109497 	&	$-$	&	 1.93$^{+0.04}_{-0.04}$ 	&	$>$17	&	$<$0.95	&	$-$	&	 0.43$^{+0.02}_{-0.02}$ 	&	244/255	\\
2MASXJ21192912+3332566   	&	$-$	&	 1.90$^{+0.03}_{-0.03}$ 	&	$>$15	&	 0.38$^{+0.36}_{-0.27}$ 	&	$-$	&	 0.45$^{+0.01}_{-0.01}$ 	&	354/344	\\
2MASXJ21355399+4728217  	&	$-$	&	1.81$^{+0.03}_{-0.03}$	&	16$^{+12}_{-4}$	&	0.40$^{+0.36}_{-0.35}$	&	$-$	&	0.43$^{+0.02}_{-0.02}$	&	291/292	\\
2MASXJ23013626-5913210 	&	$-$	&	 1.87$^{+0.04}_{-0.04}$ 	&	 13$^{+6}_{-3}$ 	&	 0.52$^{+0.65}_{-0.42}$ 	&	$-$	&	 0.30$^{+0.02}_{-0.02}$ 	&	 176/179 	\\
3C 109 	&	$-$	&	1.78$^{+0.02}_{-0.02}$	&	18$^{+8}_{-3}$	&	0.26$^{+0.17}_{-0.16}$	&	$-$	&	0.32$^{+0.01}_{-0.01}$	&	591/627	\\
3C 111	&	2.46$^{+0.39}_{-0.38}$	&	1.84$^{+0.01}_{-0.01}$	&	 37$^{+34}_{-10}$ 	&	 0.07$^{+0.07}_{-0.06}$ 	&	$-$	&	 2.46$^{+0.03}_{-0.03}$ 	&	1081/1025	\\
3C 120           	&	$-$	&	 1.85$^{+0.01}_{-0.01}$ 	&	 45$^{+19}_{-8}$ 	&	 0.23$^{+0.04}_{-0.04}$ 	&	 2.72$^{+0.11}_{-0.08}$ 	&	$-$	&	 1566/1592 	\\
3C 206           	&	$-$	&	1.84$^{+0.04}_{-0.04}$	&	$>$15	&	$<$0.52	&	$-$	&	0.63$^{+0.02}_{-0.02}$	&	271/264	\\
3C 382           	&	$-$	&	1.77$^{+0.01}_{-0.01}$	&	34$^{+18}_{-8}$	&	0.07$^{+0.05}_{-0.05}$	&	1.73$^{+0.02}_{-0.02}$	&	$-$	&	1276/1247	\\
3C 390.3         	&	 2.55$^{+0.41}_{-0.41}$ 	&	 1.85$^{+0.01}_{-0.01}$ 	&	 44$^{+49}_{-12}$ 	&	 0.17$^{+0.08}_{-0.08}$ 	&	$-$	&	 2.44$^{+0.03}_{-0.03}$ 	&	1002/1017	\\
6dFJ1254564-265702 	&	$-$	&	 1.69$^{+0.05}_{-0.04}$ 	&	 9$^{+4}_{-2}$ 	&	$<$1.09	&	$-$	&	 0.23$^{+0.01}_{-0.01}$ 	&	121/143	\\
ARK 120          	&	$-$	&	 1.96$^{+0.02}_{-0.03}$ 	&	$>$67	&	 0.50$^{+0.09}_{-0.07}$ 	&	 2.27$^{+0.06}_{-0.12}$ 	&	$-$	&	1147/1145	\\
ARK 564        	&	$-$	&	2.40$^{+0.02}_{-0.02}$	&	24$^{+10}_{-6}$	&	0.46$^{+0.29}_{-0.07}$	&	0.96$^{+0.10}_{-0.26}$	&	$<$0.08	&	1168/1165	\\
CGCG229$-$015    	&	$-$	&	 1.89$^{+0.05}_{-0.04}$ 	&	 17$^{+42}_{-6}$ 	&	 0.82$^{+0.83}_{-0.55}$ 	&	$-$	&	 0.21$^{+0.01}_{-0.01}$ 	&	168/173	\\
ESO 025$-$G002   	&	$-$	&	 1.74$^{+0.03}_{-0.03}$ 	&	$>$22	&	$<$0.25	&	$-$	&	 0.69$^{+0.02}_{-0.02}$ 	&	389/417	\\
ESO 323$-$G077   	&	 38$^{+3}_{-3}$ 	&	 1.70$^{+0.02}_{-0.02}$ 	&	 35$^{+13}_{-12}$ 	&	 2.35$^{+0.81}_{-0.66}$ 	&	$-$	&	 0.29$^{+0.02}_{-0.02}$ 	&	411/386	\\
ESO 511$-$G030   	&	$-$	&	 1.81$^{+0.04}_{-0.03}$ 	&	$>$16	&	 0.57$^{+0.55}_{-0.32}$ 	&	$-$	&	 0.22$^{+0.01}_{-0.01}$ 	&	313/289	\\
ESO381-G007 	&	$-$	&	1.80$^{+0.05}_{-0.05}$	&	$>$12	&	$<$0.77	&	$-$	&	0.25$^{+0.01}_{-0.01}$	&	165/161	\\
FAIRALL 1146 	&	$-$	&	2.05$^{+0.03}_{-0.03}$	&	$>$25	&	1.06$^{+0.53}_{-0.31}$	&	$-$	&	0.69$^{+0.02}_{-0.02}$	&	434/423	\\
Fairall 51       	&	 11.55$^{+1.60}_{-1.23}$ 	&	 2.00$^{+0.09}_{-0.17}$ 	&	$>$34	&	 4.42$^{+1.38}_{-1.04}$ 	&	 0.39$^{+0.05}_{-0.05}$ 	&	$-$	&	786/761	\\
GRS 1734-292     	&	 3.75$^{+0.46}_{-0.44}$ 	&	 1.81$^{+0.01}_{-0.01}$ 	&	 20$^{+4}_{-2}$ 	&	 0.24$^{+0.08}_{-0.09}$ 	&	$-$	&	 2.98$^{+0.05}_{-0.04}$ 	&	 818/856 	\\
H1821+643        	&	$-$	&	 1.94$^{+0.03}_{-0.03}$ 	&	$>$39	&	 0.21$^{+0.21}_{-0.17}$ 	&	$-$	&	 1.04$^{+0.03}_{-0.03}$ 	&	451/454	\\
HE 1143$-$1810   	&	$-$	&	 1.85$^{+0.02}_{-0.02}$ 	&	27$^{+19}_{-7}$ 	&	 0.26$^{+0.15}_{-0.13}$ 	&	$-$	&	 1.37$^{+0.02}_{-0.02}$ 	&	534/597	\\
HE1136$-$2304    	&	$-$	&	 1.73$^{+0.02}_{-0.02}$ 	&	 18$^{+7}_{-3}$ 	&	$<$0.24	&	$-$	&	 0.40$^{+0.01}_{-0.01}$ 	&	656/650	\\
IC 1198 	&	$-$	&	1.83$^{+0.04}_{-0.04}$	&	$>$18	&	0.91$^{+0.75}_{-0.47}$	&	$-$	&	0.21$^{+0.01}_{-0.01}$	&	193/195	\\
IC 4329A         	&	 1.89$^{+0.11}_{-0.12}$ 	&	 1.83$^{+0.003}_{-0.003}$ 	&	 64$^{+15}_{-12}$ 	&	 0.30$^{+0.03}_{-0.03}$ 	&	$-$	&	 6.27$^{+0.03}_{-0.04}$ 	&	 2245/2088 	\\
IGRJ14552$-$5133 	&	$-$	&	 1.96$^{+0.02}_{-0.02}$ 	&	$>$45	&	 0.53$^{+0.15}_{-0.13}$ 	&	$-$	&	 0.50$^{+0.01}_{-0.01}$ 	&	741/775	\\
IGRJ19378-0617   	&	$-$	&	 2.13$^{+0.05}_{-0.05}$ 	&	$>$46	&	 0.74$^{+0.33}_{-0.19}$ 	&	 0.91$^{+0.17}_{-0.19}$ 	&	$-$	&	757/786 	\\
IRAS 05589+2828  	&	$-$	&	 1.90$^{+0.11}_{-0.07}$ 	&	 43$^{+120}_{-24}$ 	&	 1.00$^{+1.03}_{-0.40}$ 	&	 1.19$^{+0.14}_{-0.27}$ 	&	$-$	&	 788/738 	\\
IRAS 09149-6206 	&	$-$	&	1.90$^{+0.10}_{-0.08}$	&	18$^{+21}_{-4}$	&	2.05$^{+0.68}_{-0.47}$	&	0.60$^{+0.02}_{-0.02}$	&	$-$	&	992/1005	\\
IRAS04124$-$0803 	&	$-$	&	1.67$^{+0.03}_{-0.03}$	&	15$^{+4}_{-3}$	&	0.39$^{+0.27}_{-0.23}$	&	$-$	&	0.58$^{+0.02}_{-0.02}$	&	292/330	\\
KUG 1141+371 	&	$-$	&	1.92$^{+0.02}_{-0.02}$	&	29$^{+79}_{-11}$	&	0.33$^{+0.22}_{-0.18}$	&	$-$	&	0.58$^{+0.01}_{-0.01}$	&	517/547	\\
MCG-06-30-15 	&	$-$	&	 1.97$^{+0.06}_{-0.04}$ 	&	 50$^{+98}_{-11}$ 	&	 0.58$^{+0.39}_{-0.12}$ 	&	 1.71$^{+0.31}_{-0.47}$ 	&	 0.89$^{+0.19}_{-0.18}$ 	&	1572/1517	\\
MCG+05-40-026    	&	$-$	&	1.84$^{+0.06}_{-0.06}$	&	$>$16	&	$<$0.78	&	$-$	&	0.27$^{+0.02}_{-0.01}$	&	162/155	\\
MCG+08-11-011 	&	$-$	&	1.88$^{+0.01}_{-0.01}$	&	56$^{+26}_{-16}$	&	0.35$^{+0.06}_{-0.05}$	&	2.92$^{+0.02}_{-0.02}$	&	$-$	&	1506/1419	\\
MR 2251$-$178    	&	$-$	&	 1.76$^{+0.01}_{-0.01}$ 	&	 22$^{+8}_{-4}$ 	&	$<$0.02	&	$-$	&	 2.59$^{+0.02}_{-0.02}$ 	&	835/859	\\
MRK 1040         	&	$-$	&	1.91$^{+0.01}_{-0.01}$	&	$>$62	&	0.60$^{+0.12}_{-0.10}$	&	$-$	&	1.48$^{+0.02}_{-0.02}$	&	1029/1007	\\
MRK 1044         	&	$-$	&	 2.15$^{+0.03}_{-0.05}$ 	&	$>$96	&	 0.53$^{+0.16}_{-0.15}$ 	&	 0.63$^{+0.16}_{-0.15}$ 	&	 0.19$^{+0.10}_{-0.11}$ 	&	1388/1286	\\
MRK 110          	&	$-$	&	 1.82$^{+0.01}_{-0.01}$ 	&	 25$^{+4}_{-3}$ 	&	$<$0.09	&	2.03$^{+0.01}_{-0.01}$	&	$-$	&	1630/1576  \\
MRK 1148         	&	$-$	&	 1.85$^{+0.02}_{-0.02}$ 	&	 25$^{+74}_{-8}$ 	&	$<$0.17	&	$-$	&	 1.04$^{+0.02}_{-0.02}$ 	&	555/532	\\
MRK 205 	&	$-$	&	1.97$^{+0.04}_{-0.04}$	&	$>$20	&	0.57$^{+0.43}_{-0.35}$	&	$-$	&	0.47$^{+0.02}_{-0.02}$	&	260/255	\\
MRK 279          	&	$-$	&	 1.68$^{+0.01}_{-0.01}$ 	&	 15$^{+2}_{-2}$ 	&	 0.13$^{+0.09}_{-0.08}$ 	&	 0.29$^{+0.02}_{-0.01}$ 	&	 0.09$^{+0.02}_{-0.02}$ 	&	1007/995	\\
MRK 290          	&	$-$	&	 1.70$^{+0.03}_{-0.03}$ 	&	$>$15	&	$<$0.51	&	$-$	&	 0.43$^{+0.01}_{-0.01}$ 	&	319/364	\\
Mrk 509          	&	$-$	&	 1.82$^{+0.02}_{-0.01}$ 	&	 23$^{+3}_{-2}$ 	&	 0.19$^{+0.03}_{-0.04}$ 	&	 2.15$^{+0.05}_{-0.04}$ 	&	$-$	&	1691/1603	\\
MRK 590          	&	 2.07$^{+0.59}_{-0.57}$ 	&	 1.82$^{+0.03}_{-0.03}$ 	&	$>$49	&	 0.20$^{+0.12}_{-0.11}$ 	&	 0.92$^{+0.19}_{-0.16}$ 	&	$-$	&	815/774	\\
MRK 704          	&	 10.24$^{+1.38}_{-1.36}$ 	&	 1.86$^{+0.03}_{-0.03}$ 	&	$>$30	&	 1.11$^{+0.50}_{-0.37}$ 	&	$-$	&	 0.62$^{+0.04}_{-0.03}$ 	&	374/342	\\
MRK 79           	&	$-$	&	 1.88$^{+0.04}_{-0.04}$ 	&	$>$47	&	 0.53$^{+0.19}_{-0.16}$ 	&	 1.26$^{+0.06}_{-0.13}$ 	&	$-$	&	994/968	\\
MRK 841          	&	$-$	&	 1.86$^{+0.02}_{-0.02}$ 	&	 33$^{+185}_{-14}$ 	&	 0.37$^{+0.23}_{-0.19}$ 	&	$-$	&	 0.89$^{+0.02}_{-0.02}$ 	&	469/508	\\
MRK 876 	&	$-$	&	1.86$^{+0.05}_{-0.04}$	&	$>$15	&	0.70$^{+0.56}_{-0.39}$	&	$-$	&	0.18$^{+0.01}_{-0.01}$	&	192/187	\\
MRK 915	&	 6.90$^{+0.98}_{-1.04}$ 	&	 1.81$^{+0.03}_{-0.03}$ 	&	$>$28	&	 0.33$^{+0.25}_{-0.16}$ 	&	$-$	&	 0.45$^{+0.01}_{-0.02}$ 	&	522/547	\\
MRK 926 	&	$-$	&	1.78$^{+0.01}_{-0.01}$	&	31$^{+7}_{-5}$	&	0.10$^{+0.02}_{-0.02}$	&	3.13$^{+0.11}_{-0.09}$	&	$-$	&	1525/1494	\\
NGC 3227 	&	2.66$^{+0.34}_{-0.33}$	&	1.76$^{+0.01}_{-0.01}$	&	26$^{+6}_{-4}$	&	0.56$^{+0.09}_{-0.09}$	&	$-$	&	1.89$^{+0.02}_{-0.02}$	&	1198/1163	\\
NGC 3783         	&	$-$	&	 1.74$^{+0.02}_{-0.02}$ 	&	 48$^{+48}_{-9}$ 	&	 0.41$^{+0.20}_{-0.16}$ 	&	 2.33$^{+0.23}_{-0.15}$ 	&	 0.76$^{+0.27}_{-0.26}$ 	&	1270/1132	\\
NGC 4579 	&	$-$	&	1.84$^{+0.03}_{-0.03}$	&	23$^{+23}_{-7}$	&	0.25$^{+0.09}_{-0.07}$	&	0.33$^{+0.04}_{-0.03}$	&	$-$	&	828/739	\\
NGC 5273 	&	$-$	&	1.79$^{+0.06}_{-0.10}$	&	$>$18	&	0.98$^{+0.53}_{-0.33}$	&	1.06$^{+0.21}_{-0.20}$	&	$-$	&	589/572	\\
NGC 5548         	&	 3.32$^{+0.34}_{-0.35}$ 	&	 1.81$^{+0.01}_{-0.01}$ 	&	 35$^{+9}_{-7}$ 	&	 0.43$^{+0.10}_{-0.09}$ 	&	$-$	&	 2.23$^{+0.07}_{-0.03}$ 	&	1228/1143	\\
NGC 7469         	&	 $-$	&	 2.00$^{+0.02}_{-0.02}$ 	&	 45$^{+52}_{-17}$ 	&	 0.74$^{+0.20}_{-0.17}$ 	&	$-$	&	 1.59$^{+0.03}_{-0.03}$ 	&	 650/692 	\\
NGC 931 	&	$-$	&	1.91$^{+0.01}_{-0.01}$	&	$>$50	&	0.58$^{+0.14}_{-0.09}$	&	$-$	&	1.40$^{+0.02}_{-0.02}$	&	979/954	\\
PG0026+129  	&	$-$	&	 1.89$^{+0.01}_{-0.01}$ 	&	 22$^{+9}_{-4}$ 	&	 0.30$^{+0.12}_{-0.11}$ 	&	$-$	&	 0.39$^{+0.12}_{-0.11}$ 	&	 899/856 	\\
RBS0770 	&	$-$	&	1.78$^{+0.02}_{-0.02}$	&	15$^{+3}_{-2}$	&	0.42$^{+0.16}_{-0.13}$	&	0.75$^{+0.04}_{-0.04}$	&	$-$	&	757/713	\\
S52116+81  	&	$-$	&	1.84$^{+0.03}_{-0.03}$	&	$>$18	&	0.29$^{+0.26}_{-0.20}$	&	$-$	&	0.74$^{+0.02}_{-0.02}$	&	373/409	\\
SDSS J114921.52+532013.4 	&	$-$	&	1.76$^{+0.07}_{-0.07}$	&	7$^{+1}_{-1}$	&	1.38$^{+3.34}_{-1.26}$	&	$-$	&	0.08$^{+0.01}_{-0.01}$	&	88/75	\\
SWIFTJ2127.4+5654 	&	$-$	&	1.96$^{+0.01}_{-0.01}$	&	21$^{+3}_{-2}$	&	0.72$^{+0.10}_{-0.10}$	&	$-$	&	1.49$^{+0.01}_{-0.01}$	&	1084/1089	\\
UGC03601 	&	$-$	&	1.68$^{+0.07}_{-0.06}$	&	$>$9	&	$<$0.53	&	$-$	&	0.24$^{+0.01}_{-0.01}$	&	129/109	\\
UGC06728     	&	$-$	&	 1.73$^{+0.07}_{-0.07}$ 	&	 17$^{+25}_{-5}$ 	&	0.34$^{+0.42}_{-0.21}$	&	0.31$^{+0.58}_{-0.39}$	&	$-$	&	223/239	\\
\hline
\enddata
\end{deluxetable*} 
\end{longrotatetable}
\end{center}

\section{$\rm{kT_{e}}$: The temperature of the corona}
\label{sec:source description}
Here, we discuss the results obtained from the spectral analysis using the model  \textsc{const $\times$ tbabs $\times$ ztbabs $\times$ (xillverCP/relxillCP/(relxillCP+xillverCP))}. We also compare the best-fit values of our analysis with the previously measured values of $\rm{kT_{e}}$ from the literature, if available. Among the 42 sources for which we could constrain $\rm{kT_{e}}$, 18 sources were already discussed  by us earlier \citep{2022A&A...662A..78P, 2023MNRAS.518.2529P}.  Therefore, here we give details on the rest of the 24 sources. 

\noindent {\bf 2MASXJ23013626-5913210:} This source at a redshift $z$ = 0.150 was observed by {\it NuSTAR} once in 2017. We used Model$-$2a to estimate the coronal properties of the source. We found the source spectra to be well described with $\Gamma$ = 1.87$^{+0.04}_{-0.04}$ and $\rm{kT_{e}}$ = 13.35$^{+06.23}_{-03.48}$ keV. Previously, using similar Comptonization model \cite{2022ApJ...927...42K} found a value  of $\Gamma$ = 1.78$^{+0.14}_{-0.11}$ and $\rm{kT_{e}}$ = 11.10$^{+12.22}_{-02.87}$ keV. Our results are thus in agreement with \cite{2022ApJ...927...42K}.

\noindent {\bf 3C 120:} This is a radio-loud Seyfert 1 galaxy at $z$=0.033. {\it NuSTAR} observed the source twice on the same day in February 2013. Of the two observations, we analysed the spectrum with the highest exposure time using Model$-$2b. We obtained values of $\Gamma$ = 1.85$^{+0.01}_{-0.01}$ and  $\rm{kT_{e}}$ = 45.31$^{+18.79}_{-07.82}$ keV. Analysing the same data set using {\it relxillCP} \cite{Kang_2022} found a lower limit of $\rm{kT_{e}}$ $>$ 91 keV.

\noindent {\bf 3C 390.3:} This radio-loud Seyfert 1 galaxy at $z$=0.05613 was observed twice by {\it NuSTAR}  on the same day in May 2013. From spectral analysis of the data using Model$-$2a we obtained  $\Gamma$ = 1.84$^{+0.01}_{-0.01}$ and $\rm{kT_{e}}$ = 44.13$^{+54.75}_{-12.50}$ keV. For the same data set \cite{Kang_2022} and \cite{2022ApJ...927...42K} reported lower limits of $\rm{kT_{e}}$ $>$ 46 keV and $\rm{kT_{e}}$ $>$ 49.86 keV respectively.

\noindent {\bf ARK 564:} This source was observed by {\it NuSTAR} three times between May 2015 and November 2018. Of these, results on the observation done by {\it NuSTAR} in September 2018 is reported in this work for the first time. Fitting the observed data with Model$-$2c, we  obtained $\Gamma$ = 2.40$^{+0.02}_{-0.02}$ and  $\rm{kT_{e}}$ = 24.28$^{+13.60}_{-04.29}$ keV respectively. From analysis of the data acquired by {\it NuSTAR} in 2015,  \cite{2017MNRAS.468.3489K} determined $\rm{kT_{e}}$ = 15$\pm$2 keV arguing the source to have the coolest corona. Also, based on two epochs of data, \cite{2020MNRAS.492.3041B} reported variation in the temperature of the corona.

\noindent {\bf HE 1136$-$2304:} Active Galactic Nuclei exhibit flux variations across various timescales and across the entire electromagnetic spectrum. In the last decade, an increasing number of sources have displayed notably more pronounced changes in their flux and spectral characteristics, both in the X-ray range and the optical/UV range. These events are often referred to as changing-look AGN \citep{2022arXiv221105132R}. HE 1136$-$2304 is such a changing look AGN. It was found to change its optical spectral nature from Type 2 in 1993 to Type 1.5 in 2014 \citep{2016MNRAS.461.1927P}. It was observed by {\it NuSTAR}  twice on the same day in July 2014. Of the two, we analysed the spectrum with maximum exposure. The best-fit values obtained from fitting Model$-$2a to the source spectrum were $\Gamma$ = 1.78$^{+0.03}_{-0.02}$ and $\rm{kT_{e}}$ = 27.81$^{+78.85}_{-09.30}$ keV. From an analysis of same {\it NuSTAR} spectrum using {\it relxillCP} \cite{Kang_2022} obtained a lower limit of $\rm{kT_{e}}$ $>$ 21 keV.

\noindent {\bf IC 4329A:} This Seyfert 1 galaxy was observed six times by {\it NuSTAR}, once in 2012 and the others during August 2021. We analysed here the {\it NuSTAR} spectrum taken in 2012. Fitting the spectrum using Model$-$2a we obtained  best-fit values of $\Gamma$ and $\rm{kT_{e}}$ as 1.83$^{+0.003}_{-0.003}$ and 64.16$^{+15.41}_{-11.63}$ keV respectively. This source has been studied extensively in the past.  For example, \cite{refId0} reported $\rm{kT_{e}}$ = 37$\pm$7 keV from fitting {\it compTT} for a slab geometry. \cite{Kang_2022} estimated $\rm{kT_{e}}$ = 71$^{+37}_{-15}$ keV using {\it relxillCP} model. \cite{2022ApJ...927...42K} also found $\rm{kT_{e}}$ = 82$^{+16}_{-7}$ keV from {\it xillverCP} fit to the source spectrum.

\noindent{\bf 2MASXJ21355399+4728217:} This Seyfert galaxy was observed by {\it NuSTAR} on September 2019. From the analysis of the source spectrum \cite{refId01} reported $\rm{E_{cut}}$ = 55$^{+50}_{-19}$ keV. We analysed the same observation ID using Model$-$2a and found $\rm{kT_{e}}$ = 15.57$^{+12.24}_{-03.90}$ keV.

\noindent {\bf IRAS 04124-0803:} Analysis of the {\it NuSTAR} observations (done in September 2021) on this source is carried out for the first time. From fitting  Model$-$2a to the source spectrum, we obtained best-fit values of $\Gamma$ = 1.66$^{+0.03}_{-0.03}$ and $\rm{kT_{e}}$ = 14.88$^{+03.70}_{-02.57}$ keV.

\noindent {\bf IRAS 09149-6206:} Results on {\it NuSTAR} observations of this source are reported for the first time. This source was observed by {\it NuSTAR} twice between July and August, 2018. We modelled the Comptonized  spectrum (observed on August 2018) and estimated the best-fit value of $\rm{kT_{e}}$ using Model$-$2b. From the model fit to the spectrum we obtained $\Gamma$ = 1.90$^{+0.11}_{-0.09}$ and  $\rm{kT_{e}}$ = 18.09$^{+16.87}_{-04.07}$ keV. 

\noindent {\bf IRAS 05589+2828:} This Seyfert 1 galaxy situated at $z$=0.02940 was observed by {\it NuSTAR} in April 2020. The temperature of the corona of the source is reported for the first time. From the physical model fit to the observed spectrum, we found values of $\Gamma$ = 1.90$^{+0.11}_{-0.07}$ and $\rm{kT_{e}}$ = 42.90$^{+120.46}_{-23.92}$ keV. 

\noindent {\bf Mrk 1148:} This Seyfert 1 galaxy was observed by {\it NuSTAR} in January, 2018. We carried out the spectral analysis using Model$-$2a. The best-fit values obtained using the model fit to the spectrum are $\Gamma$ = 1.86$^{+0.02}_{-0.02}$ and $\rm{kT_{e}}$ = 24.04$^{+19.81}_{-06.76}$ keV. Recently, analysing the same spectrum, both \cite{2022ApJ...927...42K} and \cite{Kang_2022} found values of $\rm{kT_{e}}$ $>$ 18 keV. 

\noindent {\bf Mrk 509:} {\it NuSTAR} observed this source two times between April and June 2015. In this work, we analysed the spectrum taken on April 2015. From fitting Model$-$2c to the observed spectrum we obtained $\Gamma$ = 1.86$^{+0.01}_{-0.02}$ and  $\rm{kT_{e}}$ = 35.78$^{+06.78}_{-05.72}$ keV. On analysis of the same spectrum using {\it relxillCP} model \cite{Kang_2022} reported $\rm{kT_{e}}$ = 24$\pm$2 keV.

\noindent{\bf 2MASXJ18560128+1538059:} This Seyfert 1 galaxy was observed by {\it NuSTAR} in 2017, and from the analysis of the source spectrum using our Model$-$2a, we found $\rm{kT_{e}}$ = 12.32$^{+3.12}_{-2.36}$ keV. Using the same observation ID \cite{refId01} reported $\rm{E_{cut}}$ = 43$^{+20}_{-11}$ keV.

\noindent {\bf PG 0026+129:} {\it NuSTAR} observed the Seyfert 1 galaxy once in January 2021 and results on the analysis of the observation is reported for the first time. From  Model$-$2a fit to the observed spectrum we obtained best-fit values of $\Gamma$ = 1.89$^{+0.01}_{-0.01}$ and $\rm{kT_{e}}$ = 22.18$^{+08.88}_{-04.03}$ keV.

\noindent {\bf SWIFTJ2127.4+5654:} This source  classified as a narrow line Seyfert 1 galaxy, was observed by {\it NuSTAR} nine times between September 2012 and December 2018. We analysed the observations carried out by {\it NuSTAR} in September 2012 as it has the maximum exposure time. By fitting the observed spectrum using Model$-$2a, we obtained $\Gamma$ = 1.96$^{+0.01}_{-0.01}$ and $\rm{kT_{e}}$ = 20.70$^{+03.36}_{-01.94}$ keV. From an analysis of the same spectrum, \cite{2021MNRAS.502...80K} reported  a $\rm{kT_{e}}$ of 21$^{+2}_{-2}$ keV. 

\noindent{\bf IGRJ19378$-$0617:} This source is situated at z=0.0103. It was classified as a Seyfert 1 galaxy, observed six times by {\it NuSTAR} between 2015 and 2022. From fitting the source spectrum using Model$-$2a, we found $\rm{kT_{e}}$ = 49.35$^{+36.94}_{-13.04}$ keV. From the spectral analysis of the source spectrum \cite{2022ApJ...927...42K} reported $\rm{kT_{e}}$ $>$ 122 keV.

\noindent{\bf Fairall 51:} {\it NuSTAR} observed this Seyfert 1 galaxy 4 times between 2018 and 2021. We analysed the {\it NuSTAR} spectrum observed in June 2018. From fitting the source spectrum using Model$-$2a, we found $\rm{kT_{e}}$ = 19.48$^{+6.54}_{-1.83}$ keV.

\noindent{\bf Mrk 279:} This Seyfert 1 galaxy was observed 4 times by {\it NuSTAR} between 2019 and 2020. We analysed the August 2020 spectrum using Model$-$2c and found $\rm{kT_{e}}$ = 16.38$^{+1.72}_{-1.55}$ keV. By analysing the source spectrum taken in October 2019, \cite{Kang_2022} reported a lower limit of $\rm{kT_{e}}$ $>$ 84 keV. 

\noindent{\bf ESO 323$-$G077:} This source is classified as a Seyfert 1.5 galaxy \citep{1992MNRAS.257..677W}, situated at z = 0.0155. {\it NuSTAR} observed this source six times between August 2016 to February 2017. We analyzed January 2017 {\it NuSTAR} data. From the Model$-$2a fit to the source spectrum, we obtained $\rm{kT_{e}}$ = 35.21$^{+13.02}_{-11.89}$ keV. For this source, \cite{2022ApJ...927...42K} reported a lower limit of $\rm{kT_{e}}$ $>$ 34 keV. 

\noindent{\bf 3C 109:} This Seyfert galaxy was observed by {\it NuSTAR} twice in August 2017. We analysed the one with the maximum exposure time. By fitting Model$-$2a to the source spectrum, we found $\rm{kT_{e}}$ = 18.09$^{+6.91}_{-2.72}$ keV.

\noindent{\bf RBS0770:}This source was observed four times between 2012 and 2021 by {\it NuSTAR}. By fitting Model$-$2a to the source spectrum we found $\rm{kT_{e}}$ = 17.71$^{+4.30}_{-2.38}$ keV. From the analysis of the same observation, \cite{2022ApJ...927...42K} reported a lower limit for $\rm{kT_{e}}$ $>$ 24 keV. 

\noindent{\bf CGCG229$-$015:} This nearby Seyfert 1 galaxy was observed once by {\it NuSTAR} on February 2018. From an analysis of the same observation ID \cite{refId01} reported $\rm{E_{cut}}$ = 54$^{+13.02}_{-11.89}$ keV. From the Model$-$2a fit to the source spectrum, we obtained $\rm{kT_{e}}$ = 17.00$^{+41.62}_{-5.61}$ keV.

\noindent{\bf 3C 382:} This Seyfert galaxy was observed 7 times between 2012 and 2016. We analysed the 2013 spectrum and reported $\rm{kT_{e}}$ = 33.07$^{+16.81}_{-7.76}$ keV. From the analysis of the same observation \cite{2020MNRAS.495.3373E} reported $\rm{E_{cut}}$ = 132.75$^{+98.32}_{-39.98}$ keV. 

\noindent{\bf SDSS J114921.52+532013.4:} This Seyfert 1 galaxy was observed once in 2016. From the Model$-$2a fit to the source spectrum, we found $\rm{kT_{e}}$ = 6.50$^{+1.25}_{-0.97}$ keV.


\bibliography{aastex631}{}
\bibliographystyle{aasjournal}



\end{document}